\documentclass[%
 reprint,
nofootinbib,
 amsmath,amssymb,
 aps,
pra,
]{revtex4-2}

\usepackage{xurl}
\usepackage{revtex}
\usepackage{comandi_tesi}

\usepackage{hyperref}
\usepackage[capitalise]{cleveref}
\usepackage{pgfplots}
\usepgfplotslibrary{groupplots}
\usepgfplotslibrary{fillbetween}
\pgfplotsset{compat=newest}

\newcommand{\mus}[0]{\, \si{\micro \second}}
\newcommand{\MHz}[0]{\, \si{\mega \hertz}}
\newcommand{\rad}[0]{\, \si{rad}}
\newcommand{\vtheta}[0]{\boldsymbol{\theta}}
\newcommand{\hvtheta}[0]{\boldsymbol{\widehat{\theta}}}
\newcommand{\wttheta}[0]{\boldsymbol{\widetilde{\theta}}}
\newcommand{\vx}[0]{\boldsymbol{x}}
\newcommand{\vy}[0]{\boldsymbol{y}}

\newcommand{\vtau}[0]{\boldsymbol{\tau}}

\newcommand{\vlambda}[0]{\boldsymbol{\lambda}}

\newcommand{\cov}[0]{\Sigma}

\newcommand{\vpos}[0]{\boldsymbol{r}}
\newcommand{\wt}[0]{\widetilde}

\usetikzlibrary{external}
\begin{document}
\title{Applications of model-aware reinforcement learning in Bayesian quantum metrology}

\author{Federico Belliardo}
\affiliation{NEST, Scuola Normale Superiore, I-56126 Pisa, Italy}
\author{Fabio Zoratti}
\affiliation{Scuola Normale Superiore, I-56126 Pisa, Italy}
\author{Vittorio Giovannetti}
\affiliation{NEST, Scuola Normale Superiore and Istituto Nanoscienze-CNR, I-56126 Pisa, Italy}

\begin{abstract}
	An important practical problem in the field of quantum metrology and sensors is to find the optimal sequences of controls for the quantum probe that realize optimal adaptive estimation. In \href{https://arxiv.org/abs/2312.16985}{\emph{Belliardo et al.}, arXiv:2312.16985 (2023)}, we solved this problem in general, by introducing a procedure capable of optimizing a wide range of tasks in quantum metrology and estimation by combining model-aware reinforcement learning with Bayesian inference. We take a model-based approach to the optimisation where the physics describing the system is explicitly taken into account in the training through automatic differentiation. In this follow-up paper we present some applications of the framework. The first family of examples concerns the estimation of magnetic fields, hyperfine interactions, and decoherence times for electronic spins in diamond. For these examples, we perform multiple Ramsey measurements on the spin. The second family of applications concerns the estimation of phases and coherent states on photonic circuits, without squeezing elements, where the bosonic lines are measured by photon counters. This exposition showcases the broad applicability of the method, which has been implemented in the qsensoropt library released on PyPI, which can be installed with pip.
\end{abstract}

\maketitle

\section{Introduction}
\label{sec:introduction}
In recent years, the intersection of quantum mechanics and machine learning has become a focal point of exploration, with a particular emphasis on leveraging quantum technologies for various applications. This convergence holds immense potential for mutual enhancement across both domains. Quantum technologies, notably quantum computers, offer unique capabilities to tackle classical machine learning challenges, such as classification and sampling, utilizing both classical and quantum data sources~\cite{flamini_photonic_2020, broughton_tensorflow_2021, bergholm_pennylane_2022}.

Conversely, traditional machine learning methodologies can enhance quantum information tasks, including state preparation~\cite{bukov_reinforcement_2018, zhang_when_2019, niu_universal_2019, porotti_deep_2022}, optimal quantum feedback~\cite{porotti_gradient-ascent_2023}, error correction~\cite{fosel_reinforcement_2018}, device calibration~\cite{cimini_calibration_2019, ban_neural-network-based_2021, nolan_machine_2021, nolan_frequentist_2021}, characterization~\cite{nguyen_deep_2021}, and quantum tomography~\cite{palmieri_experimental_2020, quek_adaptive_2021, hsieh_direct_2022}. This work aligns with the latter category, employing model-aware reinforcement learning (RL) to derive optimized adaptive and non-adaptive control strategies for quantum metrology and estimation tasks~\cite{marquardt_machine_2021, marquardt_online_2021, krenn_artificial_2023, porotti_gradient-ascent_2023}.

Specifically, we address the challenge of optimal experimental design~\cite{fisher_design_1935}, a task already explored using machine learning techniques~\cite{foster_variational_2021, baydin_toward_2021, ballard_machine_2021, ivanova_implicit_2021, foster_deep_2021}, which we explore in the quantum realm, finding performances going beyond the current state-of-the-art~\cite{ fiderer_neural-network_2021}. Estimation processes in this context involve numerous non-differentiable steps, such as simulating measurements and resampling from posterior distributions, posing challenges to the application of model-aware RL. To overcome these obstacles, our approach incorporates techniques like importance sampling, adding log-likelihood to the loss~\cite{porotti_gradient-ascent_2023}, the reparametrization trick, and the Scibior and Wood correction~\cite{scibior_differentiable_2021}.

The methodology involves identifying tunable parameters in a given physical platform and metrological task, allowing an agent (which can be a neural network, a decision tree, or a list of trainable controls) to learn an optimal policy through gradient descent optimization. We have abstracted and packaged this procedure into the qsensoropt library, which is available on GitLab~\cite{qsensoropt_gitlab}, together with the online documentation of the classes and the examples of this paper~\cite{qsensoropt_doc}. The library can also be installed with pip on every machine without the need for cloning the environment, being it available on PyPI. This is a versatile tool for finding adaptive policies to optimize the precision of quantum sensor, which we think will be of much use for the quantum information community. By exploiting the classes and function preprogrammed in this library, users can implement the quantum mechanical model of their sensors, and optimise an adaptive or non-adaptive policy, which can be later exported and implemented in the experiments.

Demonstrating the broad applicability of our approach, we optimized examples on the nitrogen-vacancy (NV) center platform for various metrology tasks, including DC and AC magnetometry, decoherence estimation, and hyperfine coupling characterization. In the realm of photonic circuits, we explored applications such as multiphase discrimination, the agnostic Dolinar receiver~\cite{zoratti_agnostic-dolinar_2021}, and coherent states classification. Our findings showcase the superiority of model-aware RL over traditional control strategies, even outperforming model-free RL in multiple scenarios. Through this work we lay the tools to accelerate the search for optimal control policies in quantum sensing and metrology, potentially expediting the widespread industrial application of this quantum technology.

The existing literature encompasses various works addressing challenges akin to our approach, which we have categorized into four classes according to their relation with our work. The first class encompasses the competitor approaches for optimization in quantum metrology~\cite{meyer_variational_2021, zhang_quanestimation_2022, granade_qinfer_2017, mcmichael_optbayesexpt_2021, bavaresco_designing_2023}. These lack coverage of non-greedy or adaptive policies of Bayesian estimation. The second class contains those optimal control algorithms based on the optimization of the Fisher information~\cite{liu_quantum_2017, xu_generalizable_2019, rembold_introduction_2020, schuff_improving_2020, xu_generalizable_2021, liu_optimal_2022, xiao_parameter_2022, qiu_efficient_2022}, which again often lack coverage of Bayesian estimation, adaptivity through the use of neural network, or can be applied only to NV centers. The third class design a theoretical approach to optimal control in quantum sensing, but lack an implementation~\cite{baydin_toward_2021, ballard_machine_2021, krenn_artificial_2023, vedaie_framework_2023, gebhart_learning_2023}. The fourth and last class applies variational quantum circuits to specific metrological tasks~\cite{ma_adaptive_2021, kaubruegger_quantum_2021, marciniak_optimal_2022, kaubruegger_optimal_2023, kose_superresolution_2023, heras_photonic_2023, yang_variational_2022}

In this work, we give an overview of the main components of the framework that allows quantum metrology tasks to be optimized with machine learning, with several examples of interest where we achieved results that are comparable to or better than the current state of the art for each task. However, we refer to~\cite{belliardo_model-aware_2023} for a complete description of the method and a review of the literature. See also~\cite{belliardo_application_2024} for a three-pages explanation of the theory with a single example. Although we have also implemented the optimization of frequentist estimation, based on the Fisher information, the results present in this paper concern exclusively the domain of Bayesian estimation.

\subsection{Review of the framework}
\label{subsec:review_framework}
This section is meant to be a quick review of the various components of the qsensoropt framework used throughout the paper. For a rigorous treatment of the theory behind, please refer to~\cite{belliardo_model-aware_2023}. See also~\cref{sec:dynamical_model} for a more detailed review of the quantum model of a sensor, and how it is used in Bayesian estimation with different degrees of control. This section defines a common mathematical schematization to quantum metrology, in which we fit all the examples of this paper. If an optimization problem can be expressed in terms of this scheme, then it can be optimized with the qsensoropt library.

\textbf{The sensor's model:} Some elements of the Bayesian inference process in quantum metrology are universal, in the sense that they do not depend on the particular task. These are the Bayesian filtering, the repeated execution of the measurements, and the training of the agent. Together with other general aspects of the optimization routine, these are implemented in the library. To apply the framework the user needs to code the model of the specific quantum sensor of interest. This model should simulate the stochastic outcome extraction of the measurements and evaluate the probability of observing a specific outcome for the particular quantum system at hand. In this description, it should be stated which parameters are controls, that can be tuned by the experimenter, and which are the target of the estimation. These last ones, which will be indicated collectively by the symbol $\vtheta$, are unknowns related to the environment of the probe, which are codified during the probe's evolution, or they are parameters of the state of the probe if it has been encoded in a distant laboratory, on which we have no control. This encoding is formalized later within the object $\Phi_{\vtheta}$, defined in \cref{sec:dynamical_model}.

\begin{figure}[htb]
	\centering
	\includesvg[width=0.45\textwidth]{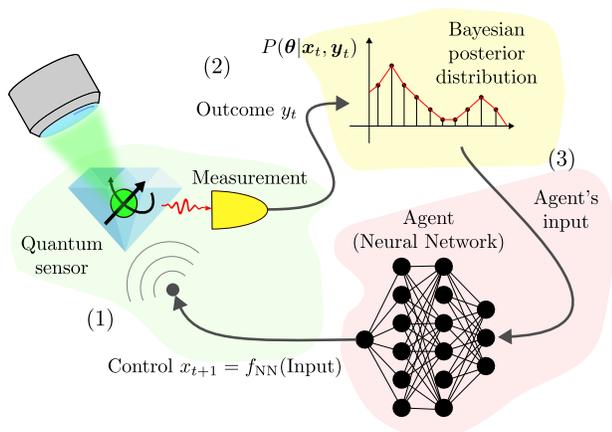}
	\caption{
		This general scheme illustrates the information flow within the measurement loop for a quantum metrology task where the quantum probe is an NV center. Refer to \cref{subsec:nv_center_pysics} for a brief description of the physics of such systems. The environment we aim to study interacts with the quantum probe and encodes it with the unknown variables $\vtheta$. This probe is then measured using a tunable instrument (1). The outcome of this measurement provides us with information about the probe's state and in turn about the environment's variables. This information is used to update the posterior Bayesian distribution on $\vtheta$ (2). Some summary information derived from the posterior is then input into an agent that decides the new control parameters for the measurement in the next iteration of the loop (3). This control is then realized through the electronics of the experiment. In this picture, the agent is a neural network.}\label{fig:pipeline}
\end{figure}

\textbf{The measurements loop:} A quantum metrology or estimation task usually involves many measurements indexed by $t$, through which the information on the target parameters is accumulated step-by-step in the Bayesian posterior. The sequence of measurements is organized in a ``measurement loop'', which is represented in \cref{fig:pipeline} and it involves three fundamental steps. An iteration of the loop starts with the simulation of the evolution and the measurement of the sensor's probe, all done thanks to the model implemented by the user. We indicate with $y_t$ the outcome of the $t$-th measurement and with $\vy_t := (y_0, y_1, \ldots, y_t)$ the list of outcomes until the $t$-th iteration (included). The second step in the loop's iteration is the processing of the outcome, which allows the internal representation of the Bayesian posterior to be updated. This is done via a particle filter, which uses an ensemble of points in the parameter space (called particles) and weights to approximate the posterior, see~\cite{belliardo_model-aware_2023, belliardo_optimizing_2022} for more details. The initial distribution is the prior $\pi(\vtheta)$, which is uniform over some interval in every application presented in this paper. At the third and last step of the iteration, some information on the state of the estimation and on the posterior is computed and fed to an agent which evaluates the controls for the next measurement on the probe, i.e. for the next iteration of the loop. We indicate with $x_t$ the controls of the $t+1$-th measurement and with $\vx_t := (x_0, x_1, \ldots, x_t)$ the list of controls until iteration $t$ (included). This agent doing the controlling will typically be a neural network (NN). This agent can leverage knowledge from past measurements to optimize the estimation task's overall performance, which makes the control strategy adaptive and non-greedy. The posterior on $\vtheta$ at the $t$-th iteration of the loop is conditioned on the full list of previous controls and outcomes, and it will be indicated with $P(\vtheta|\vx_t, \vy_t)$. Typically, during the training, a number $B>1$ of simulations of the estimation are executed in parallel with the same agent in a batch; we call $B$ the batch size. From this batch, the statistical properties of the strategy produced by the agent are estimated, like a precision figure of merit specified by the user, e.g. the mean square error or the error probability.
\begin{figure}[htb]
	\centering
	\includegraphics[width=0.45\textwidth]{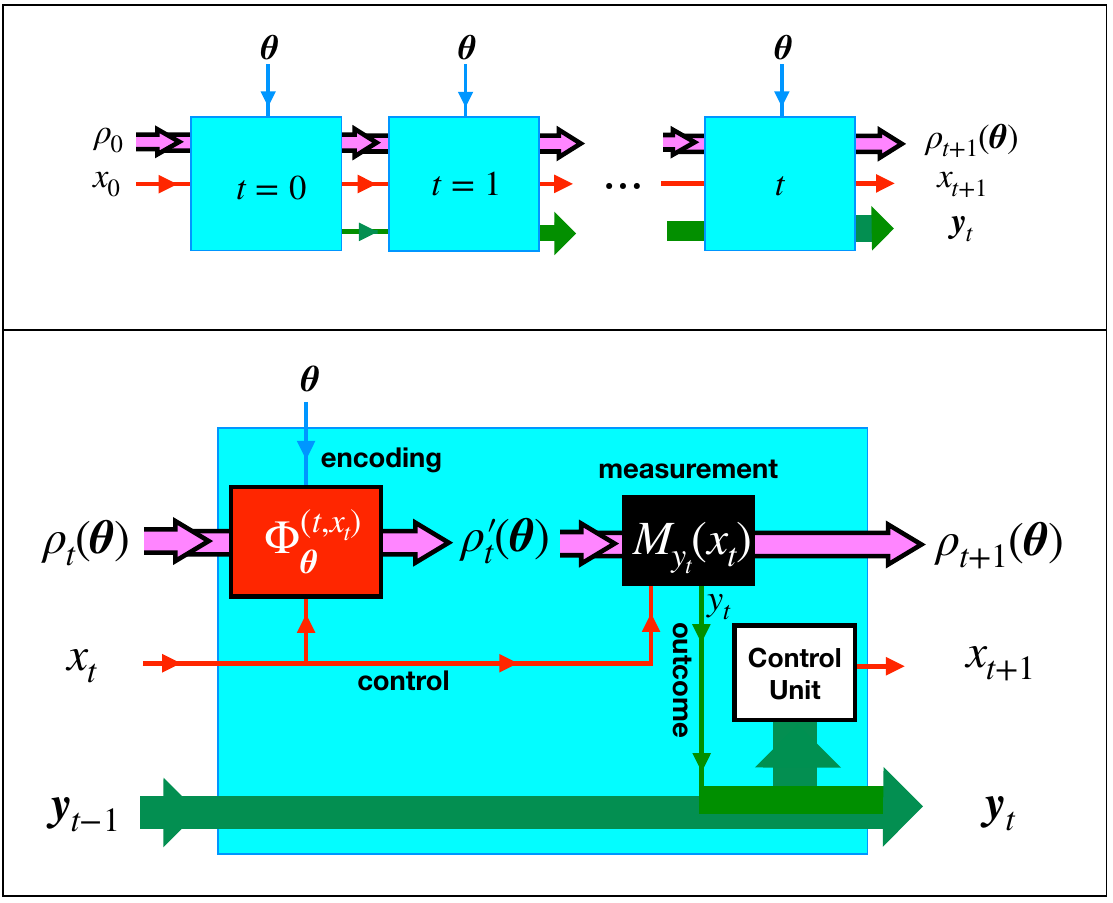}
	\caption{(Top panel) The measurement loop of Fig.~\ref{fig:pipeline} represented as an ordered sequence of concatenated events (cyan boxes). The violet arrows describe the temporal evolution of the probe state, the red lines the controls, and the green arrow the measurement outcomes that accumulate along the way. (Bottom panel) Schematic description of the processes involved in the $t$-th event: first the input state $\rho_t(\vtheta)$ of the probe undergoes to a possible new encoding stage of the parameters $\vtheta$ via the action of the LCPT map $\Phi_{\vtheta}^{(t,x_t)}$; then the transformed state $\rho'_t(\vtheta)$ gets measured producing the outcome $y_t$ and the probe emerges as the conditional state $\rho_{t+1}(\vtheta)$. As indicated by the figure all these operations are in part determined by the inputted control $x_t$. The new value of the parameter $x_{t+1}$ is determined by the control unite element elaborating the newly acquired outcome $y_t$. To connect with the scheme in \cref{fig:pipeline} observe that the control unit here encompasses both the particle filter updating the Bayesian distribution and the neural network.}
	\label{fig:event_representation}
\end{figure}

\textbf{The precision-resources paradigm:} Each iteration of the measurement loop consumes a certain amount of a specific ``resource'' such as the estimation time or the available probes. Once these resources are depleted, the measurement loop concludes. For each application, the resource must be specified, and it is as important as the definition of the precision figure of merit. The goal of the machine learning framework is to optimize the precision for a fixed maximum amount of consumed resources. In the simplest case, the resource is just the number of measurements executed in the whole estimation, which we indicate with $M\ped{max}$, which is also the number of total iterations of the measurement loop.

\textbf{Overview of the training process:} After the measurement loop ends, an estimator $\hvtheta$ for the parameters $\vtheta$ is computed from from the Bayesian posterior. The user-defined precision metric serves as the training loss to be minimized through a stochastic gradient descent procedure, which acts on the trainable variables of the agent, e.g. the biases and weights in the case of a NN, which are indicated collectively with the symbol $\vlambda$. The gradient of the loss is computed via automatic differentiation, where the derivatives flow through the sampling of stochastic variables, like the measurement outcomes, and through the physical model of the sensor. This characterizes the training as model-aware policy gradient reinforcement learning and special precautions are taken to compute an unbiased estimator of the gradient~\cite{porotti_gradient-ascent_2023, belliardo_model-aware_2023}. Each training step involves a complete execution of the measurement loop until the resources are depleted. After multiple training steps, in the order of $\mathcal{O}(10^4)$, the precision reaches a plateau and the training is stopped. At this point, the agent has been optimized. In all the applications of this paper Adam~\cite{kingma_adam_2015} has been used as an optimizer. The learning rate is set to decrease at each training step, with an in inverse square root decay law, as specified in~\cite{belliardo_achieving_2020}. The initial learning rate $\alpha_0$ has been chosen optimally for each simulation. through trial and error.

\subsection{Specifics of each experiment}
\label{subsec:user_implemented}
From the description of this machine learning approach, it is clear that in order to apply this technique on a particular sensor four pieces of information must be specified and implemented in the code by the user. These are listed in the following.
\begin{itemize}
	\item The sensor's model, represented by the code that defines the probability of each potential outcome in a measurement. This model is typically derived from the Born rule applied to the physics of the sensor, necessitating the specification of a division between control and target parameters. Further insights on how to schematize the quantum mechanics behind this aspect are detailed in \cref{sec:dynamical_model}.
	\item The nature and the amount of resources consumed in the metrological task.
	\item The input to the agent, which must be a function of the posterior distribution, and in general of the measurement outcomes for the implementation of an adaptive control policy.
	\item The error figure of merit used to gauge the precision of the estimation, such as the mean square error.
\end{itemize}

\subsection{Selection of the agent}
\label{subsec:selection of the agent}
In all the adaptive experiments the NN has by default $5$ hidden layers with $64$ neurons each, and the activation function is $\tanh$. For the non-adaptive strategy, the controls for each measurement step do non depend on the Bayesian posterior. In this case, the agent can still be a NN with reduced input, or instead, we directly optimize the values of the controls in the training by setting $\vlambda := \vx_t$. In \cref{subsec:qml_classifier} we experimented with using a decision tree to compute the controls. This means that the posterior is bypassed and the adaptive controls are directly selected based on the list of past measurement outcomes $\vy_t$. More precisely the outcome selects the tree branch at each node.

\subsection{Dynamical model} \label{sec:dynamical_model}
In this section, we present a rigorous characterization of the measurement loop of \cref{fig:pipeline}. To start with, let us observe that the whole procedure can be represented as an ordered sequence of concatenated events describing the information acquired by the agent at the various stages and the associated evolution of the probe state (see top panel of \cref{fig:event_representation}). In particular, the event $t$ takes as input the state of the probe at the beginning of the $t$-th measurement loop (orange), the control value $x_t$ synthesized by the agent in the previous event (red line), the complete list of the outcomes $\vy_{t-1}=(y_0,y_1, \cdots, y_{t-1})$ of the measurements performed up to that point, as well as a possible new encoding of the parameters $\vtheta$ into the probe state. As shown in the bottom panel of \cref{fig:event_representation}, a characterization of $t$-th event is obtained by assigning the following three elements:
\begin{itemize}
	\item a Completely Positive, Trace Preserving ( LCPT) linear map~\cite{nielsen00,Holevo_book} $\Phi_{\vtheta}^{(t,x_t)}$, describing the physical process which encodes the parameters $\vtheta$ into the state of the probe at the beginning of the event (red box of the figure); \item a collection of operators $\{ {M}_{y_t}{(x_t)}\}_{y_t}$ fulfilling the normalization condition $\sum_{y_t} {E}_{y_t}(x_t)=\mathds{1}$, ${E}_{y_t}(x_t):= {M}_{y_t}^{\dag}(x_t) {M}_{y_t}(x_t)$, describing the measurement process (black box);
	\item a control unit (white box).
\end{itemize}
This is effectively a different way to divide the iteration, more centred on the quantum mechanical evolution with respect to~\cref{fig:pipeline} and the previous section, which was more centred on the machine learning side and the flow of information. It is important to connect this two different representation, and this can be done by observing that the first two element in the above list, i.e the LCPT map and the measurement, are part of the sensor's model, while the Bayesian update and the NN are part of the last element, i.e. the control unit. The inclusion of the index $t$ in the definition of $\Phi_{\vtheta}^{(t,x_t)}$ allows for the analysis of non-uniform models where the encoding mechanism varies along the measurement loop. For instance, setting $\Phi_{\vtheta}^{(t,x_t)}$ as the identity map $\mathds{1}$ for $t\geq 1$ represents models where $\vtheta$ is imprinted on the probe only at the beginning of the first measurement loop. The explicit inclusion of the parameter $x_t$ in the definition of $\{ {M}_{y_t}{(x_t)}\}_{y_t}$ and $\Phi_{\vtheta}^{(t,x_t)}$ accommodates instead models where the agent exerts control over the measurement and possibly on the encoding mechanism (e.g. determining factors such as the duration of the probe's interaction with the external system that is responsible for the effect). Define now ${\rho}_{t}(\vtheta):= {\rho}(\vtheta, \vx_{t-1},\vy_{t-1})$ the density matrix of the probe at the beginning of the $t$-th event (i.e. the state that emerges as output from $({t-1})$-th event). According to the model, such configuration is first evolved via $\Phi_{\vtheta}^{(t,x_t)}$ through the mapping
\begin{equation}
	{\rho}_{t}(\vtheta) \mapsto {\rho}'_{t}(\vtheta):= \Phi_{\vtheta}^{(t,x_t)}[{\rho}_{t}(\vtheta)] \; ,
	\label{eq:mapsto}
\end{equation}
producing the state ${\rho}'_{t}(\vtheta) := {\rho}'(\vtheta, \vx_t,\vy_{t-1})$ that undergoes to the measurement process defined by the operators $\{ {M}_{y_t}{(x_t)}\}_{y_t}$. The probability of outcome $y_t$ follows the Born rule:
\begin{equation}
	P(y_t| \vtheta,\vx_t, \vy_{t-1}):= \mbox{Tr}[ {E}_{y_t}(x_t) {\rho}'_{t}(\vtheta)] \; .
	\label{eq:bornrule}
\end{equation}
The conditional probe state ${\rho}_{t+1}(\vtheta):= {\rho}(\vtheta, \vx_t,\vy_{t})$ after measurement is determined instead by the formula:
\begin{equation}
	{\rho}_{t+1}(\vtheta):= \frac{{M}_{y_t}(x_t) {\rho}'_{t}(\vtheta) {M}^\dag_{y_t}(x_t)}{P(y_t| \vtheta,\vx_t, \vy_{t-1})} \; .
	\label{eq:stateafterm}
\end{equation}
\cref{eq:mapsto} and \cref{eq:stateafterm} define the update of the input probe density matrix from the event $t$ to event $t+1$. The update of the control parameter $x_{t+1}$ is instead determined by the classical data processing unit of the model that uses $\vy_t$ as input information. An essential ingredient of such a procedure is the Bayesian inference formula,
\begin{equation}
	P( \vtheta|\vx_t, \vy_{t}) := \frac{P(y_t| \vtheta,\vx_t, \vy_{t-1})P( \vtheta|\vx_{t-1}, \vy_{t-1})}{\sum_{\vtheta'}
	P(y_t| \vtheta',\vx_t, \vy_{t-1})P( \vtheta'|\vx_{t-1}, \vy_{t-1})} \; ,
	\label{eq:bayesian}
\end{equation}
which allows one to update the posterior probability $P( \vtheta|\vx_{t-1}, \vy_{t-1})$ acquired at the beginning of the $t$ event to the posterior probability $P( \vtheta|\vx_t, \vy_{t})$ of the next event.

\section{Platforms for quantum metrology}
\label{sec:intro_platforms}

\subsection{Experiments on the NV center}
\label{subsec:nv_center_pysics}
The first family of examples presented in \cref{sec:nvcenter_platform} are applications on the platform of single electronic spins in diamond. The nitrogen-vacancy (NV) center is a point defect of the diamond crystal that allows for the initialization, detection, and manipulation of its electronic spin. It exhibits an exceptionally long quantum coherence time, maintaining this property even at room temperature. Consequently, it has found applications in areas such as magnetometry, thermometry, and stress sensing~\cite{doherty_quantum_2022,chen_quantum_2018, maze_rios_quantum_2010, barry_sensitivity_2020}. The $e^{-}$ spin of the NV center is a two-level system, and the Ramsey measurement, upon which the examples in this paper are based, involves the initialization of such spin in the coherent state $\ket{\psi} := (\ket{\uparrow}+\ket{\downarrow})/\sqrt{2}$ with a $\pi/2$ microwave (MW) pulse, where $\ket{\uparrow}$/$\ket{\downarrow}$ is the spin-up/spin-down state. The spin is then left to freely evolve and interact with the environment, in order to encode the target parameters. The duration $\tau$ of the time interval of free evolution is the control parameter in this platform, changed from measurement to measurement. A second $\pi/2$ MW pulse closes the encoding stage. When the NV center is excited with green light, according to the state of the $e^{-}$ spin, it has different probabilities of decaying through a radiative or a non-radiative path, which means that the number of photoluminescence photons emitted is different for the two states of the spin. This mechanism allows a reliable direct measurement of the spin even at room temperature. We can also tune an extra phase $\varphi$ of the spin evolution through the MW pulse. There are two possible choices for the resources on this platform, either we fix the maximum number of Ramsey measurements $M\ped{max}$, or we fix the maximum total free evolution time $T = \sum_{t=0}^{M\ped{max}-1} \tau_t$, which is the sum of the evolution times in each Ramsey measurement. In this second case, the measurement number $M$ is a stochastic variable. We demonstrated the applicability of our machine learning methods for single and multiparameter metrology on various estimation tasks on NV centers including both DC~\cite{fiderer_neural-network_2021} and AC magnetometry, decoherence estimation~\cite{arshad_real-time_2024}, and the characterization of the hyperfine coupling with a ${}^{13} C$ nucleus~\cite{joas_online_2021}.

\subsection{Experiments with photonic circuits}
\label{subsec:photonic_circuits_physics}
The second family of examples we study is based on photonic circuits~\cite{valeri_experimental_2020, paesani_experimental_2017, polino_experimental_2019, polino_photonic_2020, barbieri_quantum_2013}. The systems we simulated can all be realized with lasers as photon sources and passive elements, like phase plates and beam splitters (BS), together with number resolving photon counters. We will assume to have programmable elements, which means that we can dynamically change the values of the phase imprinted by a plat and of the transmissivity of a BS on a time scale faster than the time interval between the measurements. This is necessary if we want to have adaptivity in the estimation. In the examples of this paper, we range from using a single bosonic mode system to controlling ten modes. The resources for this class of experiments are either the number of input states or the number of photons in a signal. We studied multiphase estimation, the agnostic Dolinar receiver~\cite{zoratti_agnostic-dolinar_2021}, and coherent states classification, both in the case where the states are classically known and in the case they must be learned from a quantum training set. We have avoided the use of active elements in all these experiments, that would generate single or multimode squeezing. We also avoided the use of other non-classical states of light like Fock states, which would require an implementation of the physics of the sensor that goes beyond that of Gaussian systems. This is a limitation only of the presented applications, and nothing prevents our framework from being useful also for those systems.

\section{Applications on the NV center platform}
\label{sec:nvcenter_platform}
All the NV center applications share the same input to the agent, independently of the nature or the number $d$ of parameters to estimate. The construction of this input is presented in \cref{subsec:input_nn}. Similarly all these application share also the same loss~\cref{subsec:loss_nv_center}.

\subsection{Input to the neural network}
\label{subsec:input_nn}
The input to the NN is obtained by concatenating, at each iteration of the measurement loop, the estimators for the unknown parameters, their standard deviations, their correlation matrix, the total number of consumed resources up to that point, and the index of the measurement iteration. All these variables are rescaled to make them fit in the $[-1, 1]$ interval, which makes them more suitable to be the inputs of a NN. More precisely the input at the $t$-th measurement step is composed of the following elements.
\begin{itemize}
	\item Mean of the posterior given by
	\begin{equation}
		\hvtheta_t := \int \vtheta P(\vtheta| \vx_t, \vy_t) \; ,
		\label{eq:mean_pf}
	\end{equation}
	normalized to lay in the interval $[-1, 1]$, which is possible since the prior is uniform with known extrema. These inputs are $d$ scalars, where $d$ is the number of parameters to estimate, and will be indicated with the symbol $\wttheta_t$ in the following.
	\item Standard deviations around the mean for each parameter computed from the Bayesian posterior distribution. Given the covariance matrix $\cov_t$ defined as
	\begin{equation}
		\cov_t := \int (\vtheta - \hvtheta_t) (\vtheta - \hvtheta_t)^\intercal P(\vtheta|\vx_t, \vy_t) \dd \vtheta \; ,
		\label{eq:Sigma_pf}
	\end{equation}
	the next $d$ inputs to the NN are given by the vector $\wt{\boldsymbol{\sigma}}_t$, with entries
	\begin{equation}
		\wt{\sigma}_{t, j} := -\frac{2}{10} \ln \sqrt{\cov_{t, jj}} - 1 \; ,
	\end{equation}
	being $\sqrt{\cov_{t, jj}}$ the said standard deviations. This time, since we do not know in advance the admissible values for the covariance matrix, we cannot cast the standard deviation exactly in $[-1, 1]$, but we can do it approximately for standard deviations in the range $(10^{-5}, 1)$, through the above formula. These inputs are $d$ scalars.
	\item Correlation matrix $\chi_t$ between the parameters, computed as
	\begin{equation}
		\chi_{t, ij} := \frac{\cov_{t, ij}}{\sqrt{\cov_{t, ii} \cov_{t, jj}}} \; .
	\end{equation}
	This matrix doesn't need the normalization, since its entries are already in the interval $[-1, 1]$. The matrix $\chi_{t}$ is flattened and each entry is added to the input of the NN. These are $d^2$ further scalars.
	\item The iteration index of the measurement loop $t$ normalized in $[-1, 1]$, according to the maximum number of measurement steps $M\ped{max}$, fixed beforehand. This input is a single scalar, indicated with $\wt{t} := 2t/M\ped{max} - 1$.
	\item The amount of consumed resources $R_t$ normalized in $[-1, 1]$, according to the maximum amount of resources $R$, also fixed beforehand. This is a single scalar indicated with $\wt{R}_t := 2 R_t/R -1$. For the NV center applications the resources are either the number of measurements, i.e. $R_t := t$ and $R:=M\ped{max}$, or the total free evolution time, i.e. $R_t := \sum_{k=0}^t \tau_k$ and $R:=T\ped{max}$.
\end{itemize}
The total length of the NN input as a function of the number of parameters is $n_i := d^2+2d+2$. If the NN is used for the non-adaptive strategy it receives in input only the two scalars $\wt{R}_t$ and $\wt{t}$, this is the case for the decoherence estimation in \cref{subsec:nv_center_dec}.

\subsection{Definition of the precision figure of merit}
\label{subsec:loss_nv_center}
The error on the estimation task for the NV center is
\begin{equation}
	\mathcal{L} (\vlambda) := \tr [ G \cdot K (\vlambda)] \; ,
	\label{eq:loss_G_stateless}
\end{equation}
where $K (\vlambda)$ is the mean error matrix of the estimator $\hvtheta$ on the batch of $B$ parallel simulations, i.e.
\begin{equation}
	K(\vlambda) := \sum_{k=1}^B (\hvtheta - \vtheta) (\hvtheta - \vtheta)^{\intercal} \; ,
\end{equation}
$G \ge 0$ is the weight matrix used to obtain a scalar error for the multiparameter metrological problem, and $\vlambda$ are the trainable variables of the agent. The estimator $\hvtheta$ is the mean of the posterior defined in \cref{eq:mean_pf} but $K(\vlambda)$ ought not to be confused with the covariance $\Sigma_t$ defined in \cref{eq:Sigma_pf}. The latter refers to a single estimation, the former to multiples; in other words $K(\vlambda)$ is the empirical dispersion of the estimator, computed through multiple experiments, while $\cov_t$ is the uncertainty on $\hvtheta_t$ of a single estimation. The mean error in \cref{eq:loss_G_stateless} can be expanded as
\begin{equation}
	\mathcal{L} (\vlambda) = \sum_{k=1}^B \tr [G \cdot (\hvtheta_k - \vtheta_k) (\hvtheta_k - \vtheta_k)^\intercal ] \; ,
\end{equation}
from which it is clear that the error $\ell (\hvtheta_k, \vtheta_k)$ for each single estimation in the batch should be:
\begin{equation}
	\ell (\hvtheta_k, \vtheta_k) := \tr [G \cdot (\hvtheta_k - \vtheta_k) (\hvtheta_k - \vtheta_k)^\intercal ] \; ,
\end{equation}
where the subscript $k$ is the index of the simulation inside the batch. For those examples with a fixed maximum amount of free evolution time $T_{\text{max}}$ we use the cumulative loss as a figure of merit for the precision in the training:
\begin{equation}
	\mathcal{L}\ped{cum}(\vlambda) := \frac{1}{M\ped{max} B} \sum_{t=0}^{M\ped{max}-1} \sum_{k=1}^B \frac{\ell (\hvtheta_{k, t}, \vtheta_k)}{\eta (\vtheta_k, T_{t, k})} \; ,
	\label{eq:all_time_loss_h}
\end{equation}
with normalizing factor $\eta (\vtheta_k, T_k)$ given by
\begin{equation}
	\eta (\vtheta_k, T_{t, k}) := \min \left( \sum_{i=1}^d G_{ii} \frac{(b_i-a_i)}{12}, \frac{1}{T_{t, k}} \right) \; ,
\end{equation}
where the sum is over the $d$ parameters $\vtheta$, being $(a_i, b_i)$ the extrema of the uniform prior on the $i$-th parameter, and $G_{ii}$ the $i$-th diagonal entry of the weight matrix. The quantity $T_{t, k}$ is the total elapsed evolution time, i.e.
\begin{equation}
	T_{t, k} := \sum_{m=0}^t \tau_{m, k} \; ,
\end{equation}
which plays the role of the amount of consumed resources and can be different across the batch of estimations. The quantity $\tau_{t, k}$ is the evolution time at the iteration index $t$, for the $k$-th instance of the estimation in the batch. The figure of merit $\mathcal{L}\ped{cum}$ is designed to take into account also the precision of the intermediate results, and not only the error of $\hvtheta$ at the end of the estimation. For those examples referring to a measurement-limited estimation we use the logarithm loss, i.e.
\begin{equation}
	\mathcal{L}\ped{log}(\vlambda) := \frac{1}{M\ped{max}} \sum_{t=0}^{M\ped{max}-1} \log \left[ \frac{1}{B} \sum_{k=1}^B \ell (\hvtheta_{k, t}, \vtheta_k) \right] \; ,
	\label{eq:log_loss}
\end{equation}
which is a version of the cumulative loss that doesn't require the normalizing factor $\eta$ and has been found to work better for measurement-limited estimations.

\subsection{DC magnetometry with time and phase control}
\label{subsec:nv_center_dc_phase}

\subsubsection{Description of the task}
The NV center electron spin is sensitive to magnetic fields; for example, static fields determine the electron Larmour frequency, which can be measured as an accumulated phase by a Ramsey experiment. The spin projection measurement that follows has a binary outcome according to the selected spin state. Indicating with $\pm 1$ these two outcomes their probabilities are
\begin{align}
	p(\pm 1|\omega, T_2, \tau) := \frac{1}{2} \pm \frac{1}{2} e^{-\tau/T_2} \cos \left(\omega \tau + \varphi \right) \; .
	\label{eq:nv_center_model_dc}
\end{align}
This is the quantum sensor's model that has to be hard coded for the RL framework to be applicable.
We can easily identify the theoretical description of the model using the convention defined in \cref{sec:dynamical_model}. The initial state for the system is the state $\ketbra{+}{+}$, while the physical map $\Phi_{\vtheta}^{(t,x_t)}$ is the same at each step and can be divided in the usual evolution of a spin under the action of a magnetic field, with Hamiltonian
\begin{equation}
	\hat{H} := \frac{\hbar \omega }{2} \hat{\sigma}_z \; ,
\end{equation}
followed by a $\varphi$ phase rotation, and a dephasing of the state. the phase $\omega := \gamma B$ represents the unknown precession frequency to be estimated, which is proportional to the static magnetic field $B$ with $\gamma \simeq 28 \MHz/\text{mT}$. The unitary component of the evolution is
\begin{equation}
	\hat{U}(\tau, \varphi) := \exp\left[-i \left( \frac{\hat{H}\tau}{\hbar} + \frac{\varphi}{2} \sigma_z \right) \right] \; ,
\end{equation}
and we define the action on the state as
\begin{equation}
	\mathcal U_{\tau, \varphi} (\rrho) := \hat{U} (\tau, \varphi) \rrho \hat{U}  (\tau, \varphi)^\dag \; ,
\end{equation}
The action of the dephasing integrated on the evolution time $\tau$ can be described by a depolarizing dissipative term that is written in the form of an LCPT map
\begin{equation}
	\varPhi_\tau (\rrho) := \rrho e^{-\tau/T_2} + \frac{\id}{2} (1 - e^{-\tau/T_2}) \; ,
\end{equation}
so that the map $\Phi_{\vtheta}^{(t, x_t)}$ defined in the breakdown of \cref{sec:dynamical_model} is
\begin{equation}
	\Phi_{\vtheta}^{(t, x_t)} := \varPhi_{\tau_t} \circ\, \mathcal{U}_{\tau_t, \varphi_t} \; ,
\end{equation}
with $x_t := (\tau_t, \varphi_t)$. The free evolution time $\tau$ and the phase $\varphi$ are controlled by the trainable agent, while $\omega$ is the unknown parameter to be estimated. The parameter $T_2$ denotes the transverse relaxation time, serving as the time scale for the dephasing induced by magnetic noise. Mostly this is caused by the ${}^{13}C$ in the diamond lattice. In some of the examples of this section we perform multiparameter estimation, indeed the transverse relaxation time $T_2$ may or may not be an unknown in the estimation. The prior on the frequency $\omega$ is uniform in $(0, 1) \MHz$. In the simulations with unknown $T_2^{-1}$, a narrow prior, uniform in $T_2^{-1} \in (0.09, 0.11) \MHz$, was chosen. The optimization of the NV center as a magnetometer has been extensively studied in the literature with analytical tools~\cite{schmitt_optimal_2021,ferrie_how_2013}, with numerics~\cite{dushenko_sequential_2020,mcmichael_sequential_2021,granade_robust_2012,oshnik_robust_2022,craigie_resource-efficient_2021,bonato_optimized_2016,santagati_magnetic-field_2019,zohar_real-time_2023,nusran_high-dynamic-range_2012,wang_experimental_2017,dinani_bayesian_2019,bonato_adaptive_2017,ferrie_adaptive_2012}, and with Machine Learning~\cite{liu_repetitive_2020,fiderer_neural-network_2021,tsukamoto_machine-learning-enhanced_2022}, but with the model-aware RL implemented by qsensoropt, we were able to outperform these works, and redefine the state-of-the-art for the performances of DC magnetometry with NV centers.

\subsubsection{Discussion of the results}
\begin{figure*}[th]
	\centering
	\includegraphics[scale=1.0]{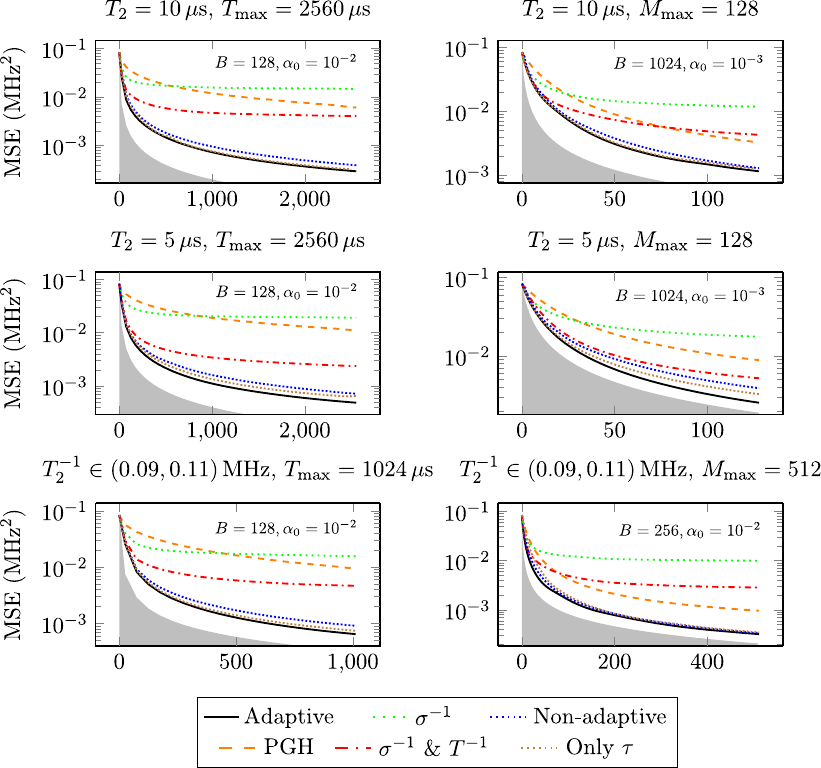}
	\caption{The performances of the optimized adaptive and non-adaptive strategies are reported in this plots for DC magnetometry on NV center with time and phase control, together with the same strategies commonly used in the literature discussed in \cref{subsec:nv_center_dc_phase}. Along with the NN optimized to control $\tau$ and $\varphi$, the performances of the NN trained to control only the evolution time $\tau$ are reported as the dotted brown line. The first two lines of plots have a fixed $T_2^{-1}$, with only the precession frequency $\omega$ being estimated. In contrast, the last line refers to the estimation of both $T_2^{-1}$ and $\omega$ simultaneously, i.e. $\vtheta = (\omega, T_2^{-1})$. In both cases $G = \id$. The shaded grey areas indicate the Bayesian Cramér-Rao bound, which is the ultimate precision bound computed from the Fisher information, that can be found in \cref{subsec:nv_center_dc_bound}. For these estimations, we have used $N = 480$ particles for the particle filter. In each plot, we reported also the batch size $B$ and the initial learning rate $\alpha_0$ used in the simulations for the ``Adaptive'' strategy.}
	\label{fig:nvcenter_comparison_phase}
\end{figure*}

We conducted multiple estimations summarized in \cref{fig:nvcenter_comparison_phase}, where we compared the performances of the optimized adaptive (NN) and non-adaptive strategies against the Particle Guess Heuristic (PGH)~\cite{wiebe_hamiltonian_2014}, which is a commonly referenced strategy in the literature. Additionally, we introduced a variant of the $\sigma^{-1}$ strategy~\cite{ferrie_how_2013}, named $\sigma^{-1}\&T^{-1}$, which accounts for the finite coherence time. According to the $\sigma^{-1} \& T^{-1}$ strategy, the evolution time $\tau_t$ is computed from the covariance matrix $\cov_t$ of the current posterior distribution as
\begin{equation}
	\tau_t = \left[ \tr (\cov_t)^{\frac{1}{2}} +\widehat{T_2^{-1}} \right]^{-1} \; ,
\end{equation}
while the standard $\sigma^{-1}$ strategy prescribes $\tau_t = \tr (\cov_t)^{-\frac{1}{2}}$. In the case of a fixed $T_2^{-1}$, this value was used instead of its estimator $\widehat{T_2^{-1}}$. For computing the controls of the PGH strategy, two particles $\boldsymbol{\theta}_1$ and $\boldsymbol{\theta}_2$ are drawn from the Bayesian posterior distribution; the evolution time is then computed as
\begin{equation}
	\tau_t = \left( ||\boldsymbol{\theta}_1-\boldsymbol{\theta}_2||_{2} + \varepsilon \right)^{-1} \; ,
\end{equation}
with $\varepsilon := 10^{-5} \MHz$. As a function of the normalized input to the NN, the controls $\tau$ and $\varphi$ are obtained from
\begin{equation}
	\begin{pmatrix} \tau_t \\ \varphi_t \end{pmatrix} = \begin{pmatrix} h \\ \pi \end{pmatrix} \cdot \lvert f_{\text{NN}} ( \wttheta_t, \wt{\boldsymbol{\sigma}}_t, \chi_{t}, \wt{R}_t, \wt{t} \, ) \lvert + \begin{pmatrix} 1 \mus \\ 0 \end{pmatrix} \; ,
	\label{eq:f_nn_phi}
\end{equation}
where $h$, which stands for ``height'', is a prefactor that is chosen to be appropriate for each simulation, and should be of the order of magnitude of the expected optimal $\tau_t$. The NN is the function $f_{\text{NN}}$. The prefactors $h$ for our experiments appear in the table below. For the time-limited estimations these prefactors are chosen to be roughly in accordance with the maximum value of the control predicted by the analysis in~\cite{belliardo_achieving_2020, cimini_experimental_2023}, for the measurement-limited case see \cref{sec:anal_meas}. The constant shift added in \cref{eq:f_nn_phi} and the absolute value are necessary to keep $\tau_t$ strictly positive, which is helpful for the convergence of the training.
%
\begin{table*}[htb]
	\centering
	\label{table:prefactor}
	\renewcommand{\arraystretch}{1.5}
	\setlength{\tabcolsep}{6pt}
	\begin{tabular}{|l|cl|cl|}
		\hline
		& Time & & Measurement & \\
		\hline
		\(T_2 = \infty\) &
		\(\frac{T\ped{max}}{20}\) &
		&
		\(\lceil 2^{\sqrt{M\ped{max}}} \mus \rceil\) &
		\\
		\hline
		\(T_2 < \infty\) &
		\(\max \Big \lbrace \frac{T\ped{max}}{20}, T_2 \Big \rbrace \) &
		&
		\(\max \Big \lbrace \lceil 2^{\sqrt{M\ped{max}}} \rceil \mus, T_2 \Big \rbrace\) &
		\\
		\hline
		\(T_2^{-1} \in (a, b)\) &
		\(\max \Big \lbrace \frac{T\ped{max}}{20}, a^{-1} \Big \rbrace \) &
		&
		\(\max \Big \lbrace \lceil 2^{\sqrt{M\ped{max}}} \rceil \mus, a^{-1} \Big \rbrace\) &
		\\
		\hline
	\end{tabular}
	\caption{Prefactor $h$ appearing in \cref{eq:f_nn_phi} for DC magnetometry on an NV center.}
\end{table*}
The NN has been pretrained to reproduce a linear ramp for the control time $\tau$, starting from $ 1 \mus$ and reaching $\tau = h$ at the end of the estimation. This behaviour has been suggested to us by the analytical results exposed in \cref{sec:anal_meas} and in~\cite{belliardo_achieving_2020}. The same initial controls are used for the non-adaptive strategy. Besides the optimized adaptive strategy with time and phase controls we reported also, under the name ``Only $\tau$'', the optimized precisions achieved through controlling the time $\tau_t$ only, also reported in~\cite{belliardo_model-aware_2023}. From the plots in \cref{fig:nvcenter_comparison_phase} we conclude that there is only a very small advantage in controlling the phase for $T_2 = 10 \mus$, if there is any at all. Similarly, with $T_2 = \infty, 100 \mus$ no advantage has been found, although the plots haven't been reported. For $T_2=10 \mus$ and $T_{\text{max}}=2560 \mus$ the phase control had converged to the constant $\varphi = \pi$, and more training could not take it out of this minimum. From the plot for $T_2 = 5 \mus$ we see that the advantage of controlling the phase $\varphi$ grows as the coherence time becomes smaller. The results of the simulations with $T_2^{-1} \in (0.09, 0.11) \MHz$ are very similar to that with $T_2=10 \mus$ because of the relatively narrow prior on $T_2^{-1}$.

\subsubsection{Future directions}
Typically, in the applications, the meaningful resource is the total time required for the estimation. This doesn't coincide however with $T = \sum_{t=0}^{M\ped{max}-1} \tau_t$, because the initialization of the NV center, the read-out, and the data processing all take time. This overhead time is proportional to the number of measurements, so, in a real experiment, we expect the actual resource to be a combination of the evolution time $T$ and the number of measurements $M$. In future work, the role of the higher moments of the Bayesian posterior distribution in the determination of the controls should be explored. In particular, it should be understood if with
\begin{multline}
	\begin{pmatrix} \tau_t \\ \varphi_t \end{pmatrix} = \begin{pmatrix} h \\ \pi \end{pmatrix} \cdot \lvert f_{\text{NN}} ( \wttheta_t, \wt{\boldsymbol{\sigma}}_t, \chi_{t}, \widehat{\gamma}_1, \wt{\mu}_4, \wt{\mu}_5, \ldots, \wt{R}_t, \wt{t} \, ) \lvert \\ + \begin{pmatrix} 1 \mus \\ 0 \end{pmatrix} \; ,
	\label{eq:higher_momenta}
\end{multline}
more precision from the adaptivity can be achieved. In this formula $\wt{\gamma}_1$ is the skewness of the posterior and $\wt{\mu}_i$ are its higher moments. A further improvement of this work would be to implement non single-shot readout of the NV center state, since at room-temperature it is the only way to measure the spin, as done by Zohar \textit{et al.}~\cite{zohar_real-time_2023}. The decoherence model we have used in this example and in all the others of the section works well for surface NV center, while for bulk centers a better model is
\begin{equation}
	p(\pm 1|\omega, T_2, \tau) := \frac{1}{2} \pm \frac{1}{2} e^{-(\tau/T_2)^2} \cos \left(\omega \tau + \varphi \right) \; .
	\label{eq:nv_center_model_dc_phase}
\end{equation}
The majority of the theoretical works, especially those papers involving ML~\cite{fiderer_neural-network_2021}, refer to \cref{eq:nv_center_model_dc}, and we wanted to be consistent with this trend in order to facilitate the comparisons. A further study could be treating the exponent of the dephasing term as a nuisance parameter.
%

\subsection{AC magnetometer}
\label{subsec:nv_center_ac}
\subsubsection{Description of the task}
In this section, we study the estimation of the intensity of an oscillating magnetic field of known frequency with an NV center used as an AC magnetometer. The NV center spin precesses in the magnetic field of intensity $B$ and frequency $\omega$, then its state is observed. The model for the binary outcome of the Ramsey measurement is
\begin{equation}
	p(\pm 1|\Omega, T_2, \tau) := \frac{1}{2} \pm \frac{1}{2} e^{-\tau/T_2} \cos \left[ \frac{\Omega}{\omega} \sin (\omega \tau) \right] \; .
	\label{eq:nv_center_model_ac}
\end{equation}
These probabilities can be found with a physical setup very similar to the one described in \cref{subsec:nv_center_dc_phase}, where we need to change the Hamiltonian of the system, which now oscillates with frequency $\omega$, i.e.
\begin{equation}
	\hat{H} := \frac{\hbar \Omega}{2}  \cos (\omega \tau )\hat{\sigma}_z \; ,
\end{equation}
in this situation we also neglect the controllable phase $\varphi$, i.e. we set $\varphi = 0$ for all the measurement steps. The evolution time $\tau$ is controlled by the trainable agent, while $\Omega := \gamma B$ is the unknown parameter to be estimated. The parameter $T_2^{-1}$ may or may not be an unknown in the estimation; in all cases $G=\id$. The prior on $\Omega$ is uniform in $(0, 1) \MHz$ for all the examples. The formula for $\tau_t$ is
\begin{equation}
	\tau_t = 1 \mus \cdot \lvert f_{\text{NN}} ( \wttheta_t, \wt{\boldsymbol{\sigma}}_t, \chi_{t}, \wt{R}_t, \wt{t} \, ) \lvert + 1 \mus \; .
	\label{eq:f_nn_dec_ac}
\end{equation}
The weights and biases of the NN are randomly initialized, and so are the controls for the non-adaptive strategy.

\subsubsection{Discussion of the results}
\begin{figure*}[htb]
	\centering
	\includegraphics[scale=1.0]{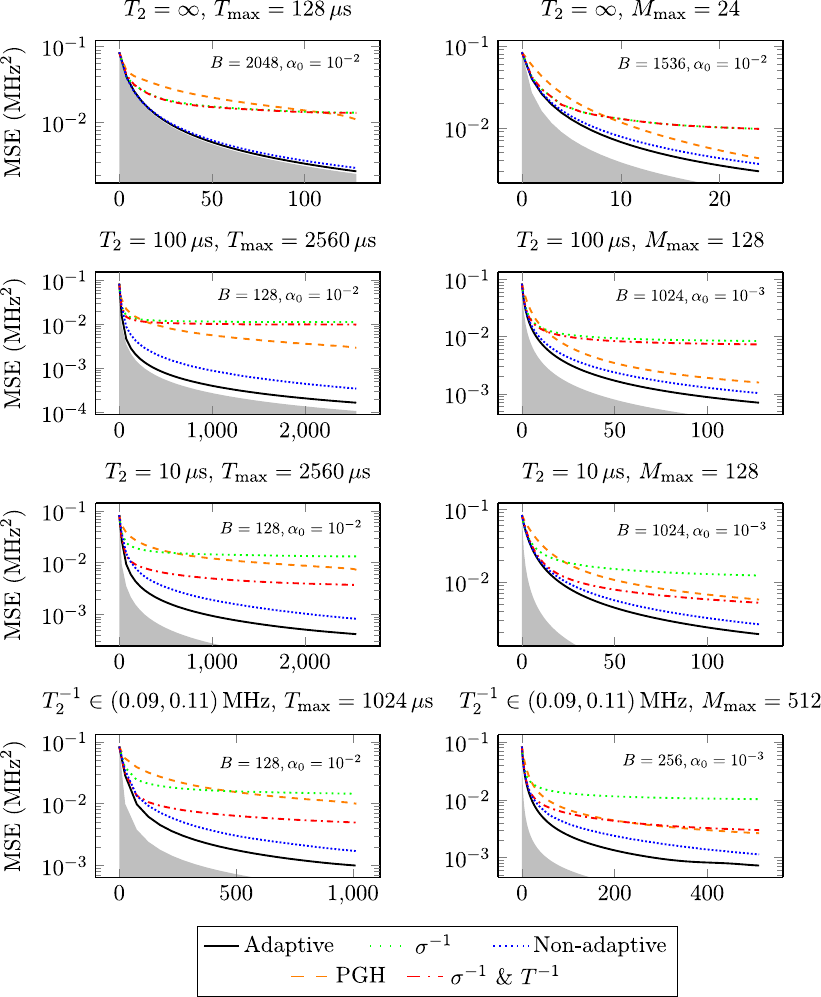}
	\caption{This figure refers to the problem of AC magnetometry on an NV center with known frequency $\omega = 0.2 \MHz$, see \cref{eq:nv_center_model_ac} for the model. The MSE of the optimized NN and static strategies are plotted, together with the performances of the $\sigma^{-1}$, the $\sigma^{-1}\&T^{-1}$, and the PGH strategies, discussed in \cref{subsec:nv_center_dc_phase}. On the left, the precisions of the time-limited simulations are reported, while the plots on the right refer to the measurement-limited estimation. For all the plots $G=\id$. The shaded grey areas indicate the Bayesian Cramér-Rao bound, which is the ultimate precision bound computed from the Fisher information, that can be found in \cref{subsec:nv_center_ac_bound}. For these estimations, we have used $N = 480$ particles for the particle filter. In each plot, we reported also the batch size $B$ and the initial learning rate $\alpha_0$ used in the simulations for the ``Adaptive'' strategy.}
	\label{fig:nvcenter_comparison_ac}
\end{figure*}

The results of the strategy optimization for this model are reported in~\cref{fig:nvcenter_comparison_ac}. Remarkably the time-limited estimations with long coherence time can saturate the bound set by the Fisher information. The model used in this example is formally equivalent to \cref{eq:nv_center_model_dc} with $\varphi=0$, where the adaptive strategy gives only small advantages, at difference with the results obtained here for AC-magnetometry. This is because the AC model maps to the DC one in a very different region of parameters with respect to the region we have explored in \cref{subsec:nv_center_dc_phase}.
%

\subsubsection{Future directions}
In AC magnetometry the technique of dynamical decoupling is often used to improve the sensitivity with respect to $\Omega$ and to increase the coherence time $T_2$. This consists of a series of $\pi$-pulses that reverse the sign of the accumulated phase. Given $t_i, \, 0<i<L+1$, the times at which the instantaneous $\pi$-pulse are applied, the outcome probabilities for a Ramsey measurement on a noiseless NV center is
\begin{widetext}
	\begin{equation}
		p(\pm 1|\Omega, T_2, \tau) := \frac{1}{2} \pm \frac{1}{2} \cos \left[ \frac{\Omega}{\omega} \sum_{i=1}^{L+1} (-1)^i \left( \sin(\omega t_{i})-\sin(\omega t_{i-1}) \right) \right] \; ,
		\label{eq:nv_center_model_ac_dd}
	\end{equation}
\end{widetext}
with $L$ being the number of pulses, and $t_0$ and $t_{L+1}$ being respectively the initialization and the measurement times. In the current simulations we haven't used $\pi$-pulses, but optimizing their application is the natural extension of this example, left for future works. For controlling the pulses there are three possibilities, listed in the following.
\begin{itemize}
	\item The interval between all the pulses is fixed to $\tau = t_{i+1}-t_i$, which is produced by a NN together with the number of pulses $L$. In this case, we have two controls, one of them $L$ being discrete.
	\item The controls are the $L$ time intervals $\tau_i = t_{i}-t_{i-1}$. The number of pulses is fixed but they can be made ineffective with $\tau_i = 0$.
	\item The control is the free evolution time $\tau$ together with a boolean variable, that tells whether a pulse or a measurement has to be applied after the free evolution. This would require a stateful model for the NV center.
\end{itemize}
The problem of estimating the magnetic field knowing the frequency is complementary to the protocol for the optimal discrimination of frequencies~\cite{schmitt_optimal_2021}, which has been applied to distinguish two chemical species in a sample. In this work the authors put forward an optimal strategy for the discrimination of two frequencies $\omega$ and $\omega+\Delta \omega$, knowing the intensity of the field. If $\Delta \omega \ll \omega$ and the field intensity is unknown, then a two stage approach to the problem is possible. We first estimate the intensity of the field as done in this example, while considering the frequency $\omega$ to be fixed and known, then we proceed with the optimal frequency discrimination based on the intensity just estimated. The error probability of the second stage depends on the precision of the first. Given a maximum total time for the discrimination, the assignment of time to the first and the second stages can be optimized to minimize the final error. It is important for the first stage to have low sensitivity to variations in the frequency, and for $\Delta \omega \simeq 0$ we expect this two stage protocol to be close to optimality. For optimizing the frequency discrimination with the field intensity as a nuisance parameter in a fully integrated protocol the introduction of $\pi$-pulses is necessary, as they have been used in~\cite{schmitt_optimal_2021}. The natural extension of frequency discrimination is the estimation of both the intensity and the frequency of the magnetic field optimally, both starting from a broad prior. All these improvements are left for future work.

\subsection{Decoherence estimation}
\label{subsec:nv_center_dec}

\subsubsection{Description of the task}
In this example, we study an NV center subject to variable decoherence of the dephasing type, that we want to characterize. The model for the binary outcome of the Ramsey measurement is
\begin{equation}
	p(\pm 1|T, \beta, \tau) := \frac{1 \pm e^{-\left( \tau/T\right)^{\beta}}}{2} \;,
	\label{eq:nv_center_model_dec}
\end{equation}
that can be obtained with a dynamical model similar to the one described in \cref{subsec:nv_center_dc_phase}, where we need to set the magnetic field $B$ or equivalently the frequency $\omega$ to zero and change slightly the noise model, using
\begin{equation}
	\varPhi_\tau (\rrho) := \rrho e^{-(\tau/T)^\beta} + \frac{\id}{2} \left[ 1 - e^{-(\tau/T)^\beta} \right] \, .
\end{equation}
The evolution time $\tau$ is controlled by the trainable agent, while $T^{-1}$ and $\beta$ are the two unknown characteristic parameters of the dephasing noise, i.e. respectively the transverse relaxation time and the decay exponent. In the applications, the coherence time of the noise encodes some useful information about the environment, like the concentration of radicals in a biological sample or the transport property of a material. The priors on $T^{-1}$ and $\beta$ are uniform in $(0.01, 0.1) \MHz$ and $(1.5, 4.0)$ respectively.

\subsubsection{Discussion of the results}
\begin{figure*}[htb]
	\centering
	\includegraphics[scale=1.0]{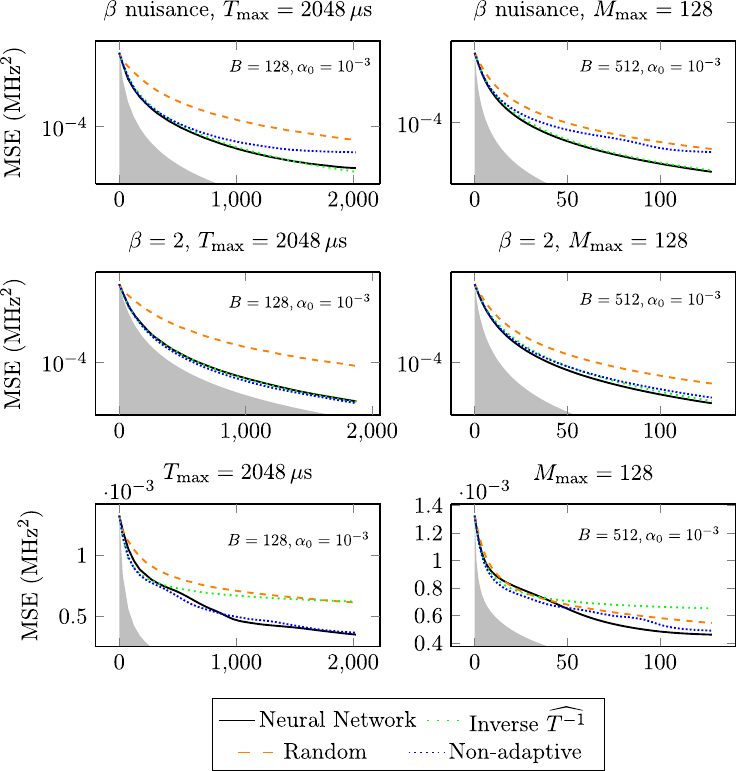}
	\caption{In these plots the MSE for the decoherence estimation on an NV center is reported as a function of the consumed resources, with the time-limited estimations on the left and the measurement-limited ones on the right. In the first line of plots, the decaying exponent $\beta$ is unknown and it is treated as a nuisance parameter. Therefore, only the precision on the inverse decay time appears in the loss. The plots on the second line refer to the case where $\beta$ is known and fixed ($\beta=2$), and only $T^{-1}$ is estimated. In the third line of plots $\beta$ and $T^{-1}$ are unknown and are both parameters of interest. In this case, the weight matrix is chosen to be $G = \text{diag} (1, 1/800 \MHz^2)$, to compensate for the different order of magnitude of the values of the parameters $\beta$ and $T^{-1}$. The shaded grey areas indicate the Bayesian Cramér-Rao bound, which is the ultimate precision bound computed from the Fisher information, see \cref{subsec:nv_center_dec_bound} for details. For these estimations, we have used $N = 2048$ particles for the particle filter. In each plot, we reported also the batch size $B$ and the initial learning rate $\alpha_0$ used in the simulations for the ``Neural Network'' strategy.}
	\label{fig:nvcenter_comparison_dec}
\end{figure*}

In this section, we compare the results of various strategies to control $\tau$, which are reported in the following.
\begin{itemize}
	\item Optimized adaptive strategy with a trained NN, that outputs the control according to
	\begin{equation}
		\tau_t = 100 \mus \cdot \lvert f_{\text{NN}} ( \wttheta_t, \wt{\boldsymbol{\sigma}}_t, \chi_{t}, \wt{R}_t, \wt{t} \, ) \lvert + 1 \mus \; ,
		\label{eq:f_nn_dec}
	\end{equation}
	\item Static strategy implemented with a NN that receives in input only $\wt{R}_t$ and $\wt{t}$, i.e.
	\begin{equation}
		\tau_t = 100 \mus \cdot \lvert f_{\text{NN}} (\wt{R}_t, \wt{t} \, ) \lvert + 1 \mus \; ,
		\label{eq:f_nn_dec_non_adaptive}
	\end{equation}
	\item Random strategy, where the inverse of the evolution time $\tau^{-1}$ is chosen randomly and uniformly in the support of the prior for $T^{-1}$, i.e. $(0.01, 0.1) \MHz$.
	\item Inverse time strategy. In this case, the evolution time is $\tau := \alpha_M^{1/\widehat{\beta}} \widehat{T^{-1}}$ for the measurement-limited estimation and $\tau := \alpha_T^{1/\widehat{\beta}} \widehat{T^{-1}}$ for the time-limited estimations, where $\alpha_M = 0.79681$ and $\alpha_M = 0.43711$. These numerical coefficients come from the optimization of the Fisher information reported in \cref{subsec:nv_center_dec_bound}. For the simulations where $\beta$ is treated as a nuisance parameter, $\alpha_M$ is used also for the time-limited estimation. This strategy not only depends on the current estimate $\widehat{T^{-1}}$ for the decoherence time but also on the estimator for the decay coefficient $\widehat{\beta}$, and it is, of course, adaptive.
\end{itemize}
Notice that, differently from the other NV center examples, the non-adaptive strategy for this application has been implemented not as a list of controls for each individual measurement but as a NN. The results of the strategy optimization for this model are reported in~\cref{fig:nvcenter_comparison_dec}. Both the adaptive and non-adaptive strategies have been pretrained to reproduce a linear ramp that reaches the maximum $\tau = 100 \mus$ at the end of the estimation.
%
From these simulations, we conclude that there is no advantage in using a NN instead of the strategy that optimizes the Fisher information, except when we are interested in the estimation of $\beta$.

\subsubsection{Future directions}
Identifying the decay exponent has applications in distinguishing the type of NV center, surface of bulk, that we are dealing with. Such a problem would be formulated in terms of a discrimination task between $\beta = 1, 2$, with $T$ as the nuisance parameter.

\subsection{Hyperfine coupling estimation}
\label{subsec:nv_center_hyperfine}
\subsubsection{Description of the task}
In this example, we study the measurement of an NV center strongly coupled to a ${}^{13}C$ nuclear spin in the diamond lattice. Such nuclear spin is not hidden in the spin bath of nuclei that are responsible for the dephasing noise, instead, it causes a relatively large split of the energy levels of the NV center, according to the hyperfine interaction strength, that can be measured in an experiment. The precession frequency of the NV center in a magnetic field is determined by the state of the nuclear spin. In the experiment of T. Joas \textit{et al.}~\cite{joas_online_2021} multiple incoherent nuclear spin flips happen during the read out, so that the nuclear spin is in each eigenstate approximately half of the time. This motivates the choice for model probability of the Ramsey measurement:
\begin{multline}
	p(\pm 1|\omega_0, \omega_1, T_2, \tau) \\ := \frac{1}{2} \pm \frac{1}{4} e^{-\tau/T_2} \left[ \cos \left(\omega_0 \tau \right) + \cos \left(\omega_1 \tau \right) \right] \; ,
	\label{eq:nv_center_model_hyperfine}
\end{multline}
which we take directly from~\cite{joas_online_2021}. In such a model, $\omega_0$ and $\omega_1$ are the two precession frequencies to be estimated, split by the hyperfine interaction, $T_2$ is the coherence time, while $\tau$ and $\varphi$ are the controls, that are respectively the evolution time and the phase. This model is completely symmetric under permutation of the two precession frequencies, therefore only those weight matrices $G$ that are permutationally invariant in the two parameters should be considered for this estimation. The prior on $(\omega_0, \omega_1)$ is the uniform distribution over the triangle in the $(\omega_0, \omega_1)$ plane identified by the points $(0, 0) \MHz$, ($0, 1) \MHz$, and $(1, 1) \MHz$, since we have decided to cancel this permutation symmetry by fixing $\omega_1 > \omega_0$. An important observation to be made is that the Fisher information matrix (FI) $I(\tau, \varphi)$ for $\omega_0$ and $\omega_1$ of the model in \cref{eq:nv_center_model_hyperfine} is singular and remains singular even for multiple measurements with different $\tau$ and $\varphi$. This means that the Fisher information is of no use in the optimization of the strategy. The controls $\tau$ and $\varphi$ are computed according to \cref{eq:f_nn_phi} with the prefactor
\begin{equation}
	h = \min \left( 40, \frac{T_2}{2} \right) \; .
\end{equation}
The means square errors of the two frequencies are weighted equally with $G = \id$. The frequency difference is the component of the hyperfine interaction parallel to the NV center quantization axis, i.e. $A_{||} = | \omega_1 - \omega_0 |$.

\subsubsection{Discussion of the results}
\begin{figure*}[htb]
	\centering
	\includegraphics[scale=1.0]{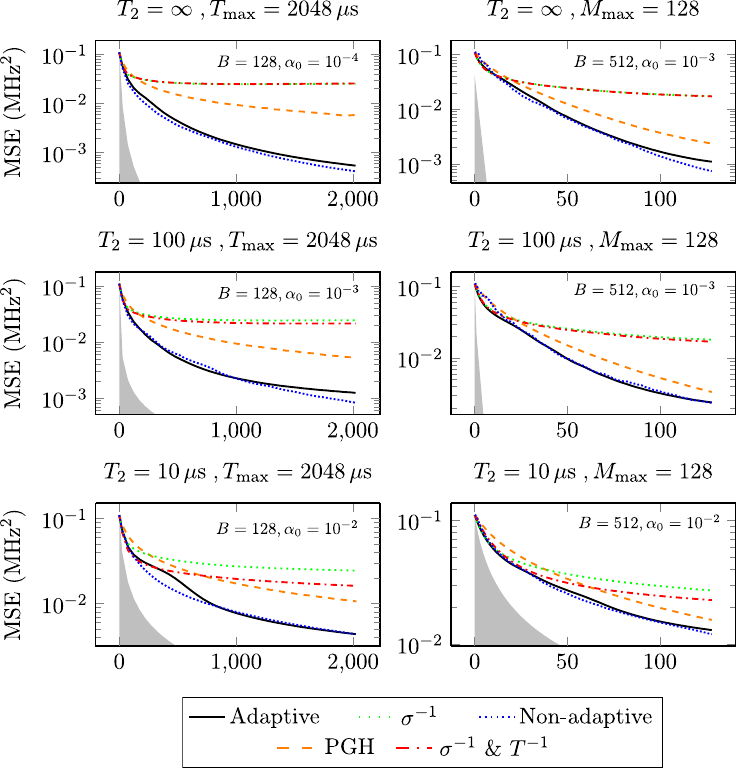}
	\caption{Sum of the mean square errors on the frequencies $\omega_0$ and $\omega_1$ for the optimized adaptive (NN) and non-adaptive strategies (static), compared to other common strategies used in the literature~\cite{joas_online_2021} and described in \cref{subsec:nv_center_dc_phase}. The shaded grey areas in the above plot indicate the Bayesian Cramér-Rao bound, which is the ultimate precision bound computed from the Fisher information, see \cref{subsec:nv_center_hyperfine_bound} for details. For these estimations, we have used $N = 4096$ particles for the particle filter. In each plot, we reported also the batch size $B$ and the initial learning rate $\alpha_0$ used in the simulations for the ``Adaptive'' strategy.}
	\label{fig:nvcenter_comparison_double}
\end{figure*}

Besides the NN and the static strategies, the performances of the particle guess heuristic (PGH) and of the $\sigma^{-1}$ strategies have been tested. The results are reported in~\cref{fig:nvcenter_comparison_double}. The NN is pretrained to reproduce a linear ramp for the control time, that reaches its maximum $\tau = h$ at the end of the estimation, while the phase control is random. Similarly, the non-adaptive strategy is initially a linear ramp in $\tau$ and is random in $\varphi$.
%
From these simulations, there seems to be no significant advantage in using an adaptive strategy for the simultaneous estimation of the two precession frequencies, neither for large nor for small coherence times $T_2$. Nevertheless optimizing with model-aware Reinforcement Learning still gives us an advantage with respect to other simpler approaches.

\subsubsection{Future directions}
There is also another way of expressing the outcome probability of the measurement: instead of defining the model in terms of $\omega_0$ and $\omega_1$ we could define it in terms of the frequencies sum $\Sigma = \omega_0 + \omega_1$ and difference $\Delta = \omega_1 - \omega_0$, thus writing
\begin{widetext}
	\begin{equation}
		p(\pm 1|\Sigma, \Delta, T_2, \tau) := \frac{1}{2} \pm \frac{1}{4} e^{-\tau/T_2} \left[ \cos \left(\frac{\Sigma+\Delta}{2} \tau \right) + \cos \left(\frac{\Sigma-\Delta}{2} \tau \right) \right] \; .
		\label{eq:alt_double}
	\end{equation}
\end{widetext}
In this form there is no permutational invariance in $\Sigma$ and $\Delta$, instead, the model is invariant under the transformation $\Delta \rightarrow - \Delta$. Using this expression, we should choose a prior having $\Delta>0$, which means $\omega_1>\omega_0$, and we should impose the positivity of the frequencies by requiring $\Sigma>\Delta$ in the prior. Since we are interested in the difference $\Delta$, the sum of the frequencies $\Sigma$ would be treated as a nuisance parameter. The absence of an advantage of the adaptive strategy over the non-adaptive one is probably also due to the hyper-simplified information passed to the NN. We are in fact approximating a complex 2D posterior, with many peaks and valleys with a Gaussian. A better approach would be to train an autoencoder to compress the information contained in the posterior and pass it to the NN. The autoencoder will be trained to compress the class of distributions that are produced by the likelihood of the \cref{subsec:nv_center_hyperfine}. In uture work, we plan to extend the estimation to the detection of multiple nuclear spins surrounding the NV center. For the estimation of $n$ frequencies in the model
\begin{equation}
	p(\pm 1| \lbrace \omega_i \rbrace_{i=0}^n, T_2, \tau) := \frac{1}{2} \pm \frac{1}{2 n} e^{-\tau/T_2} \sum_{i=0}^{n-1} \cos (\omega_i \tau) \; ,
	\label{eq:nv_center_model_nanonmr}
\end{equation}
an appropriate precision figure of merit would be
\begin{equation}
	\ell (\hvtheta, \vtheta) := \min_{\pi \in S_n} \sum_{i=0}^{n-1} (\widehat{\omega}_i - \omega_i)^2 \; ,
\end{equation}
with $\vtheta = (\omega_0, \omega_1, \ldots, \omega_{n-1})$ and $\hvtheta = (\widehat{\omega}_0, \widehat{\omega}_1, \ldots, \widehat{\omega}_{n-1})$, and $S_n$ being the permutation group of $n$ elements. If we impose the condition $\omega_i \le \omega_{i+1}$, then we can get rid of the minimization in the permutation. In this case, it is interesting to notice that the $n$-dimensional volume of the parameter space for a uniform prior on $\omega_i$ which is null outside of $(a, b)$ is reduced by a factor $n!$ due to the symmetry of the parameters, i.e.
\begin{equation}
	\mathcal{V}_n \le \frac{(b-a)^n}{n!} \; .
\end{equation}
This may reduce the number of particles necessary to represent the Bayesian posterior and make the optimization of this application accessible.

\section{Applications on the photonic circuits platform}
\label{sec:photonic_circuits_platform}
In this section, we report all the information and data related to the examples on the photonic platform, i.e. the agnostic Dolinar receiver, the classification of known and unknown coherent states, and multiphase discrimination in a four arms interferometer.

\subsection{Agnostic Dolinar receiver}
\label{subsec:dolinar}
\subsubsection{Description of the task}
\begin{figure*}[htb]
	\centering
	\includesvg[width=0.8\textwidth]{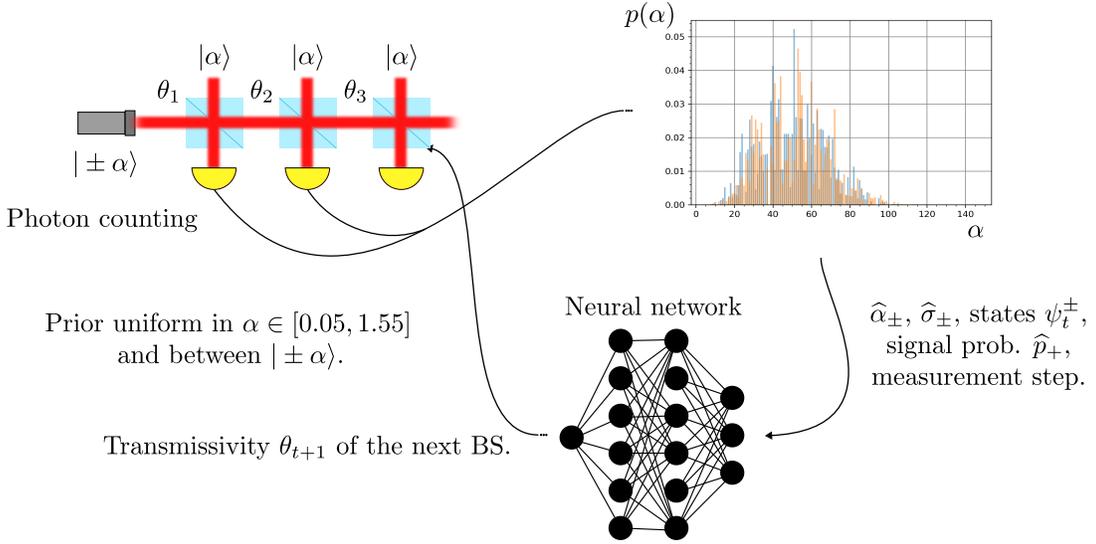}
	\caption{Representation of the measurement loop for the Dolinar receiver. The signal state $\ket{\pm\alpha}$ is mixed with the reference states and measured with photon counting. The posterior distribution on $\alpha$ has two components, corresponding to the signs $s=\pm1$. The intensity of the intermediate signal $\psi_t^\pm$, the mean and the variance of the posterior, are fed to the NN that computes the next transmissivity.}
	\label{fig:dolan_physical}
\end{figure*}

In~\cite{belliardo_model-aware_2023} we already discussed the Dolinar receiver, and in this section we briefly review the physics of this system and the task we are trying to solve. The goal consists of distinguishing between the two coherent states $\ket{-\alpha}$ and $\ket{\alpha}$ with $\alpha$ unknown using a single copy of the signal $\ket{\pm \alpha}$. The Dolinar receiver, known for its good performance, traditionally uses a local oscillator (LO) synchronized with the sender's laser. Recent studies introduced an alternative, the agnostic Dolinar receiver, eliminating the need for the LO. Instead, the new device utilizes $n$ reference states $\ket{\alpha}$ sent alongside the signal. In this setup, classical knowledge about $\ket{\alpha}$ is absent, treating it as an unknown parameter. The device, as shown in \cref{fig:dolan_physical}, utilizes $\ket{\alpha}^{\otimes n}$ to discriminate the sign of the signal. The signal is combined with a reference state on a programmable BS with transmissivity $\theta_t$. The result of the photon counting at each BS is used to update the Bayesian posterior on $\alpha$ and the signal's sign. A NN determines the angle $\theta_{t+1}$ for the next BS. This task involves continuous (the signal's amplitude $\alpha \in \mathbb{R}$) and discrete (signal sign) parameters. The performance of the receiver is evaluated based on the error probability in the classification of the signal's sign, while the amplitude $\alpha$ is a nuisance parameter. With this schematization, we can identify the parameters of the dynamical model defined in \cref{sec:dynamical_model}. Differently from all the variations of the NV center studied until now, in this case, there is no dynamical evolution, meaning that the LCPT map $\Phi^{t,x_t}_{\vtheta}$ is always equal to the identity. In this case, all the control is in the collection of operators $\{\hat{M}_{y_t}(x_t)\}$, which means that the initial state of the probe is already encoded, as it comes from a different laboratory on which we don't have any control. The input state for this procedure is the pure state $\ket{\alpha}^{\otimes n} \otimes \ket{\pm \alpha}$. At the $k$-th step, the measurement operators $\hat{M}_{y_k}(x_k)$ can be written as the product of a BS operation, implemented by $\hat{U}^{(k, n+1)}(\theta_k)$, that mixes the $n+1$-th state, namely the one to discriminate, with the reference state in position $k$, followed by a measurement on the $k$-th component using photon counting. In equations,
\begin{align}
 \hat{M}_{m_k}(\theta_k) = |m_k\rangle_k\langle m_k| \hat{U}^{(k, n+1)}(\theta_k) \; ,
\end{align}
where we revised slightly the notation to make it closer to the standard one for a beam splitter, using $\theta_k$ instead of $x_k$ as the mixing angle, and using $m_k$ instead of $y_k$ as the integer number that represents how many photons have been counted. This formula holds for all the steps except the last one, where two photon-counting operators are present, instead of one.

\subsubsection*{Input to the neural network}
The input to the NN is the concatenation of the following nine scalar values.
\begin{itemize}

	\item The signal intensity $\psi^{+}_t$ after the $t$-th measurements, assuming that the initial state was $\ket{+\alpha}$.

	\item The posterior estimator for the initial signal intensity $\widehat{\alpha}_+$, assuming the signal's sign was $s=+$.

	\item The variance of the posterior distribution $\widehat{\sigma}_+$ for the initial signal intensity, assuming the sign was $s=+$.

	\item The signal intensity $\psi^{-}_t$ after the $t$-th measurements, assuming that the initial state was $\ket{+\alpha}$.

	\item The posterior estimator for the initial signal intensity $\widehat{\alpha}_-$, assuming the signal's sign was $s=-$.

	\item The variance of the posterior distribution $\widehat{\sigma}_-$ for the initial signal intensity, assuming the sign was $s=-$.

	\item The posterior probability $\widehat{p}_{+}$ that the original signal had a $+$ sign.

	\item The index of the current measurement step, normalized in the interval $[-1, +1]$ with respect to the total number of measurements, indicated with $\wt{t} := 2 t/M\ped{max} -1$.
\end{itemize}
The reflectivity $\theta_t$ is produced by the NN through the following formula
\begin{widetext}
	\begin{equation}
		\theta_t = f\ped{NN} (\psi^{+}_t, \widehat{\alpha}_+, \widehat{\sigma}_+, \psi^{-}_t, \widehat{\alpha}_-, \widehat{\sigma}_-, \widehat{p}_{+}, \wt{t}) - \pi \cdot \left( n\ped{Phot} \bmod 2 \right) \; .
	\end{equation}
\end{widetext}
The symbol $n\ped{Phot}$ indicates the total number of photons measured up to the point where the NN is called, which controls the addition of an extra phase $-\pi$ to the transmissivity. The same mechanism is implemented for the non-adaptive strategy.

\subsubsection*{Definition of the precision figure of merit}
The loss in the agnostic Dolinar receiver measures the error in guessing the sign of the signal $\ket{\pm \alpha}$, after it has been completely measured. The lower bound on the error probability given $n$ copies of the reference states is $p_H(\alpha)$, and can be found in \cref{subsec:helstrom_bound}. We have implemented two possible policies for guessing the sign $s$, that we indicate respectively with $\widehat{s}_{\text{Bayes}}$ and $\widehat{s}_{\text{Parity}}$. They are respectively
\begin{equation}
	\widehat{s}_{\text{Bayes}} (\widehat{p}_{+}) := \begin{cases}
		+1 \quad \text{if} \quad
		\widehat{p}_{+} > 0.5 \; ,\\
		-1 \quad \text{if} \quad
		\widehat{p}_{+} \le 0.5 \; ,\\
	\end{cases}
\end{equation}
and
\begin{equation}
	\widehat{s}_{\text{Parity}} (n_{\text{Phot}}) := \begin{cases}
		+1 \quad \text{if} \quad
		n_{\text{Phot}} \bmod 2 = 0 \; ,\\
		-1 \quad \text{if} \quad
		n_{\text{Phot}} \bmod 2 = 1 \; .\\
	\end{cases}
\end{equation}
The Kronecker delta function
\begin{equation}
	\delta (x, y) := \begin{cases}
		1 \quad \text{if} \quad x=y \; , \\
		0 \quad \text{if} \quad x \neq y \; ,
	\end{cases}
\end{equation}
is necessary to introduce the losses, which are
\begin{itemize}
	\item \texttt{loss=0}: $\ell (\widehat{p}_{+}, s) := 1-\delta(\widehat{s}_{\text{Bayes}}, s)$,
	\item \texttt{loss=1}: $\ell (\widehat{p}_{+}, s) := 1-\widehat{p}_{s}$,
	\item \texttt{loss=2}: $\ell (n_{\text{Phot}}, s) := 1-\delta(\widehat{s}_{\text{Parity}}, s)$,
	\item \texttt{loss=3}: $\ell (\widehat{p}_{+}, s, \alpha) := 1-\delta(\widehat{s}_{\text{Bayes}}, s) - p_H(\alpha)$,
	\item \texttt{loss=4}: $\ell (\widehat{p}_{+}, s, \alpha) := 1-\widehat{p}_{s} - p_H(\alpha)$,
	\item \texttt{loss=5}: $\ell (n_{\text{Phot}}, s, \alpha) := 1-\delta(\widehat{s}_{\text{Parity}}, s) - p_H(\alpha)$,
	\item \texttt{loss=6}: $\ell (\widehat{p}_{+}, s, \alpha) := \frac{1-\delta(\widehat{s}_{\text{Bayes}}, s)}{p_H(\alpha)}$,
	\item \texttt{loss=7}: $\ell (\widehat{p}_{+}, s, \alpha) := \frac{1-\widehat{p}_{s}}{p_H(\alpha)}$,
	\item \texttt{loss=8}: $\ell (n_{\text{Phot}}, s, \alpha) := \frac{1-\delta(\widehat{s}_{\text{Parity}}, s)}{p_H(\alpha)}$,
\end{itemize}
All these losses, when averaged, converge to the probability of a wrong classification. It is not obvious a priori which loss is the best one.

\subsubsection*{Discussion of the results}
\begin{figure*}[htb]
	\centering
	\includegraphics[scale=1.0]{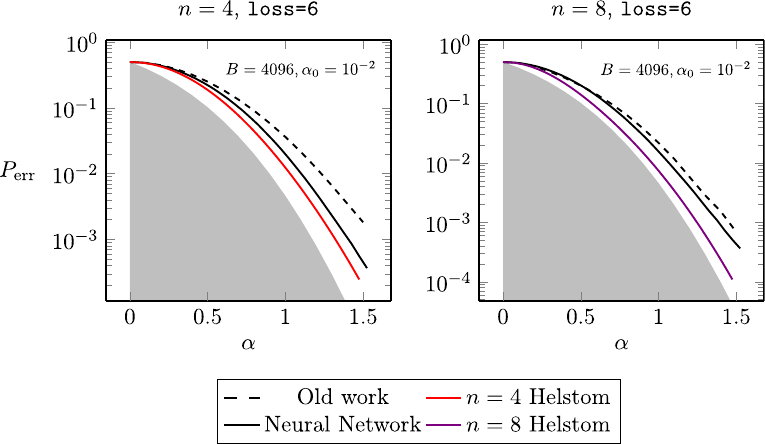}
	\caption{Comparison of error probabilities for various strategies with different numbers of copies of $\ket{\alpha}$, specifically $n=4$ and $n=8$, for \texttt{loss=6}. The shaded gray area is the region excluded by the Helstrom bound~\cite{helstrom_quantum_1969, holevo_statistical_1973}, which is the lowest error probability theoretically achievable when assuming having an infinite number of reference states ($n=\infty$) at disposal. The solid red and violet lines are the Helstrom bound calculated for a finite number of copies of $\ket{\alpha}$~\cite{zoratti_agnostic-dolinar_2021}, respectively $n=4$ and $n=8$. For the details on the computation of the Helstrom bound see \cref{subsec:helstrom_bound}. The black dashed line showcases the lowest error found in the old work~\cite{zoratti_agnostic-dolinar_2021}, without Machine Learning, while the black solid line is the performance achieved using the NN. The performances of the optimal non-adaptive strategies haven't been reported since they can't rival the ones of the NN. For both the training and the performance evaluation we used $N=512$ particles. In each plot, we reported also the batch size $B$ and the initial learning rate $\alpha_0$ used in the simulations for the ``Neural Network'' strategy.}
	\label{fig:dolinar_confronto}
\end{figure*}

The simulation results are presented in \cref{fig:dolinar_confronto}. While in~\cite{belliardo_model-aware_2023} we trained the optimized strategies with \texttt{loss=3} in this paper we used \texttt{loss=6}, which seems however to deliver slightly worst results. In both cases, the plots represent \texttt{loss=0}, averaged over many executions of the estimation task.

\subsection{Quantum Machine Learning classification of states}
\label{subsec:qml_classifier}
In this example, we put forward a quantum Machine Learning (QML)-based classifier able to distinguish between three coherent states $\ket{\alpha_0}$, $\ket{\alpha_1}$, $\ket{\alpha_2}$, with $\alpha_0, \alpha_1, \alpha_2 \in \mathbb{C}$, given $n$ copies of each of them, which constitute the quantum training set. The signal state is a single copy of a coherent state, promised to be one of the three training states, i.e. $\ket{\alpha_s}$ with $s=0, 1, 2$. The priors on the components of the training states $\ket{\alpha_0}$, $\ket{\alpha_1}$, $\ket{\alpha_2}$ are uniform in the harmonic oscillator phase space in a square of side $2\alpha$ centered at the origin, that is
\begin{equation}
	\alpha_i^R := \Re (\alpha_i) \in (-\alpha, +\alpha) \; , \alpha_i^I := \Im (\alpha_i) \in (-\alpha, +\alpha) \; ,
\end{equation}
for $i=1, 2, 3$. Also the prior on the signal classes $s=0, 1, 2$ is uniform. What makes this task difficult is that the complex numbers $\alpha_0, \alpha_1, \alpha_2$ are not precisely known, not at least beyond the information coming from the prior. There are therefore seven parameters to be estimated: the real and imaginary components of these three complex amplitudes (six continuous parameters in total), plus the class of the signal $s$, which is a discrete parameter.
\begin{figure*}[htb]
	\centering
	\includesvg[width=0.8\textwidth]{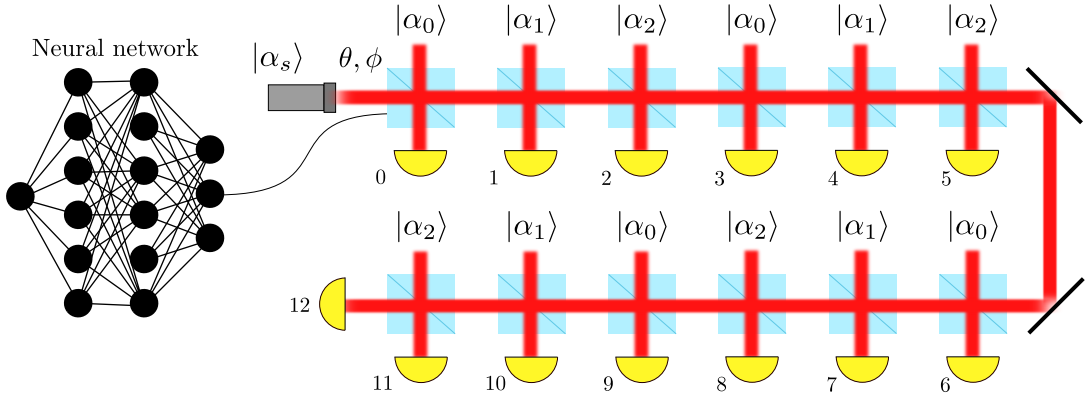}
	\caption{Schematic representation of the quantum Machine learning discrimination device controlled by the NN. The training set states enter sequentially the ports of the BS from above, where they are mixed with the signal. The outcome of the photon counting measurement is then used to update the posterior distribution on the values of $\alpha_0$, $\alpha_1$, $\alpha_2$, and $s$. The order in which the measurements are performed is indicated near the symbols for the photon counters, from which we deduce $M=13$. All the outcomes of the previous measurements contribute to the determination of the next controls through the PF and the NN.}
	\label{fig:dolinar_three_serial}
\end{figure*}
The device that performs this task is made of programmable BS only, of which we control the transmissivity $\theta$ and the phase $\varphi$, and is represented in \cref{fig:dolinar_three_serial}. The resource consumed in a single execution of the discrimination task is the average number of photons in the device, i.e.
\begin{equation}
	N_{\text{ph}} := n \left( |\alpha_0|^2+|\alpha_1|^2+|\alpha_2|^2 \right) + |\alpha_s|^2 \; .
\end{equation}
In this application, we experimented also with using a ternary decision tree as an agent instead of a NN. The device presented in this section can also be interpreted as a generalization of the agnostic Dolinar receiver of \cref{subsec:dolinar}.

\subsubsection*{Input to the agent}
\begin{figure}[htb]
	\centering
	\includesvg[scale=0.8]{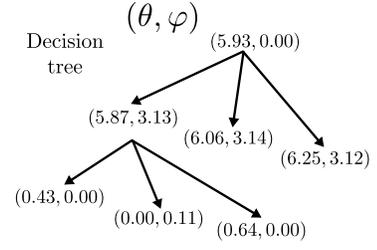}
	\caption{Example of a decision tree used in the quantum machine learning classifier. A tuple of controls $(\theta, \varphi)$ is associated with each node in the tree. The value $\widetilde{y}_t$ determines the path chosen in the tree at each node in an experiment.}
	\label{fig:decision_tree}
\end{figure}

Let us define the posterior probabilities of the signal being respectively $s=0, 1, 2$ as $\widehat{p}_0$, $\widehat{p}_1$, and $\widehat{p}_2$. This distribution is computed as the marginal of the Bayesian posterior by integrating over the continuous variables, and can be computed on the particle filter. In case the agent is a NN, then it receives as input the following 19 scalars.
\begin{itemize}
	\item The average state of the signal after the $t$-th photon counting measurements. These are two scalar values, i.e. the real and imaginary parts of the coherent state, and will be indicated collectively as $\widehat{\psi}_t$. The average is taken over the posterior distribution.

	\item The mean posterior estimator for the real and imaginary components of the complex amplitudes of the three training states at the $t$-th step, collected in the tuple $\boldsymbol{\widehat{\alpha}}_t := (\widehat{\alpha}_0^R, \widehat{\alpha}_0^I, \widehat{\alpha}_1^R, \widehat{\alpha}_1^I, \widehat{\alpha}_2^R, \widehat{\alpha}_2^I)$.

	\item The standard deviations from the mean of the amplitudes of the reference states, collected in the tuple $\boldsymbol{\widehat{\sigma}}_t := (\widehat{\sigma}_0^R, \widehat{\sigma}_0^I, \widehat{\sigma}_1^R, \widehat{\sigma}_1^I, \widehat{\sigma}_2^R, \widehat{\sigma}_2^I)$. The scalars passed to the NN are actually $-\frac{1}{10} \log \boldsymbol{\widehat{\sigma}}_t$.

	\item The posterior probability for the initial state to be $s=0$, normalized in $[-1, +1]$, i.e. $\wt{p}_0 := 2 \widehat{p}_{0}-1$.

	\item The posterior probability for the initial state to be $s=1$, normalized in $[-1, +1]$, i.e. $\wt{p}_1 := 2 \widehat{p}_{1}-1$.

	\item The posterior probability for the initial state to be $s=2$, normalized in $[-1, +1]$, i.e. $\wt{p}_2 := 2 \widehat{p}_{2}-1$

	\item The index of the current photon counting measurement normalized in $[-1, +1]$, indicated with $\widetilde{t} := 2 t/M\ped{max} - 1$.

	\item $\bar{t} = t \bmod 3 -1$. This tells whether the current photon counting is performed by mixing the signal with $\ket{\alpha_0}$, $\ket{\alpha_1}$, or $\ket{\alpha_2}$.
\end{itemize}
The two controls for the beam splitter are obtained directly from the NN as
\begin{equation}
	\begin{pmatrix} \theta_t \\ \varphi_t \end{pmatrix} = f\ped{NN} \left( \widehat{\psi}_t, \boldsymbol{\widehat{\alpha}}_t, -\frac{1}{10} \log \boldsymbol{\widehat{\sigma}}_t, \wt{p}_0, \wt{p}_1, \wt{p}_2, \wt{t}, \bar{t} \right) \; .
\end{equation}
Now we focus on the case the agent is a ternary decision tree, see~\cref{fig:decision_tree}. In this case, the input is based directly on the string of outcomes $\vy_t$ instead of going through the Bayesian posterior distribution. The outcome $y_t$ of each measurement, i.e. the number of photon observed in the photon counter, is classified into one of the three classes according to the relation with the mean number of photons $\lfloor \alpha^2 \rceil$. This is realized by defining the variable $\wt{y}_t$, i.e.
\begin{equation}
	\wt{y}_t := \begin{cases}
		0 \quad \text{if} \quad y_t \le \lfloor \alpha^2 \rceil - 1 \; , \\
		1 \quad \text{if} \quad y_t = \lfloor \alpha^2 \rceil \; , \\
		2 \quad \text{if} \quad y_t \ge \lfloor \alpha^2 \rceil + 1 \; .
	\end{cases}
\end{equation}
The infinite possibilities for the outcomes get reduced to only three classes without losing too much information. The modified outcomes up to the $t$-th measurement are used to decide which branch to follow at each node of a ternary decision tree, and are collected in the tuple $\boldsymbol{\widetilde{y}}_t := (\wt{y}_0, \wt{y}_1, \ldots, \wt{y}_t)$. The path in the tree (and the whole controls trajectory) is completely identified by $\boldsymbol{\widetilde{y}}_t$. At each node there are a couple of controls $(\theta, \varphi)$ to be used in the next measurement, which are the trainable variables of the agent, and are returned in the call to the strategy, i.e.
\begin{equation}
	\begin{pmatrix} \theta_t \\ \varphi_t \end{pmatrix} = f\ped{Tree} (\boldsymbol{\widetilde{y}}_t, t) \; .
\end{equation}
where $f\ped{Tree}$ represents the decision tree.

\subsubsection*{Definition of the precision figure of merit}
The precision is the estimated error probability of a wrong classification. The estimator for the signal class is
\begin{equation}
	\widehat{s} ({\widehat{\boldsymbol{p}}}) := \text{argmax} (\widehat{p}_0, \widehat{p}_1, \widehat{p}_2) \; ,
\end{equation}
where $\widehat{\boldsymbol{p}} := (\widehat{p}_0, \widehat{p}_1, \widehat{p}_2)$, and the loss for each instance of the task reads
\begin{equation}
	\ell (\boldsymbol{\widehat{p}}, s) := 1 - \delta (\widehat{s}, s) \; ,
\end{equation}
which is one if an error is made and zero if the guess is right. Averaged on many estimations, it converges to the error probability for the classification task.

\subsubsection*{Discussion of the results}
\begin{figure*}[htb]
	\centering
	\includegraphics[scale=1.0]{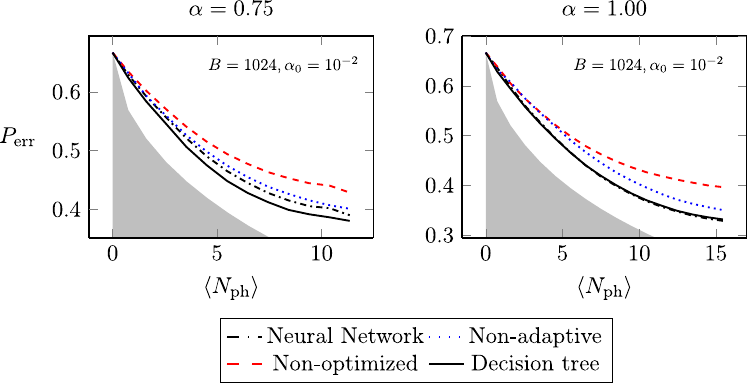}
	\caption{Error probabilities as a function of the average number of photons in the QML device for the classification of coherent states, reported for different strategies. Both ``Neural Network'' and ``Decision Tree'' are optimized adaptive strategies. The ``Non-optimized'' strategy refers to a device where the phase imprinted by the BS are all zero and the reflectivity is chosen, so that the same fraction of the signal reaches each one of the photon counters. The shaded grey areas indicate the performances of the pretty good measurement (PGM), computed for $2 n$ copies of each training state, see \cref{subsec:qml_classifier_bound} for details. This is not the ultimate achievable precision for the classification problem, but it is a reasonable reference value not achievable with linear optics. Both the weights of the NN and the controls in the non-adaptive strategies have been initialized randomly. For these estimations, we have used $N = 512$ particles to represent the posterior. In each plot, we reported also the batch size $B$ and the initial learning rate $\alpha_0$ used in the simulations for the ``Neural Network'' strategy.}
	\label{fig:qml_results}
\end{figure*}

\begin{figure*}[htb]
	\centering
	\includegraphics[scale=1.0]{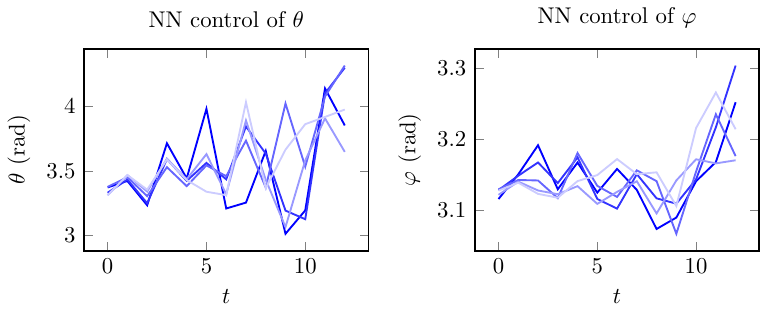}
	\caption{Five trajectories for the neural network controls $\theta$ and $\phi$ are showed as a function of the measurement index $t$. The colour shade indicates the five different executions of the experiment. The parameters $\alpha_0, \alpha_1, \alpha_2$ has been selected at random within the prior interval, like the signal class $s=0, 1, 2$ for these five instances estimation. There is a tendency for the control trajectories to diverge in time (this can be seen well in the plot for $\theta$), since having already done some measurements means that the controls can be tailored to the particular instance of the task. This behaviour is trained in the NN but is hard-coded in the decision tree strategy.}
	\label{fig:example_controls_qml}
\end{figure*}

We have trained the QML classifier for $\alpha=0.75$ and $\alpha=1.00$. All the estimations have been performed for $n=4$. We studied the precision of the discrimination device in the quantum regime, i.e. with a small number of photons, since for $\alpha \gg 1$ the coherent states are perfectly distinguishable. With $\alpha=0.75$, the signal contains on average $0.5$ photons, we should therefore expect relatively large discrimination errors even with optimal strategies. Nevertheless, the point of the optimization is to extract every last bit of information from the device, even in a regime where the errors are relatively frequent. The performances of the NN and the decision tree are compared to that of the optimized non-adaptive strategy and the precision of a non-optimized strategy, for which each photon counter receiver the same fraction of the signal $\ket{\alpha_s}$. The ultimate precision limit is represented by the pretty good measurement (PGM), discussed in \cref{sec:photonic_platform_bound}. This last cannot be realized with linear optics and photon counting, and assumes a perfect \textit{a priori} knowledge of the training states $\ket{\alpha_0}$, $\ket{\alpha_1}$, and $\ket{\alpha_2}$ which we don't have. To make a more fair comparison, we assumed the PGM consumes $2 n$ training states where this QML device consumes only $n$. The $2n$ copies are intended to be sufficient to perfectly identify $\alpha_i$, in order to later perform the corresponding PGM on the signal $\ket{\alpha_s}$. Still, the PGM error probability is not expected to be achievable. The performances of the QML classifier for different controls are reported in \cref{fig:qml_results}.
%
There is a gap between the performances of the non-adaptive controls and the two adaptive strategies (NN and decision tree). This gap starts from zero and grows with the average number of photons. This is intuitive since with very few photons we don't expect the adaptive measurements to be better than the non-adaptive one, because almost always the measurement outcomes of the photon counters will be zero, which doesn't give any information about the training states. On the other hand, with many photons there is more space for learning. This can be also understood considering the number of possible trajectories of the Bayesian posterior during the estimation, which is only one for $\alpha = 0$ (no update is made, the outcomes are always zero), and grows exponentially with the number of photons. More possible trajectories means having potentially different controls on each one of them in order to improve the precision, which is by definition adaptivity. In \cref{fig:example_controls_qml} we report some examples of trajectories.

In our experiments, we observed that the convergence of the NN to the optimal strategy is slow and doesn't always happen (see the case $\alpha=0.75$ in \cref{fig:qml_results}), while the decision tree is shown to be a superior control for this device. In general, for a problem with many parameters and a relatively small space of possible outcome trajectories, the decision tree is superior to the NN. On the other hand, with few parameters and a large space of outcome trajectories (like in the examples on NV centers), the decision tree is unusable in the form we presented here and the NN is superior. For a large number of average photons $\langle N_{\text{ph}} \rangle$ the error probability curves tend to saturate. This is because even with infinite $n$, and therefore perfect knowledge of the training states, a single copy of the signal can't be classified unambiguously. In this application, we have used $N = 512$ particles to represent the posterior distribution, which, conventional wisdom says are too few for the estimation of seven parameters. We have observed though that using $N = 1024$ particles doesn't improve significantly the precision. Notice that in this device we are performing the simultaneous estimation of multiple quantum incompatible parameters~\cite{belliardo_incompatibility_2021}, that are the real and imaginary components of the states amplitudes, whose corresponding generators $x$ and $p$ don't commute. We have also tried to implement the discrimination on a parallel device, instead of the serial one represented in \cref{fig:dolinar_three_serial}. In this case, the original signal is first equally split on three wires, i.e. $|\alpha_s\rangle \rightarrow |\frac{\alpha_s}{\sqrt{3}}\rangle^{\otimes 3}$, and on each line the signal is mixed with the $n$ copies of the same training state. The performances of this parallel device, however, turned ou to be always worst then the serial one and are not reported here.
%

\subsubsection{Future directions}
The controls from the decision tree have the advantage of being faster to compute with respect to those of the NN, but storing the tree requires more memory. A future direction of research could be to trim the decision tree after the training by keeping only those paths that correspond to the most probable trajectories.

\subsection{Multiphase discrimination on a photonic circuit}
\label{subsec:photonic_circuit}
\subsubsection{Description of the task}
In this section, we examine an example of multiphase estimation. We consider an interferometer consisting of four arms~\cite{cimini_deep_2023}, with two balanced quarters serving as the opening and closing elements respectively. A quarter generalizes a beam splitter to have four inputs and four outputs. After passing through the closing quarter, all wires are measured using photon counters. Three of the four arms experience phase shifts with unknown phases $\varphi_i$ for $i=0, 1, 2$. Immediately after they are subjected to three other phase shifts $c_i \in [0, 2 \pi)$, which are the controls in our experiment, managed by the agent.
%
\begin{figure*}[htb]
	\centering
	\includesvg[width=0.8\textwidth]{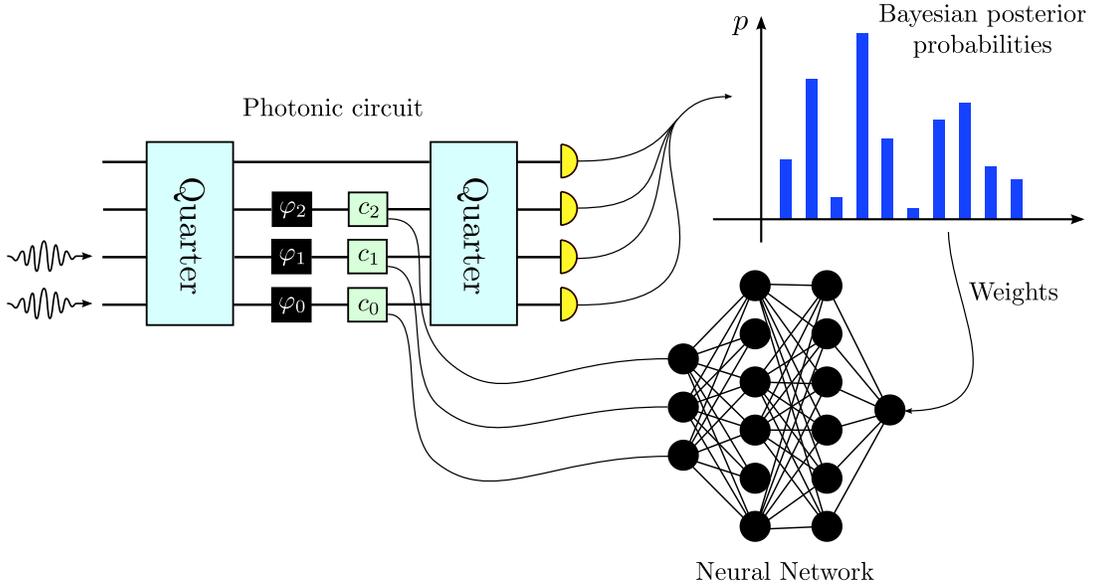}
	\caption{Four modes interferometer for the simultaneous estimation of three phases. The interferometer has two balanced quarter as opening and closing elements, which are the generalization of the balanced BS for four modes in input and four modes in output. The three unknown phases are called $\varphi_0, \varphi_1, \varphi_2 \in \lbrace 0, 1 \rbrace \rad$, while the control phases are $c_0, c_1, c_2$. The input of the NN are the eight probabilities constituting the posterior distribution on the values of the three phases.}
	\label{fig:photonic_circuit_picture}
\end{figure*}
In this interferometer, represented in \cref{fig:photonic_circuit_picture}, we simultaneously test each line to determine the presence or absence of a phase shift, which can only assume the two values $\varphi_i \in \lbrace 0, 1 \rbrace \rad$. In some practical applications, these phase shifts could result from the presence on the line of a specific object, a chemical species, or it could represent a piece of encoded information that we want to retrieve (think of an optical memory device). This photonic application can also be considered as a very rudimental form of adaptive imaging with few photons. During the experiment a total of $M_{\text{max}}$ identical copies of an input state are fed sequentially to the interferometer and measured independently on the interferometer's end. Based on the outcomes of the four photon counting measurements, the correct values for the tuple $\boldsymbol{\varphi} := (\varphi_0, \varphi_1, \varphi_2)$ is inferred using Bayesian inference. The agent is a NN that outputs the tuple of control phases $\boldsymbol{c} := (c_0, c_1, c_2)$ to be applied in the next round of measurements. The schematization of the dynamical model defined in \cref{sec:dynamical_model} can be applied also in this case. The LCPT map $\Phi^{(t,x_t)}_{\boldsymbol{\varphi}}$ is unitary and thus can be written in the form
\begin{equation}
	\Phi^{(t,x_t)}_{\boldsymbol{\varphi}} (\rrho) := \hat{U}_{\boldsymbol{\varphi}} \rrho \hat{U}_{\boldsymbol{\varphi}}^\dagger \; ,
\end{equation}
where we have introduced the unitary operator $\hat{U}_{\boldsymbol{\varphi}} (\boldsymbol{c}_t)$, which is a simple composition of the unitary matrices representing the quarter and the phase shifts shown in \cref{fig:photonic_circuit_picture} on each bosonic line:
\begin{align}
	\hat{U}_{\boldsymbol{\varphi}} := \hat{U}(\varphi_2) \hat{U}(\varphi_1) \hat{U}(\varphi_0)\hat{U}\ped{quarter}.
\end{align}
The measurement operators are similar, and can be written as a product of unitaries implementing teh controlled phase shifts $c_{i}$, then the same quarter unitary operator and, lastly, a series of photon-counting measurements on all the lines, that can be written as projectors on the base that diagonalize the number operators for each of the lines. Notice that the control $\boldsymbol{c}$ can be inserted as a part of the dynamical model $\Phi^{(t,x_t)}_{\boldsymbol{\varphi}}$ as well as a part of the measurement.

\subsubsection*{Input to the neural network}
The input to the NN is build from the concatenation of the following $10$ scalar quantities.
\begin{itemize}
	\item The Bayesian posterior probabilities for the hypotheses on the values of the phases $\varphi_i$, i.e. $\lbrace w_j \rbrace_{j=1}^8$ with $\sum_{j=1}^8 w_j = 1$.

	\item The index of the measurement step normalized in $[-1, +1]$, i.e. $\wt{t} := 2 t/M\ped{max} -1$, where $M\ped{max}$ is the number of measurements. A single use of the interferometer, in which we sent an input state and read all the four photon counters is treated as a single measurement.

	\item The normalized average number of photons consumed in all the previous measurements, indicated with $\widetilde{N}\ped{ph}$.
\end{itemize}
The three controls phases $c_i$ are computed as
\begin{equation}
	\begin{pmatrix} c_0 \\ c_1 \\ c_2 \end{pmatrix} = 2 \pi \cdot f\ped{NN} \left( \lbrace w_j \rbrace_{j=1}^8, \wt{t}, \wt{N}\ped{ph} \right) \; .
\end{equation}

\subsubsection*{Definition of precision figure of merit}
On each estimation in the batch the error is zero if the correct phases are guessed correctly and one if the guess is wrong. Let us indicate with the symbol $h \in \lbrace 1, 2, \ldots, 8 \rbrace$ the index of the correct value, then the estimator for this discrete parameter is
\begin{equation}
	\widehat{h} \left( \lbrace w_j \rbrace_{j=1}^{8} \right) := \text{argmax} (w_1, w_2, \dots w_8) \; ,
\end{equation}
and the loss is given by
\begin{equation}
	\ell(\widehat{h}, h) := 1 - \delta (\widehat{h}, h) \; ,
\end{equation}
which, averaged on the batch, is the probability of a wrong classification.

\subsubsection*{Discussion of the results}
\begin{figure*}[htb]
	\centering
	\includegraphics[scale=1.0]{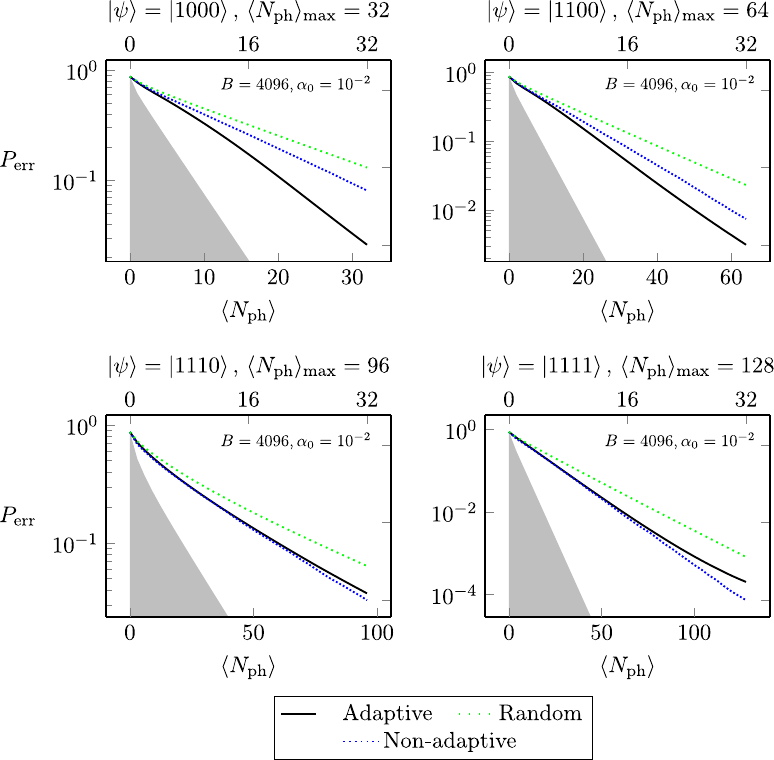}
	\caption{These plots display the probability of incorrectly guessing the value of the unknown phases $(\varphi_0, \varphi_1, \varphi_2)$ as a function of the average number of photons consumed during the estimation, shown on the lower x-axis. The title of each plot indicates the input state and the maximum average number of photons $\langle N_\text{ph} \rangle_{\text{max}}$ used for this values discrimination task. The upper x-axis represents the number of measurements $t$, i.e., the number of input states $\ket{\psi}$ utilized. The optimized adaptive and non-adaptive strategies are compared to the random strategy, where the controls $c_i$ are randomly selected uniformly in the interval $[0, 2 \pi)$. The shaded grey areas indicate the performances of the pretty good measurement (PGM), computed for multiple copies of the encoded states, see \cref{subsec:multiphase_discrimination_bound} for the details. This is not the ultimate precision bound regarding this discrimination problems, but it is a reasonable reference value not achievable with linear optics. In each plot we reported also the batch size $B$ and the initial learning rate $\alpha_0$ used in the simulations for the ``Adaptive'' strategy.}
	\label{fig:photonic_circuit}
\end{figure*}

In our simulations the input state always takes the form of a product of coherent states on the four bosonic wires, i.e. $\ket{\psi\ped{in}} = \ket{\alpha_0 \alpha_1 \alpha_2 \alpha_3}$. We investigate the performance of both optimized adaptive and non-adaptive strategies for the following four inputs having different average number of photons $\langle n \rangle$.
\begin{itemize}
	\item $\ket{\psi\ped{in}} = \ket{1000}$: $\alpha_0=1$, $\alpha_1=\alpha_2=\alpha_3=0$, $\langle n \rangle = 1$:
	\item $\ket{\psi\ped{in}} = \ket{1100}$: $\alpha_0=\alpha_1=1$, $\alpha_2=\alpha_3=0$, $\langle n \rangle = 2$:
	\item $\ket{\psi\ped{in}} = \ket{1110}$: $\alpha_0=\alpha_1=\alpha_2=1$, $\alpha_3=0$, $\langle n \rangle = 3$:
	\item $\ket{\psi\ped{in}} = \ket{1111}$: $\alpha_0=\alpha_1=\alpha_2=\alpha_3=1$, $\langle n \rangle = 4$:
\end{itemize}
The total number of measurements for each estimation remains fixed to $M_{\text{max}} = 32$ in all the plots of \cref{fig:photonic_circuit}, however, the maximum number of photons is given by $\langle N_{\text{ph}} \rangle_{\text{max}} := \langle n \rangle M_{\text{max}}$, and varies according to the input state.
%
From \cref{fig:photonic_circuit}, we infer that for the input states $\ket{\psi\ped{in}} = \ket{1000}, \, \ket{1100}$, the adaptive strategy offers some advantage over the non-adaptive one, while for states with a higher number of photons the non-adaptive strategy is optimal. The most efficient input, in terms of the number of photons consumed to achieve a specific error probability, is $\ket{\psi\ped{in}} = \ket{1000}$. The state with three photons doesn't perform as well as the other inputs. Regarding the damage to the sample or the energy consumed, both of which are proportional to the total number of photons, it is preferable to conduct multiple measurements, each involving fewer photons, rather than fewer measurements with a larger number of photons. Thus, minimizing energy consumption or potential sample damage requires extending the total measurement time, which may become impractical beyond a certain limit. For a set number of measurements, and hence a fixed estimation time, states with a higher photon number generally outperform those with fewer photons, with $\ket{\psi\ped{in}} = \ket{1110}$ being an exception. The observation that either the states with high or low photon count are optimal, depending on the resources we consider, underlines the importance of agreeing on the nature of the expensive resource while discussing metrology. For the states $\ket{\psi} = \ket{1111}$ the NN is not able to perfectly reproduce the performances of the optimal non-adaptive strategy, most probably because it converges to a local minimum of the loss during the training. Both the NN and the non-adaptive strategies have been randomly initialized in $[0, 2\pi)$ before the training.

\subsection{Non-adaptive linear classifier of coherent states}
\label{subsec:lin_multiclassifier}
\subsubsection{Description of the task}
\begin{figure*}[htb]
	\centering
	\includesvg[width=0.8\textwidth]{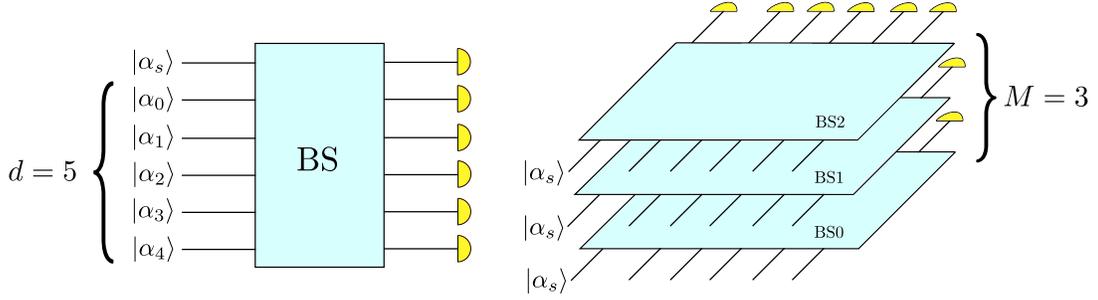}
	\caption{On the left we represent the programmable interferometer with the BS network, the input being the signal state $\ket{\alpha_s}$ and the fixed reference states $\ket{\alpha_i}$ with $i=0, 1, \ldots, d-1$. On the right we have stacked $M$ layers of interferometers, each receiving a copy of the signal. The classification is done on the basis of the outcomes of the photodetectors.}
	\label{fig:network_beam_splitter}
\end{figure*}

In the example of this section the goal is to classify a coherent state $\ket{\alpha_s}$, which we are promised is one of the states $\lbrace \ket{\alpha_i} \rbrace_{i=1}^{d-1}$, which are classically known, at difference with \cref{subsec:qml_classifier}. For doing this we will employ a network of beam splitters and phase plates (BS network) that receives in input the signal $|\alpha_s\rangle$ to be classified. The other inputs of the network are in order $\ket{\alpha_0}, \ket{\alpha_1}, \ldots, \ket{\alpha_{d-1}}$, that is, the reference states corresponding to the possible values of signal, which are fixed for each execution of the task. It follows that the network, represented in \cref{fig:network_beam_splitter}, must have $d+1$ bosonic wires.
%
The output of the network is measured with individual photon counters on each wire. The signal $|\alpha_s\rangle$ will be classified according to the photon counters outcomes. If more than one signal state is available multiple BS networks can be stacked on one another and the classification will be based on the measurements from all the layers. Each layer can have a different network. This is the only example in which we don't use a NN, that means we don't have adaptivity. Instead, we directly optimized the parametrized BS. For the same reason, it is very simple to describe the dynamical model with the framework defined in \cref{sec:dynamical_model}. There is no encoding of the state, meaning that the LCPT map $\Phi_{\vtheta}^{(t, x_t)}$ is the identity, and the measurement operators can be written as a product of the unitary matrix implementing the BS network followed by projectors on the Fock states, similarly to what has been done for the Dolinar receiver in \cref{subsec:dolinar}. We now briefly comment on the parametrization of the BS network. The initial state of the light is Gaussian, just like the linear circuit is a Gaussian operation~\cite{brask_gaussian_2022}. We can keep track of the Gaussian state of $d+1$ modes by following the evolution of the displacement $\vpos \in \mathbb{R}^{2 (d+1)}$ and of the covariance matrix $\Sigma \in \mathbb{R}^{2 (d+1)} \times \mathbb{R}^{2 (d+1)}$, which in general is
\begin{equation}
	\vpos \rightarrow \vpos' := S \vpos + \boldsymbol{d} \; , \quad \text{and} \quad \Sigma \rightarrow \Sigma' := S \Sigma S^\intercal \; ,
\end{equation}
where $S$ is a real symplectic matrix. In this context $\Sigma$ is not the covariance matrix of the posterior distribution but the covariance of Gaussian state of the modes. For a BS network, which is made of passive elements, we have no additional displacement, i.e. $\boldsymbol{d} = 0$. Furthermore the energy is conserved, which means $\tr \left( \Sigma' \right) = \tr \left( \Sigma S^\intercal S \right)$, that implies $S^\intercal S = \id$, since this condition must hold for every $\Sigma$. We have arrived at the results, that the action of a BS network is represented by a real symplectic and orthogonal matrix $S$. These two conditions are equivalent to the matrix $S$ being in the form
\begin{equation}
	S = \begin{pmatrix}
		\Re U & \Im U \\
		-\Im U & \Re U
	\end{pmatrix} \; , \quad \text{with} \quad U \in U(d+1) \; .
\end{equation}
The unitary matrix $U$ can then be parametrized as $U = e^{i (A+A^\dagger)}$, where $A \in GL(d+1, \mathbb{C})$ is a complex matrix, hose entries are the trainable variables of the agent, i.e. the BS network in this case. Despite not being an adaptive experiment, this is still a Bayesian estimation, since we start from a uniform prior over the hypotheses and apply the Bayes' rule to incorporate the outcomes of the photon counters in the posterior.

\subsubsection*{Definition of the precision figure of merit}
After all the measurements have been performed we obtain the discrete Bayesian posterior probabilities on the possible states, i.e. $\lbrace \widehat{p}_i \rbrace_{i=0}^{d-1}$, from which we define the estimator for the signal's class
\begin{equation}
	\widehat{s} = \text{argmax} \left( \widehat{p}_0, \widehat{p}_1, \ldots, \widehat{p}_{d-1} \right) \; ,
\end{equation}
and the loss for an instance of the task
\begin{equation}
	\ell(\widehat{s}, s) := 1 - \delta (\widehat{s}, s) \; ,
\end{equation}
which averaged over a batch of tests converges to the error probability of classification.

\subsubsection*{Discussion of the results}
%
\begin{figure*}[htb]
	\centering
	\includegraphics[scale=1.0]{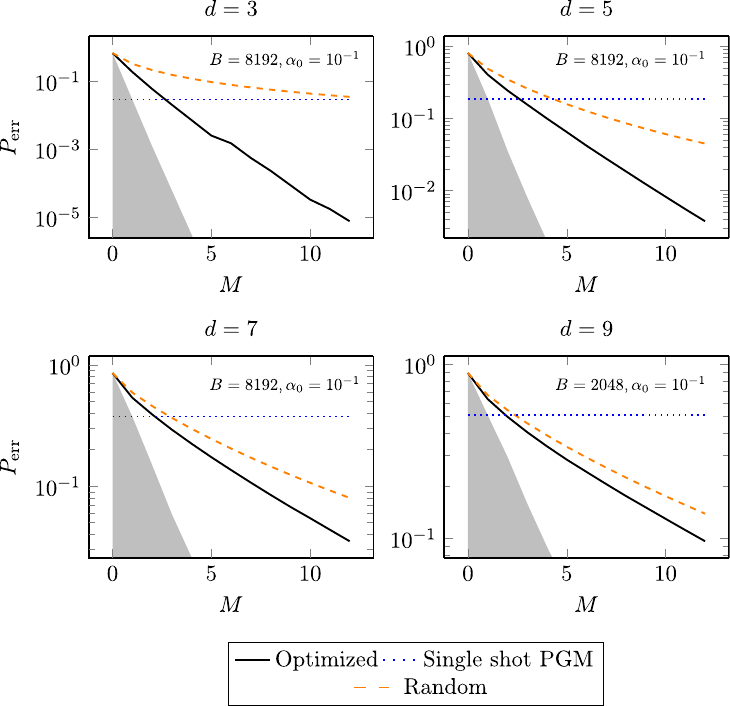}
	\caption{Probability of error in the classification of a coherent state as a function of the number $M$ of copies of the signal $\ket{\alpha_s}$ used, which is also the number of layers of BS networks used in the classifier. The dashed orange line is the average performance of randomly extracted BS networks. The dotted blue line is the optimal error probability for a single copy of the signal, while the shaded grey area indicates the optimal error probability for $M$ copies of the signal, both computed with the PGM. More details can be found in \cref{subsec:linear_multi_bound}. In each plot we reported also the batch size $B$ and the initial learning rate $\alpha_0$ used in the simulations for the ``Optimised'' strategy.}
	\label{fig:linear_classifier}
\end{figure*}
The linear classifier has been trained and tested for some symmetric configurations of the $\alpha_i$. In particular the roots of the unit for $d=3, 5, 7, 9$ have been chosen. The results are reported in \cref{fig:linear_classifier}. For a symmetric configuration of states the PGM is optimal, but it can’t be achieved with linear optics, see \cref{subsec:pgm_bound} and \cref{subsec:linear_multi_bound} for details. The optimality of the PGM means that a single layer BS network can’t achieve its performances, however this raises the interesting question of how many layers $M$, and copies of the signal $|\alpha_s\rangle$, are necessary to match or surpass the error probability of the PGM. Three copies of the signal are always sufficient in these examples to match the PGM error probability. Beside that, the performances of the trained BS networks are also compared to that of randomly extracted BS networks. An interesting observation is that the advantage over the random strategy that the trained network has, decreases as the number of hypotheses grows. For $d \rightarrow \infty$ we expect to have no gain. An adaptive strategy may be able to restore the gap with the untrained BS networks.

\section{Conclusions}

Our findings indicate that model-aware RL outperforms traditional control strategies in multiple scenarios, surpassing even model-free RL. Thee fact that many problems can be solved by minimal changes to the examples could pave the way for researchers to expedite the search for optimal controls in quantum sensors, potentially accelerating the advent of their widespread industrial application. With this work, we believe we have solved the problem of optimal controls for quantum sensors that are simple enough to be efficiently simulated. Since the sensor's model must be simulated on a computer with the overhead of automatic differentiation, we do not expect to be able to scale this procedure to sensors more complex than a handful of qubits in the near future. This limitation excludes the application of model-aware RL to domains where complex entangled states are used. However, given their limited experimental use at the moment, this constraint on efficient simulability may not significantly impact the utility of model-aware RL. It is notable that the most relevant platform for quantum metrology today is the single spin in diamond, which is straightforward to simulate. While large quantum systems remain inaccessible, we believe that scaling the procedure to a large number of unknown parameters is feasible. Although the number of particles and, consequently, the memory requirements must increase for a growing number of parameters, we have already successfully optimized a seven-parameters problem in \cref{subsec:qml_classifier}. We believe it is possible to find even more efficient representations of the Bayesian posterior distribution than the particle filter, which will enable us to optimize problems with tens of parameters. While we acknowledge that experimentally relevant problems can only rarely be approached analytically, the results obtained in this way can often be valuable in setting up and guiding more powerful numerical optimization, just as the analytical study of the Fisher information in \cref{sec:nv_center_bound} and of phase estimation in~\cite{belliardo_achieving_2020} and \cref{sec:anal_meas} have been useful in setting the initial state of the neural network and of the adaptive controls for the NV centers, and in defining the normalization of the network's output.

\section{Acknowledgments}
We gratefully acknowledge the computational resources of the Center for High Performance Computing (CHPC) at SNS. F. B. thanks F. Marquardt for useful discussions. We aknowledge finantial support by MUR (Ministero dell'Istruzione, dell'Università e della Ricerca) through the following projects: PNRR MUR project PE0000023-NQSTI, PRIN 2017 Taming complexity via Quantum Strategies: a Hybrid Integrated Photonic approach (QU-SHIP) Id. 2017SRN-BRK. Lastly, this work was supported by the Open Access Publishing Fund of the Scuola Normale Superiore.

\bibliographystyle{naturemag}
\bibliography{bibliografia.bib}

\begin{thebibliography}{10}
\expandafter\ifx\csname url\endcsname\relax
  \def\url#1{\texttt{#1}}\fi
\expandafter\ifx\csname urlprefix\endcsname\relax\def\urlprefix{URL }\fi
\providecommand{\bibinfo}[2]{#2}
\providecommand{\eprint}[2][]{\url{#2}}

\bibitem{flamini_photonic_2020}
\bibinfo{author}{Flamini, F.} \emph{et~al.}
\newblock \bibinfo{title}{Photonic architecture for reinforcement learning}.
\newblock \emph{\bibinfo{journal}{New Journal of Physics}}
  \textbf{\bibinfo{volume}{22}}, \bibinfo{pages}{045002}
  (\bibinfo{year}{2020}).
\newblock
  \urlprefix\url{https://iopscience.iop.org/article/10.1088/1367-2630/ab783c}.

\bibitem{broughton_tensorflow_2021}
\bibinfo{author}{Broughton, M.} \emph{et~al.}
\newblock \bibinfo{title}{{TensorFlow} {Quantum}: {A} {Software} {Framework}
  for {Quantum} {Machine} {Learning}} (\bibinfo{year}{2021}).
\newblock \urlprefix\url{http://arxiv.org/abs/2003.02989}.

\bibitem{bergholm_pennylane_2022}
\bibinfo{author}{Bergholm, V.} \emph{et~al.}
\newblock \bibinfo{title}{{PennyLane}: {Automatic} differentiation of hybrid
  quantum-classical computations} (\bibinfo{year}{2022}).
\newblock \urlprefix\url{http://arxiv.org/abs/1811.04968}.

\bibitem{bukov_reinforcement_2018}
\bibinfo{author}{Bukov, M.} \emph{et~al.}
\newblock \bibinfo{title}{Reinforcement {Learning} in {Different} {Phases} of
  {Quantum} {Control}}.
\newblock \emph{\bibinfo{journal}{Physical Review X}}
  \textbf{\bibinfo{volume}{8}}, \bibinfo{pages}{031086} (\bibinfo{year}{2018}).
\newblock \urlprefix\url{https://link.aps.org/doi/10.1103/PhysRevX.8.031086}.

\bibitem{zhang_when_2019}
\bibinfo{author}{Zhang, X.-M.}, \bibinfo{author}{Wei, Z.},
  \bibinfo{author}{Asad, R.}, \bibinfo{author}{Yang, X.-C.} \&
  \bibinfo{author}{Wang, X.}
\newblock \bibinfo{title}{When does reinforcement learning stand out in quantum
  control? {A} comparative study on state preparation}.
\newblock \emph{\bibinfo{journal}{npj Quantum Information}}
  \textbf{\bibinfo{volume}{5}}, \bibinfo{pages}{85} (\bibinfo{year}{2019}).
\newblock \urlprefix\url{http://www.nature.com/articles/s41534-019-0201-8}.

\bibitem{niu_universal_2019}
\bibinfo{author}{Niu, M.~Y.}, \bibinfo{author}{Boixo, S.},
  \bibinfo{author}{Smelyanskiy, V.~N.} \& \bibinfo{author}{Neven, H.}
\newblock \bibinfo{title}{Universal quantum control through deep reinforcement
  learning}.
\newblock \emph{\bibinfo{journal}{npj Quantum Information}}
  \textbf{\bibinfo{volume}{5}}, \bibinfo{pages}{33} (\bibinfo{year}{2019}).
\newblock \urlprefix\url{http://www.nature.com/articles/s41534-019-0141-3}.

\bibitem{porotti_deep_2022}
\bibinfo{author}{Porotti, R.}, \bibinfo{author}{Essig, A.},
  \bibinfo{author}{Huard, B.} \& \bibinfo{author}{Marquardt, F.}
\newblock \bibinfo{title}{Deep {Reinforcement} {Learning} for {Quantum} {State}
  {Preparation} with {Weak} {Nonlinear} {Measurements}}.
\newblock \emph{\bibinfo{journal}{Quantum}} \textbf{\bibinfo{volume}{6}},
  \bibinfo{pages}{747} (\bibinfo{year}{2022}).
\newblock \urlprefix\url{https://quantum-journal.org/papers/q-2022-06-28-747/}.

\bibitem{porotti_gradient-ascent_2023}
\bibinfo{author}{Porotti, R.}, \bibinfo{author}{Peano, V.} \&
  \bibinfo{author}{Marquardt, F.}
\newblock \bibinfo{title}{Gradient-{Ascent} {Pulse} {Engineering} with
  {Feedback}}.
\newblock \emph{\bibinfo{journal}{PRX Quantum}} \textbf{\bibinfo{volume}{4}},
  \bibinfo{pages}{030305} (\bibinfo{year}{2023}).
\newblock \urlprefix\url{https://link.aps.org/doi/10.1103/PRXQuantum.4.030305}.

\bibitem{fosel_reinforcement_2018}
\bibinfo{author}{Fösel, T.}, \bibinfo{author}{Tighineanu, P.},
  \bibinfo{author}{Weiss, T.} \& \bibinfo{author}{Marquardt, F.}
\newblock \bibinfo{title}{Reinforcement {Learning} with {Neural} {Networks} for
  {Quantum} {Feedback}}.
\newblock \emph{\bibinfo{journal}{Physical Review X}}
  \textbf{\bibinfo{volume}{8}}, \bibinfo{pages}{031084} (\bibinfo{year}{2018}).
\newblock \urlprefix\url{https://link.aps.org/doi/10.1103/PhysRevX.8.031084}.

\bibitem{cimini_calibration_2019}
\bibinfo{author}{Cimini, V.} \emph{et~al.}
\newblock \bibinfo{title}{Calibration of {Quantum} {Sensors} by {Neural}
  {Networks}}.
\newblock \emph{\bibinfo{journal}{Physical Review Letters}}
  \textbf{\bibinfo{volume}{123}}, \bibinfo{pages}{230502}
  (\bibinfo{year}{2019}).
\newblock
  \urlprefix\url{https://link.aps.org/doi/10.1103/PhysRevLett.123.230502}.

\bibitem{ban_neural-network-based_2021}
\bibinfo{author}{Ban, Y.}, \bibinfo{author}{Echanobe, J.},
  \bibinfo{author}{Ding, Y.}, \bibinfo{author}{Puebla, R.} \&
  \bibinfo{author}{Casanova, J.}
\newblock \bibinfo{title}{Neural-network-based parameter estimation for quantum
  detection}.
\newblock \emph{\bibinfo{journal}{Quantum Science and Technology}}
  \textbf{\bibinfo{volume}{6}}, \bibinfo{pages}{045012} (\bibinfo{year}{2021}).
\newblock
  \urlprefix\url{https://iopscience.iop.org/article/10.1088/2058-9565/ac16ed}.

\bibitem{nolan_machine_2021}
\bibinfo{author}{Nolan, S.}, \bibinfo{author}{Smerzi, A.} \&
  \bibinfo{author}{Pezzè, L.}
\newblock \bibinfo{title}{A machine learning approach to {Bayesian} parameter
  estimation}.
\newblock \emph{\bibinfo{journal}{npj Quantum Information}}
  \textbf{\bibinfo{volume}{7}}, \bibinfo{pages}{169} (\bibinfo{year}{2021}).
\newblock \urlprefix\url{https://www.nature.com/articles/s41534-021-00497-w}.

\bibitem{nolan_frequentist_2021}
\bibinfo{author}{Nolan, S.~P.}, \bibinfo{author}{Pezzè, L.} \&
  \bibinfo{author}{Smerzi, A.}
\newblock \bibinfo{title}{Frequentist parameter estimation with supervised
  learning}.
\newblock \emph{\bibinfo{journal}{AVS Quantum Science}}
  \textbf{\bibinfo{volume}{3}}, \bibinfo{pages}{034401} (\bibinfo{year}{2021}).
\newblock \urlprefix\url{https://avs.scitation.org/doi/10.1116/5.0058163}.

\bibitem{nguyen_deep_2021}
\bibinfo{author}{Nguyen, V.} \emph{et~al.}
\newblock \bibinfo{title}{Deep reinforcement learning for efficient measurement
  of quantum devices}.
\newblock \emph{\bibinfo{journal}{npj Quantum Information}}
  \textbf{\bibinfo{volume}{7}}, \bibinfo{pages}{100} (\bibinfo{year}{2021}).
\newblock \urlprefix\url{http://www.nature.com/articles/s41534-021-00434-x}.

\bibitem{palmieri_experimental_2020}
\bibinfo{author}{Palmieri, A.~M.} \emph{et~al.}
\newblock \bibinfo{title}{Experimental neural network enhanced quantum
  tomography}.
\newblock \emph{\bibinfo{journal}{npj Quantum Information}}
  \textbf{\bibinfo{volume}{6}}, \bibinfo{pages}{20} (\bibinfo{year}{2020}).
\newblock \urlprefix\url{http://www.nature.com/articles/s41534-020-0248-6}.

\bibitem{quek_adaptive_2021}
\bibinfo{author}{Quek, Y.}, \bibinfo{author}{Fort, S.} \& \bibinfo{author}{Ng,
  H.~K.}
\newblock \bibinfo{title}{Adaptive quantum state tomography with neural
  networks}.
\newblock \emph{\bibinfo{journal}{npj Quantum Information}}
  \textbf{\bibinfo{volume}{7}}, \bibinfo{pages}{105} (\bibinfo{year}{2021}).
\newblock \urlprefix\url{http://www.nature.com/articles/s41534-021-00436-9}.

\bibitem{hsieh_direct_2022}
\bibinfo{author}{Hsieh, H.-Y.} \emph{et~al.}
\newblock \bibinfo{title}{Direct {Parameter} {Estimations} from {Machine}
  {Learning}-{Enhanced} {Quantum} {State} {Tomography}}.
\newblock \emph{\bibinfo{journal}{Symmetry}} \textbf{\bibinfo{volume}{14}},
  \bibinfo{pages}{874} (\bibinfo{year}{2022}).
\newblock \urlprefix\url{https://www.mdpi.com/2073-8994/14/5/874}.

\bibitem{marquardt_machine_2021}
\bibinfo{author}{Marquardt, F.}
\newblock \bibinfo{title}{Machine learning and quantum devices}.
\newblock \emph{\bibinfo{journal}{SciPost Physics Lecture Notes}}
  \bibinfo{pages}{29} (\bibinfo{year}{2021}).
\newblock \urlprefix\url{https://scipost.org/10.21468/SciPostPhysLectNotes.29}.

\bibitem{marquardt_online_2021}
\bibinfo{author}{Marquardt, F.}
\newblock \bibinfo{title}{Online {Course}: {Advanced} {Machine} {Learning} for
  {Physics}, {Science}, and {Artificial} {Scientific} {Discovery}}
  (\bibinfo{year}{2021}).

\bibitem{krenn_artificial_2023}
\bibinfo{author}{Krenn, M.}, \bibinfo{author}{Landgraf, J.},
  \bibinfo{author}{Foesel, T.} \& \bibinfo{author}{Marquardt, F.}
\newblock \bibinfo{title}{Artificial intelligence and machine learning for
  quantum technologies}.
\newblock \emph{\bibinfo{journal}{Physical Review A}}
  \textbf{\bibinfo{volume}{107}}, \bibinfo{pages}{010101}
  (\bibinfo{year}{2023}).
\newblock \urlprefix\url{https://link.aps.org/doi/10.1103/PhysRevA.107.010101}.

\bibitem{fisher_design_1935}
\bibinfo{author}{Fisher, R.~A.}
\newblock \emph{\bibinfo{title}{The design of experiments}}.
\newblock The design of experiments (\bibinfo{publisher}{Oliver \& Boyd},
  \bibinfo{address}{Oxford, England}, \bibinfo{year}{1935}).

\bibitem{foster_variational_2021}
\bibinfo{author}{Foster, A.~E.}
\newblock \emph{\bibinfo{title}{Variational, {Monte} {Carlo} and policy-based
  approaches to {Bayesian} experimental design}}.
\newblock \bibinfo{type}{http://purl.org/dc/dcmitype/{Text}},
  \bibinfo{school}{University of Oxford} (\bibinfo{year}{2021}).
\newblock
  \urlprefix\url{https://ora.ox.ac.uk/objects/uuid:4a3e13ca-e6c6-4669-955e-f1a87e201228}.

\bibitem{baydin_toward_2021}
\bibinfo{author}{Baydin, A.~G.} \emph{et~al.}
\newblock \bibinfo{title}{Toward {Machine} {Learning} {Optimization} of
  {Experimental} {Design}}.
\newblock \emph{\bibinfo{journal}{Nuclear Physics News}}
  \textbf{\bibinfo{volume}{31}}, \bibinfo{pages}{25--28}
  (\bibinfo{year}{2021}).
\newblock
  \urlprefix\url{https://www.tandfonline.com/doi/full/10.1080/10619127.2021.1881364}.

\bibitem{ballard_machine_2021}
\bibinfo{author}{Ballard, Z.}, \bibinfo{author}{Brown, C.},
  \bibinfo{author}{Madni, A.~M.} \& \bibinfo{author}{Ozcan, A.}
\newblock \bibinfo{title}{Machine learning and computation-enabled intelligent
  sensor design}.
\newblock \emph{\bibinfo{journal}{Nature Machine Intelligence}}
  \textbf{\bibinfo{volume}{3}}, \bibinfo{pages}{556--565}
  (\bibinfo{year}{2021}).
\newblock \urlprefix\url{http://www.nature.com/articles/s42256-021-00360-9}.

\bibitem{ivanova_implicit_2021}
\bibinfo{author}{Ivanova, D.~R.}, \bibinfo{author}{Foster, A.},
  \bibinfo{author}{Kleinegesse, S.}, \bibinfo{author}{Gutmann, M.~U.} \&
  \bibinfo{author}{Rainforth, T.}
\newblock \bibinfo{title}{Implicit {Deep} {Adaptive} {Design}: {Policy}-{Based}
  {Experimental} {Design} without {Likelihoods}}.
\newblock In \emph{\bibinfo{booktitle}{Advances in {Neural} {Information}
  {Processing} {Systems}}}, vol.~\bibinfo{volume}{34},
  \bibinfo{pages}{25785--25798} (\bibinfo{publisher}{Curran Associates, Inc.},
  \bibinfo{year}{2021}).
\newblock
  \urlprefix\url{https://proceedings.neurips.cc/paper/2021/hash/d811406316b669ad3d370d78b51b1d2e-Abstract.html}.

\bibitem{foster_deep_2021}
\bibinfo{author}{Foster, A.}, \bibinfo{author}{Ivanova, D.~R.},
  \bibinfo{author}{Malik, I.} \& \bibinfo{author}{Rainforth, T.}
\newblock \bibinfo{title}{Deep {Adaptive} {Design}: {Amortizing} {Sequential}
  {Bayesian} {Experimental} {Design}}.
\newblock In \emph{\bibinfo{booktitle}{Proceedings of the 38th {International}
  {Conference} on {Machine} {Learning}}}, \bibinfo{pages}{3384--3395}
  (\bibinfo{publisher}{PMLR}, \bibinfo{year}{2021}).
\newblock \urlprefix\url{https://proceedings.mlr.press/v139/foster21a.html}.

\bibitem{fiderer_neural-network_2021}
\bibinfo{author}{Fiderer, L.~J.}, \bibinfo{author}{Schuff, J.} \&
  \bibinfo{author}{Braun, D.}
\newblock \bibinfo{title}{Neural-{Network} {Heuristics} for {Adaptive}
  {Bayesian} {Quantum} {Estimation}}.
\newblock \emph{\bibinfo{journal}{PRX Quantum}} \textbf{\bibinfo{volume}{2}},
  \bibinfo{pages}{020303} (\bibinfo{year}{2021}).
\newblock \urlprefix\url{https://link.aps.org/doi/10.1103/PRXQuantum.2.020303}.

\bibitem{scibior_differentiable_2021}
\bibinfo{author}{Ścibior, A.} \& \bibinfo{author}{Wood, F.}
\newblock \bibinfo{title}{Differentiable {Particle} {Filtering} without
  {Modifying} the {Forward} {Pass}} (\bibinfo{year}{2021}).
\newblock \urlprefix\url{http://arxiv.org/abs/2106.10314}.

\bibitem{qsensoropt_gitlab}
\bibinfo{author}{Belliardo, F.}, \bibinfo{author}{Zoratti, F.} \&
  \bibinfo{author}{Giovannetti, V.}
\newblock \bibinfo{title}{qsensoropt: quantum sensor optimisation}
  (\bibinfo{year}{2022}).
\newblock \urlprefix\url{https://gitlab.com/federico.belliardo/qsensoropt}.

\bibitem{qsensoropt_doc}
\bibinfo{author}{Belliardo, F.}, \bibinfo{author}{Zoratti, F.} \&
  \bibinfo{author}{Giovannetti, V.}
\newblock \bibinfo{title}{qsensoropt documentation} (\bibinfo{year}{2022}).
\newblock
  \urlprefix\url{https://qsensoropt-federico-belliardo-aafff0229087adae5a915fec60fdc5d37.gitlab.io/}.

\bibitem{zoratti_agnostic-dolinar_2021}
\bibinfo{author}{Zoratti, F.}, \bibinfo{author}{Pozza, N.~D.},
  \bibinfo{author}{Fanizza, M.} \& \bibinfo{author}{Giovannetti, V.}
\newblock \bibinfo{title}{An agnostic-{Dolinar} receiver for coherent states
  classification}.
\newblock \emph{\bibinfo{journal}{Physical Review A}}
  \textbf{\bibinfo{volume}{104}}, \bibinfo{pages}{042606}
  (\bibinfo{year}{2021}).
\newblock \urlprefix\url{http://arxiv.org/abs/2106.11909}.

\bibitem{meyer_variational_2021}
\bibinfo{author}{Meyer, J.~J.}, \bibinfo{author}{Borregaard, J.} \&
  \bibinfo{author}{Eisert, J.}
\newblock \bibinfo{title}{A variational toolbox for quantum multi-parameter
  estimation}.
\newblock \emph{\bibinfo{journal}{npj Quantum Information}}
  \textbf{\bibinfo{volume}{7}}, \bibinfo{pages}{1--5} (\bibinfo{year}{2021}).
\newblock \urlprefix\url{https://www.nature.com/articles/s41534-021-00425-y}.

\bibitem{zhang_quanestimation_2022}
\bibinfo{author}{Zhang, M.} \emph{et~al.}
\newblock \bibinfo{title}{{QuanEstimation}: {An} open-source toolkit for
  quantum parameter estimation}.
\newblock \emph{\bibinfo{journal}{Physical Review Research}}
  \textbf{\bibinfo{volume}{4}}, \bibinfo{pages}{043057} (\bibinfo{year}{2022}).
\newblock
  \urlprefix\url{https://link.aps.org/doi/10.1103/PhysRevResearch.4.043057}.

\bibitem{granade_qinfer_2017}
\bibinfo{author}{Granade, C.} \emph{et~al.}
\newblock \bibinfo{title}{{QInfer}: {Statistical} inference software for
  quantum applications}.
\newblock \emph{\bibinfo{journal}{Quantum}} \textbf{\bibinfo{volume}{1}},
  \bibinfo{pages}{5} (\bibinfo{year}{2017}).
\newblock \urlprefix\url{http://arxiv.org/abs/1610.00336}.

\bibitem{mcmichael_optbayesexpt_2021}
\bibinfo{author}{McMichael, R.~D.}, \bibinfo{author}{Blakley, S.~M.} \&
  \bibinfo{author}{Dushenko, S.}
\newblock \bibinfo{title}{Optbayesexpt: {Sequential} {Bayesian} {Experiment}
  {Design} for {Adaptive} {Measurements}}.
\newblock \emph{\bibinfo{journal}{Journal of Research of the National Institute
  of Standards and Technology}} \textbf{\bibinfo{volume}{126}},
  \bibinfo{pages}{126002} (\bibinfo{year}{2021}).
\newblock
  \urlprefix\url{https://nvlpubs.nist.gov/nistpubs/jres/126/jres.126.002.pdf}.

\bibitem{bavaresco_designing_2023}
\bibinfo{author}{Bavaresco, J.}, \bibinfo{author}{Lipka-Bartosik, P.},
  \bibinfo{author}{Sekatski, P.} \& \bibinfo{author}{Mehboudi, M.}
\newblock \bibinfo{title}{Designing optimal protocols in {Bayesian} quantum
  parameter estimation with higher-order operations} (\bibinfo{year}{2023}).
\newblock \urlprefix\url{http://arxiv.org/abs/2311.01513}.

\bibitem{liu_quantum_2017}
\bibinfo{author}{Liu, J.} \& \bibinfo{author}{Yuan, H.}
\newblock \bibinfo{title}{Quantum parameter estimation with optimal control}.
\newblock \emph{\bibinfo{journal}{Physical Review A}}
  \textbf{\bibinfo{volume}{96}}, \bibinfo{pages}{012117}
  (\bibinfo{year}{2017}).
\newblock \urlprefix\url{http://link.aps.org/doi/10.1103/PhysRevA.96.012117}.

\bibitem{xu_generalizable_2019}
\bibinfo{author}{Xu, H.} \emph{et~al.}
\newblock \bibinfo{title}{Generalizable control for quantum parameter
  estimation through reinforcement learning}.
\newblock \emph{\bibinfo{journal}{npj Quantum Information}}
  \textbf{\bibinfo{volume}{5}}, \bibinfo{pages}{1--8} (\bibinfo{year}{2019}).
\newblock \urlprefix\url{https://www.nature.com/articles/s41534-019-0198-z}.

\bibitem{rembold_introduction_2020}
\bibinfo{author}{Rembold, P.} \emph{et~al.}
\newblock \bibinfo{title}{Introduction to quantum optimal control for quantum
  sensing with nitrogen-vacancy centers in diamond}.
\newblock \emph{\bibinfo{journal}{AVS Quantum Science}}
  \textbf{\bibinfo{volume}{2}}, \bibinfo{pages}{024701} (\bibinfo{year}{2020}).
\newblock \urlprefix\url{http://avs.scitation.org/doi/10.1116/5.0006785}.

\bibitem{schuff_improving_2020}
\bibinfo{author}{Schuff, J.}, \bibinfo{author}{Fiderer, L.~J.} \&
  \bibinfo{author}{Braun, D.}
\newblock \bibinfo{title}{Improving the dynamics of quantum sensors with
  reinforcement learning}.
\newblock \emph{\bibinfo{journal}{New Journal of Physics}}
  \textbf{\bibinfo{volume}{22}}, \bibinfo{pages}{035001}
  (\bibinfo{year}{2020}).
\newblock
  \urlprefix\url{https://iopscience.iop.org/article/10.1088/1367-2630/ab6f1f}.

\bibitem{xu_generalizable_2021}
\bibinfo{author}{Xu, H.}, \bibinfo{author}{Wang, L.}, \bibinfo{author}{Yuan,
  H.} \& \bibinfo{author}{Wang, X.}
\newblock \bibinfo{title}{Generalizable control for multiparameter quantum
  metrology}.
\newblock \emph{\bibinfo{journal}{Physical Review A}}
  \textbf{\bibinfo{volume}{103}}, \bibinfo{pages}{042615}
  (\bibinfo{year}{2021}).
\newblock \urlprefix\url{http://arxiv.org/abs/2012.13377}.

\bibitem{liu_optimal_2022}
\bibinfo{author}{Liu, J.}, \bibinfo{author}{Zhang, M.}, \bibinfo{author}{Chen,
  H.}, \bibinfo{author}{Wang, L.} \& \bibinfo{author}{Yuan, H.}
\newblock \bibinfo{title}{Optimal {Scheme} for {Quantum} {Metrology}}.
\newblock \emph{\bibinfo{journal}{Advanced Quantum Technologies}}
  \textbf{\bibinfo{volume}{5}}, \bibinfo{pages}{2100080}
  (\bibinfo{year}{2022}).
\newblock \urlprefix\url{http://arxiv.org/abs/2111.12279}.

\bibitem{xiao_parameter_2022}
\bibinfo{author}{Xiao, T.}, \bibinfo{author}{Fan, J.} \& \bibinfo{author}{Zeng,
  G.}
\newblock \bibinfo{title}{Parameter estimation in quantum sensing based on deep
  reinforcement learning}.
\newblock \emph{\bibinfo{journal}{npj Quantum Information}}
  \textbf{\bibinfo{volume}{8}}, \bibinfo{pages}{1--12} (\bibinfo{year}{2022}).
\newblock \urlprefix\url{https://www.nature.com/articles/s41534-021-00513-z}.

\bibitem{qiu_efficient_2022}
\bibinfo{author}{Qiu, Y.}, \bibinfo{author}{Zhuang, M.},
  \bibinfo{author}{Huang, J.} \& \bibinfo{author}{Lee, C.}
\newblock \bibinfo{title}{Efficient and robust entanglement generation with
  deep reinforcement learning for quantum metrology}.
\newblock \emph{\bibinfo{journal}{New Journal of Physics}}
  \textbf{\bibinfo{volume}{24}}, \bibinfo{pages}{083011}
  (\bibinfo{year}{2022}).
\newblock \urlprefix\url{https://dx.doi.org/10.1088/1367-2630/ac8285}.

\bibitem{vedaie_framework_2023}
\bibinfo{author}{Vedaie, S.~S.}, \bibinfo{author}{Dalal, A.},
  \bibinfo{author}{Páez, E.~J.} \& \bibinfo{author}{Sanders, B.~C.}
\newblock \bibinfo{title}{Framework for {Learning} and {Control} in the
  {Classical} and {Quantum} {Domains}} (\bibinfo{year}{2023}).
\newblock \urlprefix\url{http://arxiv.org/abs/2307.04256}.

\bibitem{gebhart_learning_2023}
\bibinfo{author}{Gebhart, V.} \emph{et~al.}
\newblock \bibinfo{title}{Learning quantum systems}.
\newblock \emph{\bibinfo{journal}{Nature Reviews Physics}}
  \textbf{\bibinfo{volume}{5}}, \bibinfo{pages}{141--156}
  (\bibinfo{year}{2023}).
\newblock \urlprefix\url{https://www.nature.com/articles/s42254-022-00552-1}.

\bibitem{ma_adaptive_2021}
\bibinfo{author}{Ma, Z.} \emph{et~al.}
\newblock \bibinfo{title}{Adaptive {Circuit} {Learning} for {Quantum}
  {Metrology}}.
\newblock In \emph{\bibinfo{booktitle}{2021 {IEEE} {International} {Conference}
  on {Quantum} {Computing} and {Engineering} ({QCE})}},
  \bibinfo{pages}{419--430} (\bibinfo{year}{2021}).
\newblock \urlprefix\url{http://arxiv.org/abs/2010.08702}.

\bibitem{kaubruegger_quantum_2021}
\bibinfo{author}{Kaubruegger, R.}, \bibinfo{author}{Vasilyev, D.~V.},
  \bibinfo{author}{Schulte, M.}, \bibinfo{author}{Hammerer, K.} \&
  \bibinfo{author}{Zoller, P.}
\newblock \bibinfo{title}{Quantum {Variational} {Optimization} of {Ramsey}
  {Interferometry} and {Atomic} {Clocks}}.
\newblock \emph{\bibinfo{journal}{Physical Review X}}
  \textbf{\bibinfo{volume}{11}}, \bibinfo{pages}{041045}
  (\bibinfo{year}{2021}).
\newblock \urlprefix\url{https://link.aps.org/doi/10.1103/PhysRevX.11.041045}.

\bibitem{marciniak_optimal_2022}
\bibinfo{author}{Marciniak, C.~D.} \emph{et~al.}
\newblock \bibinfo{title}{Optimal metrology with programmable quantum sensors}.
\newblock \emph{\bibinfo{journal}{Nature}} \textbf{\bibinfo{volume}{603}},
  \bibinfo{pages}{604--609} (\bibinfo{year}{2022}).
\newblock \urlprefix\url{https://www.nature.com/articles/s41586-022-04435-4}.

\bibitem{kaubruegger_optimal_2023}
\bibinfo{author}{Kaubruegger, R.}, \bibinfo{author}{Shankar, A.},
  \bibinfo{author}{Vasilyev, D.~V.} \& \bibinfo{author}{Zoller, P.}
\newblock \bibinfo{title}{Optimal and {Variational} {Multi}-{Parameter}
  {Quantum} {Metrology} and {Vector} {Field} {Sensing}} (\bibinfo{year}{2023}).
\newblock \urlprefix\url{http://arxiv.org/abs/2302.07785}.

\bibitem{kose_superresolution_2023}
\bibinfo{author}{Köse, E.} \& \bibinfo{author}{Braun, D.}
\newblock \bibinfo{title}{Superresolution imaging with multiparameter quantum
  metrology in passive remote sensing}.
\newblock \emph{\bibinfo{journal}{Physical Review A}}
  \textbf{\bibinfo{volume}{107}}, \bibinfo{pages}{032607}
  (\bibinfo{year}{2023}).
\newblock \urlprefix\url{https://link.aps.org/doi/10.1103/PhysRevA.107.032607}.

\bibitem{heras_photonic_2023}
\bibinfo{author}{Heras, A. M. d.~l.} \emph{et~al.}
\newblock \bibinfo{title}{Photonic quantum metrology with variational quantum
  optical non-linearities} (\bibinfo{year}{2023}).
\newblock \urlprefix\url{http://arxiv.org/abs/2309.09841}.

\bibitem{yang_variational_2022}
\bibinfo{author}{Yang, J.}, \bibinfo{author}{Pang, S.}, \bibinfo{author}{Chen,
  Z.}, \bibinfo{author}{Jordan, A.~N.} \& \bibinfo{author}{Del~Campo, A.}
\newblock \bibinfo{title}{Variational principle for optimal quantum controls in
  quantum metrology}.
\newblock \emph{\bibinfo{journal}{Physical Review Letters}}
  \textbf{\bibinfo{volume}{128}}, \bibinfo{pages}{160505}
  (\bibinfo{year}{2022}).
\newblock
  \urlprefix\url{https://link.aps.org/doi/10.1103/PhysRevLett.128.160505}.

\bibitem{belliardo_model-aware_2023}
\bibinfo{author}{Belliardo, F.}, \bibinfo{author}{Zoratti, F.},
  \bibinfo{author}{Marquardt, F.} \& \bibinfo{author}{Giovannetti, V.}
\newblock \bibinfo{title}{Model-aware reinforcement learning for
  high-performance bayesian experimental design in quantum metrology}.
\newblock \urlprefix\url{http://arxiv.org/abs/2312.16985}.
\newblock \eprint{2312.16985 [quant-ph]}.

\bibitem{belliardo_application_2024}
\bibinfo{author}{Belliardo, F.}, \bibinfo{author}{Zoratti, F.} \&
  \bibinfo{author}{Giovannetti, V.}
\newblock \bibinfo{title}{Application of machine learning to experimental
  design in quantum mechanics}.
\newblock \emph{\bibinfo{journal}{International Journal of Quantum
  Information}}  (\bibinfo{year}{2024}).
\newblock
  \urlprefix\url{https://www.worldscientific.com/doi/abs/10.1142/S0219749924500023}.

\bibitem{belliardo_optimizing_2022}
\bibinfo{author}{Belliardo, F.} \emph{et~al.}
\newblock \bibinfo{title}{Optimizing quantum-enhanced {Bayesian} multiparameter
  estimation in noisy apparata} (\bibinfo{year}{2022}).
\newblock \urlprefix\url{http://arxiv.org/abs/2211.04747}.
\newblock \bibinfo{note}{ArXiv:2211.04747 [quant-ph]}.

\bibitem{kingma_adam_2015}
\bibinfo{author}{Kingma, D.~P.} \& \bibinfo{author}{Ba, J.}
\newblock \bibinfo{title}{Adam: {A} {Method} for {Stochastic} {Optimization}}.
\newblock In \bibinfo{editor}{Bengio, Y.} \& \bibinfo{editor}{LeCun, Y.} (eds.)
  \emph{\bibinfo{booktitle}{3rd {International} {Conference} on {Learning}
  {Representations}, {ICLR} 2015, {San} {Diego}, {CA}, {USA}, {May} 7-9, 2015,
  {Conference} {Track} {Proceedings}}} (\bibinfo{year}{2015}).
\newblock \urlprefix\url{http://arxiv.org/abs/1412.6980}.

\bibitem{belliardo_achieving_2020}
\bibinfo{author}{Belliardo, F.} \& \bibinfo{author}{Giovannetti, V.}
\newblock \bibinfo{title}{Achieving {Heisenberg} scaling with maximally
  entangled states: {An} analytic upper bound for the attainable
  root-mean-square error}.
\newblock \emph{\bibinfo{journal}{Physical Review A}}
  \textbf{\bibinfo{volume}{102}}, \bibinfo{pages}{042613}
  (\bibinfo{year}{2020}).
\newblock \urlprefix\url{https://link.aps.org/doi/10.1103/PhysRevA.102.042613}.

\bibitem{nielsen00}
\bibinfo{author}{Nielsen, M.~A.} \& \bibinfo{author}{Chuang, I.~L.}
\newblock \emph{\bibinfo{title}{Quantum Computation and Quantum Information}}
  (\bibinfo{publisher}{Cambridge University Press}, \bibinfo{year}{2000}).

\bibitem{Holevo_book}
\bibinfo{author}{Holevo, A.~S.}
\newblock \emph{\bibinfo{title}{Quantum Systems, Channels, Information: A
  Mathematical Introduction}} (\bibinfo{publisher}{De Gruyter},
  \bibinfo{address}{Berlin, Boston}, \bibinfo{year}{2019}).
\newblock \urlprefix\url{https://doi.org/10.1515/9783110642490}.

\bibitem{doherty_quantum_2022}
\bibinfo{author}{Doherty, M.~W.}, \bibinfo{author}{Du, C.~R.} \&
  \bibinfo{author}{Fuchs, G.~D.}
\newblock \bibinfo{title}{Quantum science and technology based on color centers
  with accessible spin}.
\newblock \emph{\bibinfo{journal}{Journal of Applied Physics}}
  \textbf{\bibinfo{volume}{131}}, \bibinfo{pages}{010401}
  (\bibinfo{year}{2022}).
\newblock \urlprefix\url{https://aip.scitation.org/doi/10.1063/5.0082219}.

\bibitem{chen_quantum_2018}
\bibinfo{author}{Chen, M.} \emph{et~al.}
\newblock \bibinfo{title}{Quantum metrology with single spins in diamond under
  ambient conditions}.
\newblock \emph{\bibinfo{journal}{National Science Review}}
  \textbf{\bibinfo{volume}{5}}, \bibinfo{pages}{346--355}
  (\bibinfo{year}{2018}).
\newblock \urlprefix\url{https://academic.oup.com/nsr/article/5/3/346/4430770}.

\bibitem{maze_rios_quantum_2010}
\bibinfo{author}{Maze, J.}
\newblock \emph{\bibinfo{title}{Quantum manipulation of nitrogen-vacancy
  centers in diamond: From basic properties to applications}}.
\newblock Ph.D. thesis (\bibinfo{year}{2010}).

\bibitem{barry_sensitivity_2020}
\bibinfo{author}{Barry, J.~F.} \emph{et~al.}
\newblock \bibinfo{title}{Sensitivity optimization for {NV}-diamond
  magnetometry}.
\newblock \emph{\bibinfo{journal}{Rev. Mod. Phys.}}
  \textbf{\bibinfo{volume}{92}}, \bibinfo{pages}{68} (\bibinfo{year}{2020}).

\bibitem{arshad_real-time_2024}
\bibinfo{author}{Arshad, M.~J.} \emph{et~al.}
\newblock \bibinfo{title}{Real-time adaptive estimation of decoherence
  timescales for a single qubit}.
\newblock \emph{\bibinfo{journal}{Phys. Rev. Appl.}}
  \textbf{\bibinfo{volume}{21}}, \bibinfo{pages}{024026}
  (\bibinfo{year}{2024}).

\bibitem{joas_online_2021}
\bibinfo{author}{Joas, T.} \emph{et~al.}
\newblock \bibinfo{title}{Online adaptive quantum characterization of a nuclear
  spin}.
\newblock \emph{\bibinfo{journal}{npj Quantum Information}}
  \textbf{\bibinfo{volume}{7}}, \bibinfo{pages}{1--8} (\bibinfo{year}{2021}).
\newblock \urlprefix\url{https://www.nature.com/articles/s41534-021-00389-z}.

\bibitem{valeri_experimental_2020}
\bibinfo{author}{Valeri, M.}
\newblock \bibinfo{title}{Experimental adaptive {Bayesian} estimation of
  multiple phases with limited data}.
\newblock \emph{\bibinfo{journal}{npj Quantum Information}} \bibinfo{pages}{11}
  (\bibinfo{year}{2020}).

\bibitem{paesani_experimental_2017}
\bibinfo{author}{Paesani, S.} \emph{et~al.}
\newblock \bibinfo{title}{Experimental {Bayesian} {Quantum} {Phase}
  {Estimation} on a {Silicon} {Photonic} {Chip}}.
\newblock \emph{\bibinfo{journal}{Physical Review Letters}}
  \textbf{\bibinfo{volume}{118}}, \bibinfo{pages}{100503}
  (\bibinfo{year}{2017}).
\newblock
  \urlprefix\url{https://link.aps.org/doi/10.1103/PhysRevLett.118.100503}.

\bibitem{polino_experimental_2019}
\bibinfo{author}{Polino, E.} \emph{et~al.}
\newblock \bibinfo{title}{Experimental multiphase estimation on a chip}.
\newblock \emph{\bibinfo{journal}{Optica}} \textbf{\bibinfo{volume}{6}},
  \bibinfo{pages}{288} (\bibinfo{year}{2019}).
\newblock
  \urlprefix\url{https://www.osapublishing.org/abstract.cfm?URI=optica-6-3-288}.

\bibitem{polino_photonic_2020}
\bibinfo{author}{Polino, E.}, \bibinfo{author}{Valeri, M.},
  \bibinfo{author}{Spagnolo, N.} \& \bibinfo{author}{Sciarrino, F.}
\newblock \bibinfo{title}{Photonic quantum metrology}.
\newblock \emph{\bibinfo{journal}{AVS Quantum Science}}
  \textbf{\bibinfo{volume}{2}}, \bibinfo{pages}{024703} (\bibinfo{year}{2020}).
\newblock \urlprefix\url{http://avs.scitation.org/doi/10.1116/5.0007577}.

\bibitem{barbieri_quantum_2013}
\bibinfo{author}{Barbieri, M.}, \bibinfo{author}{Datta, A.},
  \bibinfo{author}{Walmsley, I.~A.} \& \bibinfo{author}{Humphreys, P.~C.}
\newblock \bibinfo{title}{Quantum {Enhanced} {Multiple} {Phase} {Estimation}}.
\newblock \emph{\bibinfo{journal}{Physical Review Letters}}
  \textbf{\bibinfo{volume}{111}}, \bibinfo{pages}{070403}
  (\bibinfo{year}{2013}).
\newblock
  \urlprefix\url{https://link.aps.org/doi/10.1103/PhysRevLett.111.070403}.

\bibitem{schmitt_optimal_2021}
\bibinfo{author}{Schmitt, S.} \emph{et~al.}
\newblock \bibinfo{title}{Optimal frequency measurements with quantum probes}.
\newblock \emph{\bibinfo{journal}{npj Quantum Information}}
  \textbf{\bibinfo{volume}{7}}, \bibinfo{pages}{55} (\bibinfo{year}{2021}).
\newblock \urlprefix\url{http://www.nature.com/articles/s41534-021-00391-5}.

\bibitem{ferrie_how_2013}
\bibinfo{author}{Ferrie, C.}, \bibinfo{author}{Granade, C.~E.} \&
  \bibinfo{author}{Cory, D.~G.}
\newblock \bibinfo{title}{How to best sample a periodic probability
  distribution, or on the accuracy of {Hamiltonian} finding strategies}.
\newblock \emph{\bibinfo{journal}{Quantum Information Processing}}
  \textbf{\bibinfo{volume}{12}}, \bibinfo{pages}{611--623}
  (\bibinfo{year}{2013}).
\newblock \urlprefix\url{http://link.springer.com/10.1007/s11128-012-0407-6}.

\bibitem{dushenko_sequential_2020}
\bibinfo{author}{Dushenko, S.}, \bibinfo{author}{Ambal, K.} \&
  \bibinfo{author}{McMichael, R.~D.}
\newblock \bibinfo{title}{Sequential {Bayesian} {Experiment} {Design} for
  {Optically} {Detected} {Magnetic} {Resonance} of {Nitrogen}-{Vacancy}
  {Centers}}.
\newblock \emph{\bibinfo{journal}{Physical Review Applied}}
  \textbf{\bibinfo{volume}{14}}, \bibinfo{pages}{054036}
  (\bibinfo{year}{2020}).
\newblock
  \urlprefix\url{https://link.aps.org/doi/10.1103/PhysRevApplied.14.054036}.

\bibitem{mcmichael_sequential_2021}
\bibinfo{author}{McMichael, R.~D.}, \bibinfo{author}{Dushenko, S.} \&
  \bibinfo{author}{Blakley, S.~M.}
\newblock \bibinfo{title}{Sequential {Bayesian} experiment design for adaptive
  {Ramsey} sequence measurements}.
\newblock \emph{\bibinfo{journal}{Journal of Applied Physics}}
  \textbf{\bibinfo{volume}{130}}, \bibinfo{pages}{144401}
  (\bibinfo{year}{2021}).
\newblock \urlprefix\url{https://aip.scitation.org/doi/10.1063/5.0055630}.

\bibitem{granade_robust_2012}
\bibinfo{author}{Granade, C.~E.}, \bibinfo{author}{Ferrie, C.},
  \bibinfo{author}{Wiebe, N.} \& \bibinfo{author}{Cory, D.~G.}
\newblock \bibinfo{title}{Robust online {Hamiltonian} learning}.
\newblock \emph{\bibinfo{journal}{New Journal of Physics}}
  \textbf{\bibinfo{volume}{14}}, \bibinfo{pages}{103013}
  (\bibinfo{year}{2012}).
\newblock \urlprefix\url{https://dx.doi.org/10.1088/1367-2630/14/10/103013}.

\bibitem{oshnik_robust_2022}
\bibinfo{author}{Oshnik, N.} \emph{et~al.}
\newblock \bibinfo{title}{Robust magnetometry with single nitrogen-vacancy
  centers via two-step optimization}.
\newblock \emph{\bibinfo{journal}{Physical Review A}}
  \textbf{\bibinfo{volume}{106}}, \bibinfo{pages}{013107}
  (\bibinfo{year}{2022}).
\newblock \urlprefix\url{https://link.aps.org/doi/10.1103/PhysRevA.106.013107}.

\bibitem{craigie_resource-efficient_2021}
\bibinfo{author}{Craigie, K.}, \bibinfo{author}{Gauger, E.~M.},
  \bibinfo{author}{Altmann, Y.} \& \bibinfo{author}{Bonato, C.}
\newblock \bibinfo{title}{Resource-efficient adaptive {Bayesian} tracking of
  magnetic fields with a quantum sensor}.
\newblock \emph{\bibinfo{journal}{Journal of Physics: Condensed Matter}}
  \textbf{\bibinfo{volume}{33}}, \bibinfo{pages}{195801}
  (\bibinfo{year}{2021}).
\newblock
  \urlprefix\url{https://iopscience.iop.org/article/10.1088/1361-648X/abe34f}.

\bibitem{bonato_optimized_2016}
\bibinfo{author}{Bonato, C.} \emph{et~al.}
\newblock \bibinfo{title}{Optimized quantum sensing with a single electron spin
  using real-time adaptive measurements}.
\newblock \emph{\bibinfo{journal}{Nature Nanotechnology}}
  \textbf{\bibinfo{volume}{11}}, \bibinfo{pages}{247--252}
  (\bibinfo{year}{2016}).
\newblock \urlprefix\url{https://www.nature.com/articles/nnano.2015.261}.

\bibitem{santagati_magnetic-field_2019}
\bibinfo{author}{Santagati, R.} \emph{et~al.}
\newblock \bibinfo{title}{Magnetic-{Field} {Learning} {Using} a {Single}
  {Electronic} {Spin} in {Diamond} with {One}-{Photon} {Readout} at {Room}
  {Temperature}}.
\newblock \emph{\bibinfo{journal}{Physical Review X}}
  \textbf{\bibinfo{volume}{9}}, \bibinfo{pages}{021019} (\bibinfo{year}{2019}).
\newblock \urlprefix\url{https://link.aps.org/doi/10.1103/PhysRevX.9.021019}.

\bibitem{zohar_real-time_2023}
\bibinfo{author}{Zohar, I.} \emph{et~al.}
\newblock \bibinfo{title}{Real-time frequency estimation of a qubit without
  single-shot-readout}.
\newblock \emph{\bibinfo{journal}{Quantum Science and Technology}}
  \textbf{\bibinfo{volume}{8}}, \bibinfo{pages}{035017} (\bibinfo{year}{2023}).
\newblock
  \urlprefix\url{https://iopscience.iop.org/article/10.1088/2058-9565/acd415}.

\bibitem{nusran_high-dynamic-range_2012}
\bibinfo{author}{Nusran, N.~M.}, \bibinfo{author}{Momeen, M.~U.} \&
  \bibinfo{author}{Dutt, M. V.~G.}
\newblock \bibinfo{title}{High-dynamic-range magnetometry with a single
  electronic spin in diamond}.
\newblock \emph{\bibinfo{journal}{Nature Nanotechnology}}
  \textbf{\bibinfo{volume}{7}}, \bibinfo{pages}{109--113}
  (\bibinfo{year}{2012}).
\newblock \urlprefix\url{http://www.nature.com/articles/nnano.2011.225}.

\bibitem{wang_experimental_2017}
\bibinfo{author}{Wang, J.} \emph{et~al.}
\newblock \bibinfo{title}{Experimental quantum {Hamiltonian} learning}.
\newblock \emph{\bibinfo{journal}{Nature Physics}}
  \textbf{\bibinfo{volume}{13}}, \bibinfo{pages}{551--555}
  (\bibinfo{year}{2017}).
\newblock \urlprefix\url{http://www.nature.com/articles/nphys4074}.

\bibitem{dinani_bayesian_2019}
\bibinfo{author}{Dinani, H.~T.}, \bibinfo{author}{Berry, D.~W.},
  \bibinfo{author}{Gonzalez, R.}, \bibinfo{author}{Maze, J.~R.} \&
  \bibinfo{author}{Bonato, C.}
\newblock \bibinfo{title}{Bayesian estimation for quantum sensing in the
  absence of single-shot detection}.
\newblock \emph{\bibinfo{journal}{Physical Review B}}
  \textbf{\bibinfo{volume}{99}}, \bibinfo{pages}{125413}
  (\bibinfo{year}{2019}).
\newblock \urlprefix\url{https://link.aps.org/doi/10.1103/PhysRevB.99.125413}.

\bibitem{bonato_adaptive_2017}
\bibinfo{author}{Bonato, C.} \& \bibinfo{author}{Berry, D.~W.}
\newblock \bibinfo{title}{Adaptive tracking of a time-varying field with a
  quantum sensor}.
\newblock \emph{\bibinfo{journal}{Physical Review A}}
  \textbf{\bibinfo{volume}{95}}, \bibinfo{pages}{052348}
  (\bibinfo{year}{2017}).
\newblock \urlprefix\url{http://link.aps.org/doi/10.1103/PhysRevA.95.052348}.

\bibitem{ferrie_adaptive_2012}
\bibinfo{author}{Ferrie, C.}, \bibinfo{author}{Granade, C.~E.} \&
  \bibinfo{author}{Cory, D.~G.}
\newblock \bibinfo{title}{Adaptive {Hamiltonian} estimation using {Bayesian}
  experimental design}.
\newblock \emph{\bibinfo{journal}{AIP Conference Proceedings}}
  \textbf{\bibinfo{volume}{1443}}, \bibinfo{pages}{165--173}
  (\bibinfo{year}{2012}).
\newblock \urlprefix\url{https://doi.org/10.1063/1.3703632}.

\bibitem{liu_repetitive_2020}
\bibinfo{author}{Liu, G.}, \bibinfo{author}{Chen, M.}, \bibinfo{author}{Liu,
  Y.-X.}, \bibinfo{author}{Layden, D.} \& \bibinfo{author}{Cappellaro, P.}
\newblock \bibinfo{title}{Repetitive readout enhanced by machine learning}.
\newblock \emph{\bibinfo{journal}{Machine Learning: Science and Technology}}
  \textbf{\bibinfo{volume}{1}}, \bibinfo{pages}{015003} (\bibinfo{year}{2020}).
\newblock
  \urlprefix\url{https://iopscience.iop.org/article/10.1088/2632-2153/ab4e24}.

\bibitem{tsukamoto_machine-learning-enhanced_2022}
\bibinfo{author}{Tsukamoto, M.} \emph{et~al.}
\newblock \bibinfo{title}{Machine-learning-enhanced quantum sensors for
  accurate magnetic field imaging}.
\newblock \emph{\bibinfo{journal}{Scientific Reports}}
  \textbf{\bibinfo{volume}{12}}, \bibinfo{pages}{13942} (\bibinfo{year}{2022}).
\newblock \urlprefix\url{http://arxiv.org/abs/2202.00380}.

\bibitem{wiebe_hamiltonian_2014}
\bibinfo{author}{Wiebe, N.}, \bibinfo{author}{Granade, C.},
  \bibinfo{author}{Ferrie, C.} \& \bibinfo{author}{Cory, D.~G.}
\newblock \bibinfo{title}{Hamiltonian {Learning} and {Certification} {Using}
  {Quantum} {Resources}}.
\newblock \emph{\bibinfo{journal}{Physical Review Letters}}
  \textbf{\bibinfo{volume}{112}}, \bibinfo{pages}{190501}
  (\bibinfo{year}{2014}).
\newblock
  \urlprefix\url{https://link.aps.org/doi/10.1103/PhysRevLett.112.190501}.

\bibitem{cimini_experimental_2023}
\bibinfo{author}{Cimini, V.} \emph{et~al.}
\newblock \bibinfo{title}{Experimental metrology beyond the standard quantum
  limit for a wide resources range}.
\newblock \emph{\bibinfo{journal}{npj Quantum Information}}
  \textbf{\bibinfo{volume}{9}}, \bibinfo{pages}{1--9} (\bibinfo{year}{2023}).
\newblock \urlprefix\url{https://www.nature.com/articles/s41534-023-00691-y}.

\bibitem{helstrom_quantum_1969}
\bibinfo{author}{Helstrom, C.~W.}
\newblock \bibinfo{title}{Quantum detection and estimation theory}.
\newblock \emph{\bibinfo{journal}{Journal of Statistical Physics}}
  \textbf{\bibinfo{volume}{1}}, \bibinfo{pages}{231--252}
  (\bibinfo{year}{1969}).
\newblock \urlprefix\url{https://doi.org/10.1007/BF01007479}.

\bibitem{holevo_statistical_1973}
\bibinfo{author}{Holevo, A.~S.}
\newblock \bibinfo{title}{Statistical problems in quantum physics}.
\newblock In \bibinfo{editor}{Maruyama, G.} \& \bibinfo{editor}{Prokhorov,
  Y.~V.} (eds.) \emph{\bibinfo{booktitle}{Proceedings of the {Second}
  {Japan}-{USSR} {Symposium} on {Probability} {Theory}}}, Lecture {Notes} in
  {Mathematics}, \bibinfo{pages}{104--119} (\bibinfo{publisher}{Springer},
  \bibinfo{address}{Berlin, Heidelberg}, \bibinfo{year}{1973}).

\bibitem{belliardo_incompatibility_2021}
\bibinfo{author}{Belliardo, F.} \& \bibinfo{author}{Giovannetti, V.}
\newblock \bibinfo{title}{Incompatibility in quantum parameter estimation}.
\newblock \emph{\bibinfo{journal}{New Journal of Physics}}
  \textbf{\bibinfo{volume}{23}}, \bibinfo{pages}{063055}
  (\bibinfo{year}{2021}).
\newblock \urlprefix\url{https://doi.org/10.1088/1367-2630/ac04ca}.
\newblock \bibinfo{note}{Publisher: IOP Publishing}.

\bibitem{cimini_deep_2023}
\bibinfo{author}{Cimini, V.} \emph{et~al.}
\newblock \bibinfo{title}{Deep reinforcement learning for quantum
  multiparameter estimation}.
\newblock \emph{\bibinfo{journal}{Advanced Photonics}}
  \textbf{\bibinfo{volume}{5}}, \bibinfo{pages}{016005} (\bibinfo{year}{2023}).
\newblock
  \urlprefix\url{https://iris.uniroma1.it/retrieve/e0ceb934-f61a-49c3-bb0a-7e1047b6f239/Cimini_Deep-reinforcement_2023.pdf}.

\bibitem{brask_gaussian_2022}
\bibinfo{author}{Brask, J.~B.}
\newblock \bibinfo{title}{Gaussian states and operations -- a quick reference}
  (\bibinfo{year}{2022}).
\newblock \urlprefix\url{http://arxiv.org/abs/2102.05748}.

\bibitem{hausladen_pretty_1994}
\bibinfo{author}{Hausladen, P.} \& \bibinfo{author}{Wootters, W.~K.}
\newblock \bibinfo{title}{A ‘{Pretty} {Good}’ {Measurement} for
  {Distinguishing} {Quantum} {States}}.
\newblock \emph{\bibinfo{journal}{Journal of Modern Optics}}
  \textbf{\bibinfo{volume}{41}}, \bibinfo{pages}{2385--2390}
  (\bibinfo{year}{1994}).
\newblock \urlprefix\url{https://doi.org/10.1080/09500349414552221}.

\bibitem{eldar_designing_2003}
\bibinfo{author}{Eldar, Y.~C.}, \bibinfo{author}{Megretski, A.} \&
  \bibinfo{author}{Verghese, G.~C.}
\newblock \bibinfo{title}{Designing {Optimal} {Quantum} {Detectors} {Via}
  {Semidefinite} {Programming}}.
\newblock \emph{\bibinfo{journal}{IEEE Transactions on Information Theory}}
  \textbf{\bibinfo{volume}{49}}, \bibinfo{pages}{1007--1012}
  (\bibinfo{year}{2003}).
\newblock \urlprefix\url{http://arxiv.org/abs/quant-ph/0205178}.

\bibitem{eldar_quantum_2006}
\bibinfo{author}{Eldar, Y.~C.} \& \bibinfo{author}{Forney, G.~D.}
\newblock \bibinfo{title}{On quantum detection and the square-root
  measurement}.
\newblock \emph{\bibinfo{journal}{IEEE Transactions on Information Theory}}
  \textbf{\bibinfo{volume}{47}}, \bibinfo{pages}{858--872}
  (\bibinfo{year}{2006}).
\newblock \urlprefix\url{https://doi.org/10.1109/18.915636}.

\bibitem{izumi_displacement_2012}
\bibinfo{author}{Izumi, S.} \emph{et~al.}
\newblock \bibinfo{title}{Displacement receiver for phase-shift-keyed coherent
  states}.
\newblock \emph{\bibinfo{journal}{Physical Review A}}
  \textbf{\bibinfo{volume}{86}}, \bibinfo{pages}{042328}
  (\bibinfo{year}{2012}).
\newblock \urlprefix\url{https://link.aps.org/doi/10.1103/PhysRevA.86.042328}.

\end{thebibliography}

\appendix

\section{Lower bounds for NV centers}
\label{sec:nv_center_bound}
In this section, we apply the Bayesian Cramér-Rao bound to the estimation of various parameters on the NV center platform. This bounds will be based on the Fisher information~\cite{fisher_design_1935}. Consider a stochastic variable $y$, which is extracted from a probability distribution $p (y|\theta)$, where $\theta$ is a parameter we want to estimate. This is a model for an experiment leading to a stochastic outcome. The information on $\theta$ available from the knowledge of $y$ can be measured by the Fisher information (FI), defined as
\begin{equation}
	I(\theta) := \mathbb{E}_y \left[ \left( \frac{\partial \log p (y|\theta)}{\partial \theta} \right)^2 \right] \; ,
	\label{eq:single_parameter_fi}
\end{equation}
where the expectation value is taken over the distribution $p (y|\theta)$. There is also a multiparameter version of the FI, called the Fisher information matrix (FI matrix), defined as
\begin{equation}
	I_{ij} (\vtheta) := \mathbb{E}_y \left[ \frac{\partial \log p (y|\vtheta)}{\partial \theta_i} \frac{\partial \log p (y|\vtheta)}{\partial \theta_j} \right] \; .
	\label{eq:multiple_parameter_fi_app}
\end{equation}
If the experiment allows to be controlled through the parameter $x$, then the outcome probability is $p(y|x, \theta)$ and the FI inherits such dependence, i.e. we write $I(\theta|x)$. In this paper the control parameter $x$ is computed from a strategy $h$, that stands for ``heuristic'', which could be the Particle Guess Heuristic or a neural network for example. In this case, we indicate it explicitly in symbol for the control $x_h$.

\subsection{Bayesian Cramér-Rao bound}
\label{subsec:bcrb}
Given $\theta$ a single parameter to estimate, we call $I(\theta|\vx_h)$ the Fisher information of a sequence of measurements with controls $\vx_h = (x_0^h, x_1^h, \cdots, x_{M-1}^h)$, which are computed from a strategy $h$. The quantity $I(\theta|\vx_h)$, together with the Fisher information of the prior $\pi(\theta)$, i.e. $I(\pi)$, defines a lower bound on the precision $\Delta^2 \widehat{\theta}$ of whatever estimator $\widehat{\theta}$, that contains the expectation value of $I(\theta|\vx_h)$ on $\pi(\theta)$, and is optimized on the strategy $h$. This lower bounds reads
\begin{equation}
	\Delta^2 \widehat{\theta} \ge \frac{1}{\sup_{h} \mathbb{E}_{\theta} \left[ I(\theta|\vx_h) \right] + I(\pi)} \; .
\end{equation}
This definition appears in the work of Fiderer \textit{et al.}~\cite{fiderer_neural-network_2021}. For the NV center the controls are the evolution time $\tau$ and the phase $\varphi$, this last however doesn't play any role in the computation of the lower bound, and it will be omitted in the following. For the multiparameter bound we will avoid the complication of introducing the Fisher information matrix and consider instead the sum of the bounds for two independent estimations, thus writing instead
\begin{equation}
	\tr[ G \cdot (\hvtheta - \vtheta)^\intercal (\hvtheta - \vtheta) ] \ge \sum_{k=1}^d \frac{1}{\sup_{h} \mathbb{E}_{\theta_k} \left[ I(\theta_k|\vtau_h) \right] + I(\pi_k)} \; .
\end{equation}
The Fisher information of a sequence of measurements is always additive, even if the quantum probe is only measured weakly, but in dealing with projective measurements, as it is the case for NV center, the advantage is that the measurements are uncorrelated, and the same expression for the Fisher information applies to all of them, independently on the results of the previous measurements, i.e.
\begin{equation}
	I(\theta|\boldsymbol{\tau}) = \sum_{t=1}^M I(\theta|\tau_t) \le M \sup_{\tau} I(\theta|\tau) \; ,
\end{equation}
where $M$ is the total number of measurements. The optimization of the single measurement FI gives directly the precision bound for the measurement-limited estimation:
\begin{align}
	\Delta^2 \widehat{\theta} &\ge \frac{1}{\sup_{h} \mathbb{E}_{\theta} \left[ I(\theta|\vtau_h) \right] + I(\pi)} \nonumber \\ &\ge \frac{1}{M \, \mathbb{E}_{\theta} \left[ \sup_{\tau} I(\theta|\tau) \right] + I(\pi)} \; . \label{eq:lower_meas}
\end{align}
If the total evolution time is the limiting resource, then, the expression for the total FI is
\begin{equation}
	I(\theta|\boldsymbol{\tau}) = T \sum_{t=1}^M \frac{\tau_t}{T} \left[ \frac {I(\theta|\tau_t)}{\tau_t} \right] \le T \sup_{\tau} \frac{I(\theta|\tau)}{\tau} \; ,
\end{equation}
with $\sum_{t=1}^M \tau_t = T$. In this expression the total FI is the weighted sum of the renormalized FI of each measurement, i.e. $\frac{I(\theta|\tau_t)}{\tau_t}$, and can be manifestly upper bounded by concentrating all the weights on the supremum of the renormalized FI. This gives the lower bound for the precision of the time-limited estimation:
\begin{align}
	\Delta^2 \widehat{\theta} &\ge \frac{1}{\sup_{h} \mathbb{E}_{\theta} \left[ I(\theta|\vtau_h) \right] + I(\pi)} \nonumber \\ &\ge \frac{1}{T \, \mathbb{E}_{\theta} \left[ \sup_{\tau} \frac{I(\theta|\tau)}{\tau} \right] + I(\pi)} \; . \label{eq:lower_time}
\end{align}
In the following we will apply this general observations to the derivation of the numerical bounds for DC and AC magnetometry, for decoherence estimation, and for the measurement of the parallel hyperfine coupling.

\subsection{Evaluation of the Fisher information}
\label{subsec:eval_fisher_bound}
Since the measurement outcome in the NV center is binary, we can compute the Fisher information for a parameter $\theta$, given the control $\tau$, as
\begin{align}
	I(\theta|\tau) &= \mathbb{E} \left[ \left( \frac{\partial \log p ({\pm 1}|\theta, \tau)}{\partial \omega} \right)^2 \right] \nonumber \\
	&= \frac{\left( \frac{\partial p}{\partial \theta} \right)^2}{p(1-p)} = \frac{\left( \frac{\partial p}{\partial \theta} \right)^2}{\frac{1}{4} - (p-\frac{1}{2})^2} \; , \label{eq:fisher_simplified}
\end{align}
where we have used the definition in \cref{eq:single_parameter_fi}, and where $p := p(+1|\theta, \tau)$. For example, for a decoherence free estimation of the precession frequency $\omega$ we have $p:=\cos^2 \left( \frac{\omega \tau}{2} \right)$, from which $\frac{\partial p}{\partial \theta} = \tau \sin(\frac{\omega \tau}{2}) \cos(\frac{\omega \tau}{2})$, and finally $I(\omega|\tau) = \tau^2$.

\subsection{DC magnetometry}
\label{subsec:nv_center_dc_bound}
The lower bound on the estimation of the frequency $\omega$ (and of the inverse of the decoherence time $T_2^{-1}$) are reported in the table below, and are represented in \cref{fig:nvcenter_comparison_phase} and \cref{fig:nvcenter_comparison_phase}. The left column contains the bound for a finite number of measurements $M$, while right column refers to the estimation with a fixed total evolution time $T$. The first row refers to the estimation of $\omega$ with perfect coherence, the second row refers to the estimation of $\omega$ with a finite and know $T_2$, while the last row refers to the simultaneous estimation of $\omega$ and $T_2^{-1}$ treated on equal footing, i.e. $G=\id$. The symbols $I(\omega)$ and $I(T_2^{-1})$ indicate the FI of the prior of the precession frequency and of the decoherence time respectively.
\begin{table*}[htb]
	\centering
	\label{table:dc_lower_bounds}
	\renewcommand{\arraystretch}{1.5}
	\setlength{\tabcolsep}{6pt}
	\begin{tabular}{|l|cl|cl|}
		\hline
		& Time & & Measurement & \\
		\hline
		\(T_2 = \infty\) &
		\(\frac{1}{T^2+I(\omega)}\) &
		\refstepcounter{equation}\theequation \label{eq:T2_infty_Time} &
		\(\frac{2^{-2(M+1)}}{3} \MHz^2\) &
		\refstepcounter{equation}\theequation \label{eq:T2_infty_Meas} \\
		\hline
		\(T_2 < \infty\) &
		\(\frac{1}{0.5 \, T T_2 + I(\omega)}\) &
		\refstepcounter{equation}\theequation \label{eq:T2_finite_Time} &
		\(\max \Big \lbrace \frac{2^{-2(M+1)}}{3} \MHz^2, \frac{1}{\mu M T_2^2 + I(\omega)} \Big \rbrace\) &
		\refstepcounter{equation}\theequation \label{eq:T2_finite_Meas} \\
		\hline
		\(T_2^{-1} \in (a, b)\) &
		\(\frac{1}{0.5 \, T \mathbb{E}\left[ T_2\right] + I(\omega)} + \frac{1}{ 0.5 \,T \mathbb{E} \left[T_2 \right] + I(T_2^{-1})}\) &
		\refstepcounter{equation}\theequation \label{eq:T2_interval_Time} &
		\(\frac{1}{\mu M \mathbb{E} \left[ T_2^2 \right] + I(\omega)} + \frac{1}{\mu M \mathbb{E} \left[ T_2^2 \right]+I(T_2^{-1})}\) &
		\refstepcounter{equation}\theequation \label{eq:T2_interval_Meas} \\
		\hline
	\end{tabular}
	\caption{Lower bounds for the precision of the frequency and decoherence time estimation in DC magnetometry on an NV center.}
\end{table*}
The numerical values of the quantities appearing in the table, for $\omega \in (0, 1) \MHz$ and $T_2^{-1} \in (0.09, 0.11) \MHz$, are: $\mu = 0.1619$, $\mathbb{E}[T_2^2] = 101.01 \mus^2$, $\mathbb{E} \left[ T_2\right] = 10.0336 \mus$, $I(\omega) = 12 \mus^2$, $I(T_2^{-1}) = 3 \cdot 10^{4} \mus^2$. In the following we derive these bounds.
\begin{itemize}
	\item The Fisher information for the precession frequency $\omega$ is given by $I(\omega|\tau) = \tau^2$, so that $\sup_{\tau} I(\omega|\tau) = \infty$ and the analysis based on the Cramér-Rao bound doesn't gives a useful bound. \cref{eq:T2_infty_Meas} can be found by observing that each measurement gives at most one bit of information about the value of $\omega$, because it has a binary outcomes~\cite{fiderer_neural-network_2021}. This bound is applied to all the measurement-limited estimations in the table in addiction to the one coming from the Fisher information.
	\item With a finite decoherence time $T_2 < \infty$ the FI for the frequency $\omega$ is
	\begin{equation}
		I(\omega|\tau, T_2) = \frac{1}{\mathcal{D}} \tau^2 e^{-\frac{2 \tau}{T_2}} \cos^2 \left( \frac{\omega \tau}{2}\right) \sin^2 \left( \frac{\omega \tau}{2}\right) \; ,
	\end{equation}
	with
	\begin{multline}
		\mathcal{D} := \left[ e^{-\frac{\tau}{T_2}} \cos^2 \left( \frac{\omega \tau}{2} \right) + \frac{1-e^{-\frac{\tau}{T_2}}}{2} \right] \cdot \\ \left[ e^{-\frac{\tau}{T_2}} \sin^2 \left( \frac{\omega \tau}{2} \right) + \frac{1-e^{-\frac{\tau}{T_2}}}{2} \right]
	\end{multline}
	which, by defining $C := \cos^2 \left( \frac{\omega \tau}{2} \right)$, can be bounded in the following way
	\begin{align}
		I(\omega|\tau, T_2) &= \frac{\tau^2 e^{-\frac{2 \tau}{T_2}} C (1-C)}{\left[ \frac{1}{4} - e^{-\frac{2 \tau}{T_2}} (C-\frac{1}{2})^2 \right]} \\
		&\le \frac{\tau^2 e^{-\frac{2 \tau}{T_2}}}{1 - e^{-\frac{2 \tau}{T_2}}} = T_2^2 \frac{x^2 e^{-2x}}{1-e^{-2x}} ; ,
	\end{align}
	where $x = \frac{\tau}{T_2}$. The maximization in $x \in \mathbb{R}_+$ gives $\sup_{\tau} I(\omega|\tau, T_2) = \mu T_2^2$ with $\mu = 0.1619$. Inserting this expression in \cref{eq:lower_meas} gives the first term in the maximum of \cref{eq:T2_finite_Meas}, the second term was explained in the previous point.
	\item We now turn to the estimation of an unknown $T_2^{-1}$, with prior uniform in $(0.09, 0.11)$, alongside $\omega$. The total MSE is $\Delta^2 \widehat{\omega} + \Delta^2 \widehat{T_2^{-1}}$. The supremum of the FI for $\omega$ is always $\sup_{\tau} I(\omega|\tau, T_2) = \mu T_2^2$, but this time the expectation value $\mathbb{E}_\theta$ is not trivial, i.e. $\mathbb{E}_\theta \left[ \sup_{\tau} I(\omega|\tau, T_2) \right] = \mu \mathbb{E} \left[ T_2^2 \right]$. The derivative of the probability with respect to $T_2^{-1}$, used to compute the FI is
	\begin{equation}
		\frac{\partial p}{\partial T_2^{-1}} = - \tau e^{-\frac{\tau}{T_2}} \left[ \cos^2 \left( \frac{\omega \tau}{2} \right) - \frac{1}{2} \right] \; .
	\end{equation}

	With the same notation for the cosine we write the FI for the inverse of the decoherence time as
	\begin{align}
		I(T_2^{-1}|\omega, \tau) &= \frac{\tau^2 e^{-\frac{2 \tau}{T_2}} \left(C-\frac{1}{2} \right)^2}{\left[ \frac{1}{4} - e^{-\frac{2 \tau}{T_2}} (C-\frac{1}{2})^2 \right]} \nonumber \\
		&\le T_2^2 \frac{x^2 e^{-2x}}{1-e^{-2x}} \; .
	\end{align}

	The maximization of this expression gives $\sup_{\tau} I(T_2^{-1}|\omega, \tau) = \mu T_2^2$, which is the same expression of $I(\omega|T_2, \tau)$, that again has a non trivial expectation value on the parameters. Putting everything together and adding the prior information on $T_2^{-1}$ gives \cref{eq:T2_interval_Meas}.
	\item Regarding the time-constrained lower bounds, for $T_2 = \infty$, the total FI is maximized by performing a single measurement of time duration $\tau = T$, which gives \cref{eq:T2_infty_Time}, through the application of \cref{eq:lower_time}.
	\item For $T_2 < \infty$ we have to maximize the normalized FI in $x \in \mathbb{R}_+$, i.e.
	\begin{align}
		\frac{I(\omega|\tau, T_2)}{\tau} &\le \frac{\tau e^{-\frac{2 \tau}{T_2}}}{1 - e^{-\frac{2 \tau}{T_2}}} \le T_2 \frac{x e^{-2x}}{1-e^{-2x}} \le \frac{TT_2}{2} \; , \label{eq:time_fixed_T2_omega}
	\end{align}
	from which \cref{eq:T2_finite_Time} follows from \cref{eq:lower_time}.
	\item In the last case we have to estimate both $\omega$ and $T_2^{-1} \in (0.09, 0.11) \MHz$. Regarding the estimation of $\omega$ the Fisher information is always given by \cref{eq:time_fixed_T2_omega}, only that this time the expectation value on $T_2$ is non trivial, and produces the first addend of \cref{eq:T2_interval_Time}. For the estimation of $T_2^{-1}$ we have a similar expression for the normalized FI:
	\begin{align}
		\frac{I(T_2^{-1}|\tau, \omega)}{\tau} &\le \frac{\tau e^{-\frac{2 \tau}{T_2}}}{1 - e^{-\frac{2 \tau}{T_2}}} \le T_2 \frac{x e^{-2x}}{1-e^{-2x}} \le \frac{TT_2}{2}\; , \label{eq:time_fixed_T2_T2}
	\end{align}
	which similarly needs the expectation value appearing in \cref{eq:lower_time} to give the second piece of \cref{eq:T2_interval_Time}.
\end{itemize}

\subsection{AC magnetometry}
\label{subsec:nv_center_ac_bound}
The lower bound on the precision of the frequency $\Omega$ and of the inverse of the decoherence time $T_2^{-1}$, are reported in the table below, and are represented in \cref{fig:nvcenter_comparison_ac} of \cref{subsec:nv_center_ac}. The meaning of the rows and columns here are the same as in \cref{table:dc_lower_bounds}. The symbols $I(\Omega) = 12 \mus^2$ and $I(T_2^{-1}) = 3 \cdot 10^4 \mus^2$ indicate respectively the FI of the prior of the frequency and of the inverse of the decoherence time.
\begin{table*}[htb]
	\centering
	\label{table:ac_lower_bounds}
	\renewcommand{\arraystretch}{1.5}
	\setlength{\tabcolsep}{6pt}
	\begin{tabular}{|l|cl|cl|}
		\hline
		& Time & & Measurement & \\
		\hline
		\(T_2 = \infty\) &
		\(\frac{1}{\frac{\gamma T}{\omega}+I(\Omega)}\) &
		\refstepcounter{equation}\theequation \label{eq:T2_infty_Time_ac} &
		\(\max \Big \lbrace \frac{2^{-2(M+1)}}{3} \MHz^2, \frac{1}{\frac{M}{\omega^2}+I(\Omega)} \Big \rbrace\) &
		\refstepcounter{equation}\theequation \label{eq:T2_infty_Meas_ac} \\
		\hline
		\(T_2 < \infty\) &
		\(\frac{1}{\frac{\gamma T}{\omega}+I(\Omega)}\) &
		\refstepcounter{equation}\theequation \label{eq:T2_finite_Time_ac} &
		\(\max \Big \lbrace \frac{2^{-2(M+1)}}{3} \MHz^2, \frac{1}{\frac{M}{\omega^2}+I(\Omega)} \Big \rbrace\) &
		\refstepcounter{equation}\theequation \label{eq:T2_finite_Meas_ac} \\
		\hline
		\(T_2^{-1} \in (a, b)\) &
		\(\frac{1}{\frac{M}{\omega^2}+I(\Omega)} + \frac{1}{ 0.5 \,T \mathbb{E} \left[T_2 \right] + I(T_2^{-1})}\) &
		\refstepcounter{equation}\theequation \label{eq:T2_interval_Time_ac} &
		\(\frac{1}{\frac{\gamma T}{\omega}+I(\Omega)} + \frac{1}{\mu M \mathbb{E} \left[ T_2^2 \right]+I(T_2^{-1})}\) &
		\refstepcounter{equation}\theequation \label{eq:T2_interval_Meas_ac} \\
		\hline
	\end{tabular}
	\caption{Lower bounds on the estimation precision of the magnetic field intensity and decoherence time in AC magnetometry on an NV center.}
\end{table*}
In the following we derive each of these bounds.
\begin{itemize}
	\item The decoherence free FI for $\Omega$ in the measurement-limited case is obtained from \cref{eq:fisher_simplified} through the derivative
	\begin{equation}
		\frac{\partial p}{\partial \Omega} = \frac{\sin (\omega \tau)}{\omega} \cos \left[ \frac{B}{2 \omega} \sin(\omega \tau) \right] \sin \left[ \frac{B}{2 \omega} \sin(\omega \tau) \right] \; ,
	\end{equation}
	and reads
	\begin{equation}
		I(\Omega|\tau) = \frac{\sin^2{\omega \tau}}{\omega^2} \le \frac{1}{\omega^2} \; .
	\end{equation}
	Inserting this expression in \cref{eq:lower_time} gives immediately the first part of \cref{eq:T2_infty_Meas_ac} in the table. The second operand of the $\max$ come from analysis on the number of bits~\cite{fiderer_neural-network_2021}, which we already discussed in the previous section.

	\item For the time-limited estimation of $\Omega$ we have to maximize the normalized FI in $x = \omega \tau \in \mathbb{R}_+$:
	\begin{equation}
		\frac{I(\Omega|\tau)}{\tau} = \frac{1}{\omega} \frac{\sin^2 (\omega \tau)}{\omega \tau} = \frac{1}{\omega} \frac{\sin^2 x}{x} \le \frac{\gamma}{\omega} \; ,
	\end{equation}
	where $\gamma = 0.724611$. By inserting this result in \cref{eq:lower_time} we get \cref{eq:T2_infty_Time_ac}.

	\item For the case $T_2 < \infty$ we don't compute any new bound, instead we use the one for the decoherence free estimation, which must be valid also for the noisy model. Accordingly, \cref{eq:T2_finite_Meas_ac} is equal to \cref{eq:T2_infty_Meas_ac} and \cref{eq:T2_finite_Time_ac} is equal to \cref{eq:T2_infty_Time_ac}.
	\item For the simultaneous estimation of $\Omega$ and of $T_2^{-1}$ we use the same bound computed for DC magnetometry for the precision on $T_2^{-1}$ and obtain \cref{eq:T2_interval_Meas_ac} and \cref{eq:T2_interval_Time_ac} for the measurement-limited and time-limited estimation respectively.
\end{itemize}

\subsection{Decoherence estimation}
\label{subsec:nv_center_dec_bound}
In this section, we report the bounds based on the FI for the estimation of the inverse of the decoherence time $T_2^{-1}$ and the exponent $\beta$ on the NV center platform, which are reported in \cref{fig:nvcenter_comparison_dec} of \cref{subsec:nv_center_dec}.
\begin{table*}[htb]
	\centering
	\label{table:dec_lower_bounds}
	\renewcommand{\arraystretch}{1.5}
	\setlength{\tabcolsep}{4pt}
	\begin{tabular}{|l|cl|cl|}
		\hline
		& Time & & Measurement & \\
		\hline
		\(\beta\) nuis. &
		\(\frac{1}{\delta T \mathbb{E} [T_2] \mathbb{E}[\beta^2] + I(T_2^{-1})}\) &
		\refstepcounter{equation}\theequation \label{eq:beta_nuis_Time} &
		\(\frac{1}{\mu M \mathbb{E} [T_2^2] \mathbb{E} [\beta^2] + I(T_2^{-1})}\) &
		\refstepcounter{equation}\theequation \label{eq:beta_nuis_Meas} \\
		\hline
		\(\beta=2\) &
		\(\frac{1}{4 \varepsilon T \mathbb{E} [T_2] + I(T_2^{-1})}\) &
		\refstepcounter{equation}\theequation \label{eq:beta_fixed_Time} &
		\(\frac{1}{4 \mu M \mathbb{E} [T_2^2] + I(T_2^{-1})}\) &
		\refstepcounter{equation}\theequation \label{eq:beta_fixed_Meas} \\
		\hline
		Both &
		\(\frac{1}{\psi T \mathbb{E}[T_2^{-1}] \mathbb{E} [\beta^{-2}]+I(\beta)} + \frac{1}{\delta T \mathbb{E} [T_2] \mathbb{E} [\beta^2] + I(T_2^{-1})}\) &
		\refstepcounter{equation}\theequation \label{eq:both_Time} &
		\(\frac{1}{\chi M \mathbb{E}[\beta^{-2}] + I(\beta)} + \frac{1}{\mu M \mathbb{E} [T_2^2] \mathbb{E} [\beta^2] + I(T_2^{-1})}\) &
		\refstepcounter{equation}\theequation \label{eq:both_Meas} \\
		\hline
	\end{tabular}
	\caption{Lower bounds for the precision of the characterization of a dephasing noise on an NV center.}
\end{table*}
The numerical coefficients appearing in the table are $\mu = 0.16190$, $\delta = 0.24429$, $\chi = 0.23966$, $\varepsilon = 0.20687$, $\eta = 0.10582$, $\psi = 2.43013$, $\mathbb{E}[T_2^2] = 10^3 \mus^2$, $\mathbb{E}[T_2] = 25.5848 \mus$, $\mathbb{E}[T_2^{-1}] = 0.0450 \MHz^2$, $\mathbb{E} [\beta^2] = 8.08332$, $\mathbb{E} [\beta^{-2}] = 0.16666$, $I(T^{-1}) = 1481.48 \mus^2$, and $I(\beta) = 1.92$. We proceed in deriving point by point the lower bounds of the above table from the Fisher information.
\begin{itemize}
	\item The derivative of the model probability in $T_2^{-1}$ is given by
	\begin{equation}
		\frac{\partial p}{\partial T_2^{-1}} = - \frac{1}{2} e^{-\left( \frac{\tau}{T_2} \right)^\beta} \beta \tau \left( \frac{\tau}{T_2} \right)^{\beta-1} \; .
	\end{equation}
	from which we obtain the FI from \cref{eq:fisher_simplified}:
	\begin{equation}
		I(T_2^{-1}|\tau, \beta) = \beta^2 T_2^2 \frac{e^{-2 \left(\frac{\tau}{T_2} \right)^\beta} \left( \frac{\tau}{T_2} \right)^{2 \beta}}{1-e^{-2 \left( \frac{\tau}{T_2} \right)^\beta}} \; .
	\end{equation}
	By defining $x = \left( \frac{\tau}{T_2} \right)^{\beta} \in \mathbb{R}_+$ we rewrite the above expression as
	\begin{equation}
		I(T_2^{-1}|\tau, \beta) = \beta^2 T_2^2 \frac{e^{-2x} x^2}{1-e^{-2x}} \le \mu \beta^2 T_2^2 \; ,
	\end{equation}
	which inserted into \cref{eq:lower_meas}, by taking the expectation value on the parameters $\beta$ and $T_2$, gives the expression in \cref{eq:beta_nuis_Meas}.
	\item For $\beta = 2$ we can avoid the expectation value of $\beta^2$, which is replaced by the factor $4$ in \cref{eq:beta_fixed_Meas}.
	\item For the simultaneous estimation of $T_2^{-1}$ and $\beta$ we need an expression for the Fisher information on $\beta$, which can be obtained from the derivative
	\begin{equation}
		\frac{\partial p}{\partial \beta} = - \frac{1}{2} \left( \frac{\tau}{T_2} \right)^{\beta} e^{-\left( \frac{\tau}{T_2} \right)^{\beta}} \log \left( \frac{\tau}{T_2} \right) \; ,
	\end{equation}
	and reads
	\begin{align}
		I(\beta|\tau, T_2) &= \frac{1}{\beta^2} \frac{\left( \frac{\tau}{T_2} \right)^{2 \beta} e^{-2 \left( \frac{\tau}{T_2} \right)^{\beta}} \log^2 \left( \frac{\tau}{T_2} \right)^\beta}{1-e^{-2 \left( \frac{\tau}{T_2} \right)^{\beta}}} \nonumber \\
		&\le \frac{1}{\beta^2} \frac{x^2 e^{-2x} \log^2 x}{1-e^{-2x}} \le \frac{\chi}{\beta^2} \; ,
	\end{align}
	where $x = \left( \frac{\tau}{T_2} \right)^{\beta} \in \mathbb{R}_+$. The coefficient $\chi$ comes from the maximization of the function in $x$ only. Given the result $\sup_{\tau} 	I(\beta|\tau, T_2) = \chi \beta^{-2}$, we insert this expression in \cref{eq:lower_meas}, obtaining the second term of \cref{eq:both_Meas}.
	\item Regarding the time-limited estimation, the relevant quantity to optimize is the renormalized FI again. For the estimation of $T_2^{-1}$ this reads
	\begin{align}
		\frac{I(T_2^{-1}|\tau, \beta)}{\tau} &= \beta^2 T T_2 \frac{e^{-2x} x^{2-\frac{1}{\beta}}}{1-e^{-2x}} \nonumber \\
		&\le \beta^2 T T_2 \, \sup_{x, \beta} \frac{e^{-2x} x^{2-\frac{1}{\beta}}}{1-e^{-2x}} \\ &= \delta \beta^2 T T_2 \; , \label{eq:max_time_dec}
	\end{align}
	where $\delta$ is the supremum of the function in $x \in \mathbb{R}_+$ and $\beta \in (1.5, 4)$, realized for $\beta = 1.5$. Inserting the expression above in \cref{eq:lower_time} and taking the expectation values of the parameters produces \cref{eq:beta_nuis_Meas}.
	\item The $\beta=2$ bound can be obtained from \cref{eq:max_time_dec} by maximizing only on $x$. The coefficient $\varepsilon$ in \cref{eq:beta_fixed_Meas} is defined as $\varepsilon = \sup_{x} \frac{e^{-2x} x^{\frac{3}{2}}}{1-e^{-2x}}$. After adding the FI from the prior we get \cref{eq:beta_fixed_Meas}.
	\item For the estimation of $\beta$ we again maximize the normalized FI in $x \in \mathbb{R}_+$:
	\begin{align}
		\frac{I(\beta|\tau, T_2)}{\tau} &= \frac{1}{T_2 \beta^2} \frac{x^{2-\frac{1}{\beta}} e^{-2x} \log^2 x}{1-e^{-2x}} \nonumber \\
		&\le \frac{1}{T_2 \beta^2} \sup_{\beta, \tau} \frac{x^{2-\frac{1}{\beta}} e^{-2x} \log^2 x}{1-e^{-2x}} \nonumber \\
		&\le \psi T_2^{-1} \beta^{-2} \; ,
	\end{align}
	which inserted in \cref{eq:lower_time} gives the second part of the bound in \cref{eq:both_Time}, the first being identical to \cref{eq:beta_nuis_Meas}.
\end{itemize}
\subsection{Hyperfine coupling estimation}
\label{subsec:nv_center_hyperfine_bound}
In this section, we report the precision lower bounds for the estimation of the parallel hyperfine coupling of the NV center electron spin with a carbon nucleus, which are plotted in \cref{fig:nvcenter_comparison_double} of \cref{subsec:nv_center_hyperfine}. These bounds are based on the one computed for DC magnetometry in \cref{subsec:nv_center_dc_bound}.
\begin{table*}[htb]
	\centering
	\label{table:hyp_lower_bounds}
	\renewcommand{\arraystretch}{1.5}
	\setlength{\tabcolsep}{4pt}
	\begin{tabular}{|l|cl|cl|}
		\hline
		& Time & & Measurement & \\
		\hline
		\(T_2 = \infty\) &
		\(\frac{2}{\frac{1}{4}T^2+I(\omega)}\) &
		\refstepcounter{equation}\theequation \label{eq:T2_infty_Time_double} &
		\(\frac{2^{-M}}{24} \MHz^2\) &
		\refstepcounter{equation}\theequation \label{eq:T2_infty_Meas_double} \\
		\hline
		\(T_2 < \infty\) &
		\(\frac{2}{\frac{1}{8} T \mathbb{E} [T_2] + I(\omega)}\) &
		\refstepcounter{equation}\theequation \label{eq:T2_finite_Time_double} &
		\( \max \Big \lbrace \frac{2^{-M}}{24} \MHz^2, \frac{2}{\frac{1}{4} \mu M T_2^2+I(\omega)} \Big \rbrace \) &
		\refstepcounter{equation}\theequation \label{eq:T2_finite_Meas_double} \\
		\hline
	\end{tabular}
	\caption{Lower bounds for the estimation precision of two precession frequencies, split by the hyperfine interaction of the NV center with a ${}^{13}C$ nucleus.}
\end{table*}
The numerical coefficients in the table are $\mu = 0.16190$ and $I(\omega) = 18.18181 \mus^2$. The true values of both frequencies $\omega_0$ and $\omega_1$ are extracted uniformly in $(0, 1) \MHz$ in the simulations, however, since they are completely symmetrized in the model likelihood, the true prior is the uniform distribution over the triangle in the $(\omega_0, \omega_1)$ plane identified by the points $(0, 0) \MHz$, ($1, 0) \MHz$, and $(1, 1) \MHz$. Given $\sigma^2$ the variance of such distribution we define $I(\omega) := \frac{1}{2} \sigma^2$ the prior FI for a single frequency. In the following we give a derivation for the bounds in the table above.
\begin{itemize}
	\item Starting from \cref{eq:T2_infty_Meas}, based on the bit counting argument, we observe that $2 M$ measurements are needed for having $N$ bits of information for each of the phases $\omega_{0, 1} \in (0, 1) \MHz$. The MSE on a single phase after $N$ measurement is thus limited by $\frac{2^{-M}}{12}$. For two phases the bound becomes $\frac{2^{-M}}{6}$. Since it is not easy to implement the information on the triangular prior in this bound, we consider a square subset of the triangular prior only, i.e. we fix $\omega_0 \in (0.5, 1) \MHz$ and $\omega_1 \in (0, 0.5) \MHz$. This means dividing the bound further by a factor $4$, which gives us \cref{eq:T2_infty_Meas_double}.
	\item The FI for a single frequency is $\frac{1}{4}$ that of the DC magnetometry example of \cref{eq:T2_infty_Time}, i.e. $I(\omega_{0, 1}) = \frac{T^2}{4}$ in the infinite coherence case. This comes from the factor $\frac{1}{2}$ in the likelihood. For two parameters this gives the bound of \cref{eq:T2_infty_Time_double}.
	\item The same argument can be made to go from \cref{eq:T2_finite_Meas} to \cref{eq:T2_finite_Meas_double} and from \cref{eq:T2_finite_Time} to \cref{eq:T2_finite_Time_double}.
\end{itemize}
\section{Lower bounds for photonic circuits}
\label{sec:photonic_platform_bound}
In this section, we present the precision lower bound for the photonic based applications. These are the agnostic Dolinar receiver, the multiphase discrimination task on the four bosonic lines interferometer, the linear classifier of multiple states, and the QML classifier of three coherent states. All these examples are based on the identification of the correct hypothesis among a finite number of them. Accordingly, the figure of merit for their performances is the probability of a wrong classification. The bounds presented here for this family of tasks are based either on the Helstrom bound~\cite{helstrom_quantum_1969} and on its generalizations, or on the pretty good measurement (PGM), which gives good results for the one shot classification of states~\cite{hausladen_pretty_1994}. This approach doesn't really give a lower bound for the precision, in all cases but one, nevertheless we use it as a reference value, and given the fact that performing the PGM requires in general entangled measurements we consider its precision to be unachievable by means of photon counting and linear optics only.

\subsection{Helstrom bound}
\label{subsec:helstrom_bound}
Consider a pure state $\ket{\psi}$ in the Hilbert space $\mathcal{H}$. We are assured that this state is either $\ket{\psi_0}$ or $\ket{\psi_1}$, both of which also reside in $\mathcal{H}$. Our objective is to correctly identify which state it is. If we are free to perform whatever POVM measurement on $\ket{\psi}$, then the error probability is bounded as
\begin{equation}
	P_e \ge P_e^{\text{H}} := \frac{1}{2} - \frac{1}{2} \sqrt{1 - \left| \langle \psi_0, \psi_1 \rangle \right|^2} \; .
\end{equation}
This is called the Helstrom bound~\cite{helstrom_quantum_1969} and for two orthogonal states is $P_e^{\text{H}}=0$. In the Dolinar receiver, where the task is to discriminate $\ket{\pm \alpha}$, the Helstrom bound reads
\begin{align}
	P_e^{\text{H}} &= \frac{1}{2} - \frac{1}{2} \sqrt{1 - \left| \langle -\alpha, +\alpha \rangle \right|^2 } \nonumber \\
	&= \frac{1}{2} - \frac{1}{2} \sqrt{1 - e^{-4 |\alpha|^2}} \; .
\end{align}
This is represented by the shaded gray region on the precision plots of the Dolinar receiver in \cref{fig:dolinar_confronto} of \cref{subsec:dolinar}. In this work we study however the agnostic Dolinar receiver, where the task is to discriminate the states $\ket{\alpha} \otimes \ket{\alpha}^{\otimes n}$ and $\ket{-\alpha} \otimes \ket{\alpha}^{\otimes n}$, without having any classical knowledge on the value of $\alpha$. It can be proved~\cite{zoratti_agnostic-dolinar_2021} that the error lower bound in this case is
\begin{equation}
	P_e^{\text{H}} := \frac{1}{2} \left( 1- \frac{1}{2} \sum_{k=0}^\infty \mathcal{P} (k; \sqrt{n+1}\alpha) \sqrt{1-\left( \frac{n-1}{n+1} \right)^{2 k}} \right) \; ,
\end{equation}
where $\mathcal{P} (k; \mu)$ is the probability distribution of a Poissonian variable, i.e.
\begin{equation}
	\mathcal{P} (k; \mu) := \frac{\mu^k e^{-\mu}}{k!} \; .
\end{equation}
The Helstrom bound with limited $n$ is plotted as the red lines of \cref{fig:dolinar_confronto}.

\subsection{Pretty good measurement}
\label{subsec:pgm_bound}
Suppose we are given a state $\ket{\psi} \in \mathcal{H}$ and we are promised it belongs to the set of states $\lbrace \ket{\psi_j} \rbrace_{j=1}^m$, which are classically known. The \textit{a priori} probabilities for the states are $p_j$ for $j=1, \dots, m$. We can perform whatever measurement on $\ket{\psi}$ and the task is to identify the state in the set. A good approach to this problem, which doesn't require solving a semidefinite program, is the PGM~\cite{hausladen_pretty_1994}, which prescribes performing the POVM measurement $\mathcal{M}\ped{PGM} := \lbrace M_j \rbrace_{j=1}^m$, whose operators are
\begin{equation}
	M_j := p_j S^{-\frac{1}{2}} \ket{\psi_j} \! \bra{\psi_j} S^{-\frac{1}{2}} \; ,
\end{equation}
where $S := \sum_{j=1}^{m} p_j \ket{\psi_j} \! \bra{\psi_j}$. If in the measurement the outcome corresponding to the $j$-th operator is observed, then the guess for the state classification is $\ket{\psi_j}$. The error probability of the PGM is therefore
\begin{equation}
	P^{\text{PGM}}_e = 1 - \sum_{j=1}^{m} p_j \text{Tr} \left( M_j \ket{\psi_j} \! \bra{\psi_j} \right) \; .
\end{equation}
In our applications to photonic circuits such measurements are not optimal and in general their performance cannot be achieved with linear optics and photon measurement. They provide nevertheless a useful reference value for the estimation precision. Throughout the thesis the computation of $P_e^{\text{PGM}}$ has been carried out numerically.

\subsection{Multiphase discrimination}
\label{subsec:multiphase_discrimination_bound}
In this work we studied the problem of multiphase discrimination on a four arms interferometer, for four different input states. In order to give a reference value on the precision of the discrimination we apply the PGM. Let us break the interferometer and remove the control phases $\lbrace c_i \rbrace_{i=1}^3$, the closing quarter, and the photodetectors. The photon states after the encoding with the phases $\lbrace \varphi_i \rbrace_{i=1}^3$ are respectively
\begin{enumerate}
	\item \begin{equation} \Big | \frac{1}{2} e^{-i \varphi_0}, \frac{1}{2} i e^{-i \varphi_1}, \frac{1}{2} i e^{-i \varphi_2}, -\frac{1}{2} \Big \rangle \; ,
	\end{equation}
	\item \begin{multline}
		\Big | \frac{1}{2} (1+i) e^{-i \varphi_0},
		\frac{1}{2} (-1+i) e^{-i \varphi_1}, \\
		\frac{1}{2} (1+i) e^{-i \varphi_2},
		\frac{1}{2} (-1+i) \Big \rangle \; ,
	\end{multline}
	\item \begin{equation}
		\Big | \frac{1}{2} + i e^{-i \varphi_0}, \frac{1}{2}i e^{-i \varphi_1},
		\frac{1}{2} i e^{-i \varphi_2}, -\frac{1}{2} + i \Big \rangle \; ,
	\end{equation}
	\item \begin{equation}
		\Big | i e^{-i \varphi_0}, i e^{-i \varphi_1}, i e^{-i \varphi_2},
		i \Big \rangle \; ,
	\end{equation}
\end{enumerate}
for the input states with one, two, three, and four photons on average. We are allowed to consume $n$ input states in order to produce $n$ copies of the encoded coherent states reported above. These are then sent to an adder, which perform the operation
\begin{equation}
	U \ket{\alpha}^{\otimes n} = \ket{\sqrt{n} \alpha} \; ,
	\label{eq:adder}
\end{equation}
on the $n$ coherent states for each of the four bosonic wires. On this resulting state of the interferometer the PGM is then executed. The error probability can be expressed as a function of the number of input states $n$ or as a function of the total number of employed photons, and is the gray area represented in \cref{fig:photonic_circuit} of \cref{subsec:photonic_circuit}. Although this is not rigorously a lower bound on the precision it serves as a reference value, which is most certainly impossible to saturate with the device at our disposal.

\subsection{Quantum Machine Learning classifier}
\label{subsec:qml_classifier_bound}
In the QML task we have a device that identifies the class $s$ of a signal $\ket{\alpha_s}$, given the quantum training set $\ket{\alpha_0}^{\otimes n} \otimes \ket{\alpha_1}^{\otimes n} \otimes \ket{\alpha_2}^{\otimes n}$. In order to perform the PGM we must know classically the states that we aim to distinguish and this is not possible with a finite quantum training set. We suppose that by consuming double the amount of reference states, i.e. $2n$ for each class, we can estimate the complex numbers $\alpha_0, \alpha_1, \alpha_2$ exactly and then apply the optimal PGM. For many instances of the classification task, with the reference states $\ket{\alpha_0}, \ket{\alpha_1}, \ket{\alpha_2}$ and the signal $\ket{\alpha_s}$ extracted according to their prior, we compute the error probability of the PGM, i.e. $P_e^{\text{PGM}}$, and associate to it the average number of consumed photons
\begin{equation}
	N\ped{ph} = 2 n \left( |\alpha_0|^2+|\alpha_1|^2+|\alpha_2|^2\right)+|\alpha_s|^2 \; .
\end{equation}
We then average the data points $(N\ped{ph}, P_e^{\text{PGM}})$ in order to compute the expected error probability $\langle P_e^{\text{PGM}} \rangle$ for a certain average number of photons. This curve defines the shaded gray area in \cref{fig:qml_results} of \cref{subsec:qml_classifier}. Although this is not rigorously a lower bound on the precision it serves as a reference value, which is most certainly impossible to saturable with the device at our disposal.

\subsection{Liner classifier of multiple states}
\label{subsec:linear_multi_bound}
The last example on the photonic circuit platform of \cref{subsec:lin_multiclassifier} concerns the classification of a signal state $\ket{\alpha_s}$, that can take values in the set of known states $\lbrace \ket{\alpha_j} \rbrace_{j=1}^m$. For this discrimination we employ only a network of beam splitters, which is statically optimized, and no adaptivity is taken into account. By virtue of the cyclic symmetry of the states to discriminate, the PGM is optimal~\cite{eldar_designing_2003, eldar_quantum_2006, izumi_displacement_2012} and its error rate defines the shaded grey area in the plots of \cref{fig:linear_classifier}. Again we are using an adder, with action reported in \cref{eq:adder}, to build the state $\ket{\sqrt{M} \alpha_s}$ out of $M$ copies of the signal, which in the device are passed to the $M$ layers of the discriminator. On the state $\ket{\sqrt{M} \alpha_s}$ the PGM is executed and its precision is plot as a function of $M$.

\section{Analytical optimization for the NV center}
\label{sec:anal_meas}
In this section, we optimize the precision of magnetic field estimation on an NV center, with the tools developed in~\cite{belliardo_achieving_2020}. In the present paper we have so far allowed the continuous tuning of the parameters $\tau$ and $\varphi$, but now we suppose that only $\tau_j = 2^{j-1}$ can be measured, for $j=1, 2, \ldots, K$, and similarly we are limited to $\varphi = 0, \frac{\pi}{2}$. We parametrize a strategy under these constraints by defining $\nu_j$ the number of measurements performed with time $\tau_{j}$. The total number of Ramsey measurements is then
\begin{equation}
	M := 2 \sum_{j=1}^{K} \nu_j \; ,
	\label{eq:max_measurements}
\end{equation}
By applying the phase algorithm presented in~\cite{belliardo_achieving_2020} and using the corresponding analytical tool to upper bound its precision we can write
\begin{equation}
	\Delta^2 \widetilde{\omega} \le \left( \frac{2 \pi}{3} \right)^2 \left( \frac{1}{4^K} + 16 \sum_{j = 1}^{K} \frac{A}{4^{j-1}} C^{-\nu_j} \right) \; ,
	\label{eq:upper_bound_omega}
\end{equation}
where $C \simeq 1.66$ and $A \simeq 0.60$ are numerical constants. We can find the optimal measurement distribution that optimizes this upper bound with the methods of Lagrangian multiplier:
\begin{widetext}
	\begin{equation}
		\mathcal{L} := \left( \frac{2 \pi}{3} \right)^2 \left( \frac{1}{4^K} + 16 \sum_{j = 1}^{K} \frac{A}{4^{j-1}} C^{-\nu_j} \right) - \lambda \left( 2 \sum_{j=1}^{K} \nu_j - M \right) \;.
		\label{eq:mainLagrangianAlt}
	\end{equation}
\end{widetext}
The optimization on $\nu_j$ gives
\begin{equation}
	\nu_j = \nu_K + \frac{2}{\log_2 C} (K-j) \; .
	\label{eq:newLinearRamp}
\end{equation}
The number of measurements $\nu_j$ should be integers, but in the following we will neglect the rounding as we are only interested in extracting the scaling of the precision of the phase estimation procedure. By substituting \cref{eq:newLinearRamp} in \cref{eq:upper_bound_omega} we get
\begin{equation}
	\Delta^2 \widehat{\theta} \le \left(\frac{2 \pi}{3} \right)^2 \left( 1 + 64 K A C^{-\nu_K} \right) \frac{1}{4^K} \; ,
	\label{eq:newPrecisionUpperBound}
\end{equation}
and the resource resummation reads
\begin{equation}
	M = K \left( \nu_K - \frac{1}{\log_2 C} \right) + \frac{K^2}{\log_2 C} \; .
	\label{eq:newResourceResummation}
\end{equation}
We now extract $\nu_K$ as a function of $M$ and $K$ from \cref{eq:newResourceResummation}, and substitute it in \cref{eq:newPrecisionUpperBound} to get
\begin{equation}
	\Delta^2 \widehat{\omega} \le \left( \frac{2 \pi}{3} \right)^2 \left( \frac{1}{4^K} + 32 K A \, \frac{2^{-\frac{M \log_2 C}{K}}}{2^K} \right) \; .
	\label{eq:elaboratedBound}
\end{equation}
Observe that $ 2^{-\frac{M \log_2 C}{K}} < 1 \, , \, \forall \, K \ge 0$ and that the right hand side of \cref{eq:elaboratedBound} tends to zero for $K \rightarrow \infty$, because the term $K$ at numerator of $32 K A \, \frac{2^{-\log_2 C \frac{M}{K}}}{2^K}$ cannot compensate the exponential at denominator. As $K \rightarrow \infty$ we have $\nu_K \rightarrow 0$, but this is a non-physical solution because it must be $\nu_K \ge 1$. Therefore the maximum allowed $K$ ($:= K^\star$) corresponds to $\nu_{K^{\star}} = 1$. In this case, the resource resummation reads
\begin{equation}
	M = K^\star \left( 1 - \frac{1}{\log_b C} \right) + \frac{{K^\star}^2}{\log_b C} \; .
\end{equation}
For $K^\star \gg 1$ the above equation gives
\begin{equation}
	K^\star = \sqrt{\log_2 C} \sqrt{M} \; .
\end{equation}
This result led us to the choice of the prefactor $h$ that corresponds the maximum $\tau$. According to this derivation, neglecting the factor $\sqrt{\log_2 C}$, which is of order one, we get the maximum free evolution time $\tau_{K^{\star}} \simeq 2^{\sqrt{M}}$. Inserting $K^\star$ in \cref{eq:elaboratedBound} and neglecting the subleading terms we arrive at the scaling
\begin{equation}
	\Delta^2 \widehat{\theta} \lessapprox \mathcal{O} \left( \frac{\sqrt{M}}{4^{\sqrt{\log_2 C} \sqrt{M}}} \right)\; .
	\label{eq:newScaling}
\end{equation}
This scaling for the precision is exponential in the square root of the number of measurements. If the free evolution time follows instead the power law $\tau_j = b^{j-1}$, then the corresponding Lagrangian for the optimization of the precision is
\begin{widetext}
	\begin{equation}
		\mathcal{L} := \left( \frac{\pi}{n \cdot b^{K-1}} \right)^2 + \sum_{j=1}^K \left( \frac{2 \pi }{n \cdot b^{j-2}} \frac{b}{b-1}\right)^2 A C^{-\nu_j} +\lambda \left(2 \sum_{j=1}^K \nu_j - M \right) \; ,
	\end{equation}
\end{widetext}
with $n = b + 1$. The optimization with respect to $\nu_j$ gives the expression
\begin{equation}
	\nu_j = \nu_K + \frac{2}{\log_b C} \left( K - j \right) \, ,
\end{equation}
and the resource resummation is
\begin{equation}
	M = K \left( \nu_K - \frac{1}{\log_b C} \right) + \frac{K^2}{\log_b C} \; .
	\label{eq:newResourceResummationb}
\end{equation}
The same analysis done before give the precision scaling
\begin{equation}
	\Delta^2 \widehat{\theta} \lessapprox \mathcal{O} \left( \frac{\sqrt{M}}{b^{2 \sqrt{\log_b C} \sqrt{M}}} \right)\; .
	\label{eq:newScalingb}
\end{equation}
We wish to emphasis that this is an achievable scaling of the precision for a fixed number of measurements $M$, which has to be compared with the unachievable lower bound on the precision, obtained through a bit-counting argument~\cite{fiderer_neural-network_2021}:
\begin{equation}
	\Delta^2 \widehat{\theta} \ge \frac{2^{-2(M+1)}}{3} \; ,
\end{equation}
which has been used as a lower bound in \cref{sec:nv_center_bound}. This bound is computed assuming that each measurement gives exactly $1$ bit of information about the phase $\theta$. In summary, while the precision lower bound for this counting of the resources scales as $\mathcal{O} (4^{-N})$, we have been able to prove the achievability of the (much worst) $\mathcal{O} (\sqrt{N} 4^{-\sqrt{N}})$ scaling.

\end{document}